\documentclass[a4paper,fleqn,usenatbib]{mnras}

\renewcommand{\footnoterule}{%
  \kern -24pt
  \hrule width 2in
  \kern 2.6pt
}

\usepackage{mathptmx}
\usepackage[T1]{fontenc}
\usepackage{ae,aecompl}

\usepackage{tikz,xcolor,hyperref}

\definecolor{lime}{HTML}{A6CE39}
\DeclareRobustCommand{\orcidicon}{%
	\begin{tikzpicture}
	\draw[lime, fill=lime] (0,0) 
	circle [radius=0.16] 
	node[white] {{\fontfamily{qag}\selectfont \tiny ID}};
	\draw[white, fill=white] (-0.0625,0.095) 
	circle [radius=0.007];
	\end{tikzpicture}
	\hspace{-4mm}
}

\foreach \x in {A, ..., Z}{%
	\expandafter\xdef\csname orcid\x\endcsname{\noexpand\href{https://orcid.org/\csname orcidauthor\x\endcsname}{\noexpand\orcidicon}}
}

\usepackage{graphicx}	% Including figure files
\usepackage{amsmath}	% Advanced maths commands
\usepackage{amssymb}	% Extra maths symbols
\usepackage{scrextend}  % For having repeated footnotes
\usepackage{microtype}  % To avoid underfull hboxes (and also overfull ones). It allows TeX to get acceptable whitespace in lines more often.
\usepackage{subfig}     % Figures that span multiple pages
\usepackage{longtable}
\title[SF in perturbed galaxies: SFHs and O Abundances]{Star Formation in CALIFA survey perturbed galaxies.\\
II. Star Formation Histories and Oxygen Abundances}

\author[A. Morales-Vargas et al.]{
A. Morales-Vargas\orcidA{}$^{1}$\thanks{E-mail: abdmoralesv@gmail.com (AMV)},
J. P. Torres-Papaqui\orcidB{}$^{1}$,
F. F. Rosales-Ortega\orcidC{}$^{2}$,
\
\newauthor
M. Chow-Mart\'{i}nez\orcidE{}$^{3}$,
J. J. Trejo-Alonso\orcidF{}$^{4}$,
R. A. Ortega-Minakata\orcidG{}$^{5}$,
\
\newauthor
A. C. Robleto-Or\'{u}s\orcidH{}$^{1}$,
F. J. Romero-Cruz\orcidI{}$^{6}$,
D. M. Neri-Larios\orcidJ{}$^{7}$
\
\newauthor
\& the CALIFA survey Collaboration
\\
$^{1}$Departamento de Astronom\'\i{}a, Universidad de Guanajuato, Apartado Postal 144, Guanajuato 36000, Mexico\\
$^{2}$Instituto Nacional de Astrof\'\i{}sica, \'{O}ptica y Electr\'{o}nica, Luis Enrique Erro 1, Tonantzintla 72840, Mexico\\
$^{3}$Instituto de Geolog\'{i}a y Geof\'{i}sica, Universidad Nacional Aut\'{o}noma de Nicaragua, Rotonda Universitaria\\
$^{\phantom{0}}$Rigoberto L\'{o}pez P\'{e}rez 150 metros al Este, Managua 663, Nicaragua\\
$^{4}$Facultad de Ingenier\'{i}a, Universidad Aut\'{o}noma de Quer\'{e}taro, Cerro de las Campanas s/n, Centro Universitario,\\ 
$^{\phantom{0}}$Santiago de Quer\'{e}taro 76010, Mexico\\
$^{5}$Instituto de Radioastronom\'{i}a y Astrof\'{i}sica (IRyA), UNAM, Apartado Postal 72-3, Morelia, Michoac\'{a}n 58089, Mexico\\
$^{6}$Instituto Tecnol\'{o}gico Superior de Guanajuato, Guanajuato 36262, Mexico\\
$^{7}$School of Physics, The University of Melbourne, Parkville, VIC. 3010, Australia\\
}

\date{Accepted XXX. Received YYY; in original form ZZZ}

\pubyear{2021}

\begin{document}
\label{firstpage}
\pagerange{\pageref{firstpage}--\pageref{lastpage}}
\maketitle

\begin{abstract}

Galaxy evolution is generally affected by tidal interactions. Firstly, in this series, we reported several effects which suggest 
that tidal interactions contribute to regulating star formation (SF). To confirm that so, we now compare stellar mass assembly histories and SF look-back 
time annular profiles between CALIFA survey \textit{tidally} and \textit{non-tidally perturbed} galaxies. We pair their respective star-forming regions 
at the closest stellar mass surface densities to reduce the influence of stellar mass. The assembly histories and annular profiles show statistically 
significant differences so that higher star formation rates characterize regions in tidally perturbed galaxies. These regions underwent a more intense 
(re)activation of SF in the last 1 Gyr. Varying shapes of the annular profiles also reflect fluctuations between suppression and (re)activation of SF. 
Since gas-phase abundances use to be lower in more actively than in less actively star-forming galaxies, we further explore the plausible presence of 
metal-poor gas inflows able to dilute such abundances. The resolved relations of oxygen (O) abundance, with stellar mass density and with total gas 
fraction, show slightly lower O abundances for regions in tidally perturbed galaxies. The single distributions of O abundances statistically validate 
that so. Moreover, from a metallicity model based on stellar feedback, the \textit{mass rate differentials} (inflows$-$outflows) show statistically 
valid higher values for regions in tidally perturbed galaxies. These differentials, and the metal fractions from the population synthesis, suggest 
dominant gas inflows in these galaxies. This dominance, and the differences in SF through time, confirm the previously reported effects of tidal 
interactions on SF.

\end{abstract}\textit{}%%%%%%%%%%%%%%%%%%%%%IMPORTANT COMMENT: TO OUR PUBLISHER, FOR AN UNKNOWN REASON, THE FILE DOES NOT COMPILE IN THE ABSENCE OF A COMMAND 
% RIGHT AFTER \end{abstract}. SORRY FOR THIS ISSUE%%%%%%%%%%%%%%%%%%%%%%%%%%%%%%%%%%%%%%%%%%%%%%%%%%%%%%%%%%%%%%%%%%%%%%%%%%%%%%%%%%%%%%%%%%%%%%%%%%%%%%%%%%%


%%%%%%%%%%%%%%%%%%%%%%%%%%%%%%%%%%%%%%%%%%%%%%%%%%

%%%%%%%%%%%%%%%%% BODY OF PAPER %%%%%%%%%%%%%%%%%%

\section{Introduction}
\label{sec:int}

In galactic evolution, the Star Formation History (SFH) is the standard to portray over time mass assembly through Star Formation (SF) \citep[\textit{e.g.}]
[]{Tin72,SeSaBa73,GaHuTu84,IssSch86,Bica88,KenTamCong94,Cid01,Pan03,Ocv06,MatChaBri06,Pan07,Cid07,Asa07,Toj11,Pla12,Tor12,Tor13,McD15,God17a,Gon17,Lop18,
Row18,San18,Bell20,Pet20,Pet21}. The extraction of SFHs uses Initial Mass Function (IMF) prescriptions, stellar population (SP) libraries and evolutionary 
tracks, all basic elements of simple SP models \citep[\textit{e.g.}][]{Wal11}. 

In early extractions of SFHs, the time resolutions of stellar libraries were much lower than those of data \citep{Asa07}. Therefore, \citet{Bica86} reduced 
the variables to only age and metallicity, enabling the test of large star cluster combinations. This act quickened the advent of improved spectral-resolution 
stellar libraries and revised evolutionary tracks as well \citep[both integrated into simple SP models, see][]{Lei99,BrCh03,Mar05,
Gon05,SaB06,Vaz07,Vaz10,Fac11,MaSt11,Vaz16}. Also, the ability to fit spectra of galaxies (spectral synthesis) was refined. Robust synthesis algorithms 
plus resolved SP models compose SP synthesis codes such as \textsc{moped} \citep{Hea00}, \textsc{starlight} \citep{Cid05,Cid13}, 
\textsc{firefly} \citep{Wil15} and \textsc{pipe3d} \citep{San16b,San16c} to only mention a few. Their use has contributed to our knowledge of galaxy 
evolution as follows.

\textsc{moped} integrates the \citet{BrCh03} models and a \citet{Sal55} IMF. By its use, \citet{Pan07} find a 
Star Formation Rate (SFR) with no conclusive peak out to redshift ($z$) $\sim$2, a mass assembly history in line with high-$z$ 
studies and galaxy formation and evolution models \citep[\textit{e.g.}][]{Cro06}, and a robust evidence of \textit{downsizing} 
independent of SP models.

\textsc{starlight} incorporates three sets of SP models \citep[see][]{Cid14,Gon14a,Gon15}. It uses either \citet{Sal55} or \citet{Chab03} IMFs. For star-forming 
galaxies (SFGs), \citet{Cid07} firstly confirmed that massive galaxies are the first ones to form most of their stars whereas less massive galaxies form theirs 
much slower \citep{Hea04}. \citet{Asa07} show that low gas-phase metallicity systems evolve much slower than high gas-phase metallicity ones. Using \textsc{starlight}, 
\citet{Per13}, \citet{SaB14} and \citet{Gon14a,Gon15} show that, regardless of morphology, galaxies assemble their stellar masses (M$_{*}$) from inside-out. From 
radial distributions of SP properties, \citet{Gon14a,Gon15} show that massive galaxies are more compact, metal-richer, older and less dusty. For disk galaxies, 
\citet{Gon14a,Gon14b} find that the stellar mass surface density ($\Sigma_{*}$) is intrinsically related to the SFH and metal enrichment. A relatively long declining 
phase of SF in Ss, and a long active phase of growth for Es/S0s, are also \textsc{starlight}-based results of \citet{Gon17}. \citet{Gar17} show that downsizing 
not only depends on M$_{*}$ but also on galaxy morphology. \textsc{starlight} assists \citet{Lop18} to find the average mass assembly of CALIFA survey galaxies 
well obeying the SF time scale. \citet{Pet20,Pet21} show that Ss have grown more notably in light than in M$_{*}$ and that the current colour (rather than the 
actual morphology of galaxies) is more reliable on describing downsizing and the SFHs.

Made for the SDSS-IV MaNGA survey \citep{Bun15}, \textsc{firefly} incorporates models of \citet{MaSt11} and a \citet{Kro01} IMF. In contrast to \citet{Zhe17}, 
\citet{God17a,God17b} find early types in an outside-in progression of SF (positive age gradients). Mergers are likely the cause since it is assumed that
they rearrange the radial distributions of SPs. However, mergers might play a relatively small role in shaping the steeper metallicity gradients of MaNGA survey 
galaxies.

Using \textsc{pipe3d}, with a \citet{Sal55} IMF and models of \citet{Gon05} and \citet{Vaz10}, \citet{Ib16} confirm that the main driver of the global SFH is 
M$_{*}$ and that early types at earlier epochs seem to be in outside-in SF mode. Either enhancement or suppression, \citet{Ell18} show that SF is regulated by 
changes in the centres of galaxies. \citet{San18} use \textsc{pipe3d} to find the cosmic quenching of SF more complex than a one-event process. At different 
spatial regimes, \citet{Ib16} and \citet{San20} find that more massive galaxies show very similar resolved SFHs whereas less massive ones show a larger variety. 
\citet{CF21} find a similar trend for the chemical enrichment histories. Finally, using \textsc{pipe3d}, \citet{MN20} show that a large variance in the chemical 
structure of galaxies is due to a complex composition of the SPs.

On the other hand, the gas-phase metal content of galaxies, intrinsically related to their formation and evolution, has been studied through the mass-metallicity 
relation \citep[MZR, \textit{e.g.}][]{Ski92,Tre04,Ros12,San13,Zah14,Gon14b,BeMaBo16,Wu16,WuP17,San17,BaBa18}. \citet{FD08} show that the evolution of the MZR, and its 
characterizing parameters as well, are constraints on the behavior of stellar winds or outflows (typically modeled by the escape/virial velocity). These, the cold gas 
content and SF are common properties in models of metal evolution \citep[\textit{e.g.}][]{TL78,DS86,OD06,Lil13}. In simulations of interacting 
galaxies, dilution due to gas inflows and enrichment due to SF modulate the central metallicity whereas gas consumption and outflows play a secondary 
role (\citealt{Torr12}). Moreover, outflows and inflows balance more shallowly the metal content in high-$z$ galaxies than if adopting closed-box models (\citealt{FD08}).

In spatially-resolved regions, the metallicity mostly depends on the local content of molecular and atomic gas \citep[\textit{e.g.}][]{Mor12,Car15,Ho15,BaBa18}. More 
studies \citep[\textit{e.g.}][]{Ros12,San13,San17,BaBa18} suggest that also the $\Sigma_{*}$ regulate the local processes that are responsible 
for gas-phase metal unbalances. This is suggested even though outflows (with their strength correlated with the intensity/efficiency of SF) 
are effective removers of metals, especially oxygen \citep[\textit{e.g.}][]{Tre04,Tor12,WuZha13,BeMaBo16,Vic16}. In contrast, shallow metallicity 
gradients, consistent with gas inflows, seem to be an average characteristic of interacting galaxies (see \citealt{Mor15} and references therein).

We connect all this background with galaxy-pair interactions. We first compare the annular star formation and specific star formation histories of non-tidally 
and tidally perturbed galaxies. It then follows, between the same galaxies, a comparison of their look-back time annular profiles of SFR intensity ($\Sigma_{\mathrm{SFR}}$, 
M$_{\odot}$\,yr$^{-1}$\,kpc$^{-2}$) and specific SFR (sSFR, $\Sigma_{\mathrm{SFR}}\,\Sigma_{*}^{-1}$, yr$^{-1}$). We expect that both, the histories and profiles, 
support the tidal interactions as contributors to regulating the SF in tidally perturbed objects (see \citealt{MorSub}, hereafter PaperI). Additionally, we look 
for proof of metal-poor gas inflows induced by tidal interactions (diluted O/H metallicities). To do so, since oxygen (O) is tightly related to recent SF, we 
conduct a comparison of the local relations of O abundance with both the $\Sigma_{*}$ and total fraction of gas.

For proper management, we use integral-field spectra synthesized by \textsc{starlight}. These are from the Calar Alto Legacy Integral Field Area 
\citep[CALIFA,][]{San12,Hus13,Gar15,San16a} survey. The CALIFA survey favourably presents the best compromise among near-by coverage ($0.005\leq\,z\,\leq0.03$), 
spatial coverage (mostly beyond 2.5 effective radii), spatial resolution ($\sim1\,$kpc), number of targets ($>\,600$), and target sampling ($\sim4\,000$ 
spectra per target).

We order this paper as follows. Section~\ref{sec:sameth} describes the methods and sample selection. Section~\ref{sec:res} shows the contrasts 
in all annular SFHs, look-back time profiles, and O abundances. We discuss these results in Section~\ref{sec:dis}. Section~\ref{sec:conc} states 
our summary and conclusions.

The cosmological set of $\mathrm{H_{0}}\,=72\,\mathrm{h_{72}^{-1}}\,\mathrm{km\,s^{-1}\,Mpc^{-1}}$, $\Omega_\mathrm{M}\,=\,0.3$ and $\Omega_\mathrm{\Lambda}\,=\,0.7$; 
a \citet{Chab03} IMF; and a 0.05 level of significance for statistics are all used throughout this work.

\section{Methods and sample selection}
\label{sec:sameth}

Among the contents of this Section, we briefly describe the selection of the star-forming regions/spaxels 
and the galaxy samples (Sections~\ref{subsubsec:syn} to \ref{subsubsec:sf} and \ref{subsec:gal} respectively). Please refer to PaperI for more details on these topics.

\subsection{Methods}
\label{subsec:met}

\subsubsection{Stellar component subtraction and emission features}
\label{subsubsec:syn}

We use \textsc{starlight} \citep{Cid05} to subtract syntheses of the stellar component from the observations. \textsc{starlight} is used with $j=$\,1, ..., 150 simple 
SPs (25 stellar ages combined with 6 metallicities) from an update of the \citet{BrCh03} models. That update is based on the Medium-resolution Isaac-Newton telescope 
Library of Empirical Spectra \citep[MILES,][]{SaB06,Fac11} with a \citet{Chab03} IMF and a \citet{Car89} extinction law ($R_{V}\,=\,3.1$). Then, on the almost pure 
nebular spectra left from the subtractions, we adapt Gaussian profiles to the emission lines of interest via a $\chi^{2}$ iteration. A variety of features are computed 
including line fluxes and uncertainties. For the mass assembly histories, we now use the mass-fraction, smoothed vector, $\mu_{j,s}$, result of synthesizing the stellar 
component. $\mu_{j,s}$ contains the model light-to-mass ratios at the SP age smoothed vector, $t_{j,s}$ (see \citealt{Cid05}, \citealt{Asa07} hereafter 
A-07).\footnote{$s$ subscript indicates the smoothed population vector (resampled in time resolution, 50 bins in this work). Such a vector reduces the degeneracies 
of the original one while retains the time-dependent information of the latter (see A-07 and references therein).} 

\subsubsection{Galaxy morphologies and galaxy colours}
\label{subsubsec:m-c}

Morphologies result from a reclassification by \citet[][hereafter W-14]{Wal14} on all CALIFA survey Mother Sample (MS, the candidates for the survey observations) 
galaxies. Galaxy colours result from colour magnitude diagrams (CMDs) which use SDSS/DR7 \citep{Aba09} photometry and symbolic cut-offs for both the 
\emph{red sequence} and \emph{blue cloud}.

\subsubsection{Star-forming regions, \textsc{starlight} output and annular profiles}
\label{subsubsec:sf}

Current star-forming regions are selected based on:

\begin{enumerate}
 \item the detectability of the H$\alpha$ emission line (S/N\,$\geq$\,3),
 \item its intensity (EW (H$\alpha$)\,$\geq$\,6\,\AA{}, which characterizes strongly ionizing young SPs, \citealt{Cid10} and \citealt{San14}),
 \item and the comparison of line ratios on the BPT \citep{BPT81} diagram (considering only spaxels with SFG gas excitation type).
\end{enumerate}

SFRs are computed only for star-forming regions. These resolved rates are measurements of $\mathrm{\Sigma_{SFR}}$. Each spaxel has an angular surface of 
1\,arcsec$^{2}$ and the correcting factor to linear surface (kpc$^{2}$) is computed per galaxy (Hubble flow). 

The \textsc{starlight} output gives total stellar masses and mean SP ages for each region independent of its type of gas excitation (BPT diagram). The sum 
of single contributions and the median of all regions are, respectively, the global M$_{*}$ and SP median age of each galaxy. Likewise SFRs, the M$_{*}$ of 
each region is a measurement of $\mathrm{\Sigma}_{*}$.

To annularly profile the properties of the star-forming regions, we use deprojected radii enclosing fixed percentages (20, 40, 60, 80 and 100\,\%) of the 
all-excitation H$\alpha$ flux in each galaxy. The boundaries of these radii delimit the annuli which are named as the 
percentages. Moreover, we control the influence of M$_{*}$ when conducting the comparisons of the star-forming region properties between tidally 
perturbed and non-tidally perturbed galaxies. Between them, we pair star-forming spaxels against one another by minimizing their $\mathrm{\Sigma}_{*}$ 
differences. See Appendix B of PaperI for more details. 

\subsubsection{Oxygen abundances}
\label{subsubsec:che}

To look for deviations from the MZR that may reveal additions of low metallicity gas to galaxies \citep[\textit{e.g.}][]{Tre04,Ros12,San13,BaBa16,Row18}, we 
estimate gas-phase O abundances for all star-forming regions. We follow the \textit{indirect} relation of \citet{Mar13}. They obtain O abundances 
for 3\,423 H\,\textsc{ii} regions \citep{San13} by using the \citet{Pil10}-ONS calibration\footnote{That calibration uses flux ratios of strong lines as 
surrogate indicators of the electron temperature (T$_{e}$).}. \citet{Mar13} find for these abundances and their respective $O3N2$ indexes (see their equation 1) 
the best linear relation of  
\begin{equation} \label{eq:2}
 12 + \mathrm{log}_{10}\mathrm{(O/H)} = 8.505[\pm0.001] - 0.221[\pm0.004] \times O3N2.
\bigskip
\end{equation}

Equation~\ref{eq:2} is our best option due to the following. First, its range of application which avoids the high metallicities 
where the [N\,\textsc{ii}]$\lambda$6583/H$\alpha$ ratio saturates (\textit{i.e.}, $O3N2 < -$0.2\footnote{In this work, fewer $O3N2$ 
measurements (9.5\,\%) obey that condition. Though the T$_{e}$-based calibration of 
\citet{Mar13} is more reliable below that regime, they find their T$_{e}$ and the \citet{Pil10}-ONS 
based abundances well correlated ($\sigma\,=\,$0.02 dex).}). Second, the fact of potential changes in the $O3N2$ 
index between tidally/non-tidally perturbed galaxies. Third, the wide ionization range that characterizes the above 
amount of H\,\textsc{ii} regions and its statistical significance against the calibration of \citet{Mar13} (the 
low T$_{e}$ measurements in a broad range of O abundances). Finally, if the abundance estimations 
use no mixture of methods, \citet{Mar13} recommend the ONS calibration over theirs.

To maintain the accuracy of Eq.~\ref{eq:2} against direct methods, we correct for extinction the emission line fluxes of the $O3N2$ indexes. Following 
the \citet{Car89} extinction law, we obtain $A(H\alpha)\,=\,0.818\,A(V)$, $A([N\,\textsc{ii}])\,=\,0.815\,A(V)$, $A([O\,\textsc{iii}])\,=\,1.120\,A(V)$ 
and $A(H\beta)\,=\,1.164\,A(V)$. As we have used already the Balmer decrement plus this extinction law to estimate $A(H\alpha)$ (PaperI), the set 
\begin{align} \label{eq:3}
 \begin{split}
 A([N\,\textsc{ii}])\,=\,0.996\,A(H\alpha)
 \\
 A([O\,\textsc{iii}])\,=\,1.369\,A(H\alpha)
 \\
 A(H\beta)\,=\,1.423\,A(H\alpha)
 \end{split}
\bigskip
\end{align}

\noindent allows us to estimate extinctions for these lines on a spaxel-by-spaxel basis.

\subsubsection{Gas mass surface densities \& gas fractions}
\label{subsubsec:gas}

We estimate total gas mass surface densities ($\Sigma\mathrm{_{gas}}\,=\,\Sigma\mathrm{_{H_{2}}}+\Sigma\mathrm{_{H_{\textsc{i}}}}$) by following \citet[][hereafter 
BB-18]{BaBa18} and \citet{Row18}. Based on a model which describes a correlation between effective extinction and molecular gas column density, BB-18 calibrate the 
relation
\begin{equation} \label{eq:4}
 \Sigma\mathrm{_{gas}} = 30 \left( \frac{A(H\alpha)/0.818}{\mathrm{mag}} \right)\,\,\,\mathrm{M}_{\odot}\,\mathrm{pc}^{-2},
\bigskip
\end{equation}

\noindent with CO interferometric\footnote{The EDGE-CALIFA survey, \citet{Bol17}.} and $A(V)$ measurements. They also assume a constant distribution of neutral gas.

Moreover, the fraction of total gas is defined as $\mu=\Sigma\mathrm{_{gas}}/(\Sigma\mathrm{_{gas}}+\Sigma_{*})$. However, an issue emerges for the cases where we 
assumed no extinction (see PaperI, equation 2). Since $\Sigma\mathrm{_{gas}}$ is zero, we use a $\mu$ of practically zero for such cases (\textit{i.e.} 1/100\,000). 

\subsection{Sample selection}
\label{subsec:gal}

\subsubsection{The tidal perturbation parameter}
\label{subsubsec:f}

The tidal perturbation parameter, \textit{f} \citep{Byrd86}, measures the tidal force exerted by a galaxy on the outskirts of another one. This parameter defines the 
isolation criterion: non-tidally-perturbed galaxies take effect if \textit{f} is\,$<\,-4.5$ \citep{Var04}. Tidally-perturbed galaxies are assigned otherwise. 
W-14 looked for neighbours of CALIFA survey MS galaxies and estimated their \textit{f} parameters.

\subsubsection{The sample selection}
\label{subsubsec:selec}

A total of 529 CALIFA survey objects belong to the MS and their spectra show the broadest wavelength range. From them, 454 have \textit{f} estimations (W-14). 
By the isolation criterion, 101 objects are non-tidally-perturbed while 353 are tidally-perturbed. The respective fractions 62/101 and 231/353 are classified 
as Emission Line Galaxies\footnote{ELGs since their spectra passed the emission line criteria of PaperI.} and contain star-forming regions (Section~\ref{subsubsec:sf}). 
The few non-tidally-perturbed ELGs, hereafter the \emph{control} sample, led to construct, from the 231 perturbed ELGs, ten sets of 62 objects each. These sets, 
hereafter referred to as the \emph{perturbed} samples (labelled \emph{tA} to \emph{tJ} in PaperI), closely resemble five fundamental properties of the control 
sample (M$_{*}$, $z$, morphological group, galaxy colour and dominant excitation source). By selecting the best candidate in the five properties for each single 
control object (avoiding common cases when possible), the perturbed samples are simultaneously filled-in. The net fraction of perturbed ELGs used is 162/231 
(70\,\%, see PaperI).

\subsection{Preliminaries}
\label{subsec:validation}

To avoid observational constraints in the data, this work ignores low $\Sigma_{*}$ star-forming regions (log$_{10}\,\Sigma_{*}\,<$\,7.5\,M$_{\odot}$\,kpc$^{-2}$) 
since they deviate from the resolved Star Formation Main Sequence \citep[see][]{Hal18,EF19,Can19}.

\subsubsection{SFR intensities: synthesis against H$\alpha$ emission}
\label{subsubsec:synVSHa}

\begin{figure*}\centering
   \mbox{\includegraphics[width=.865625\columnwidth]{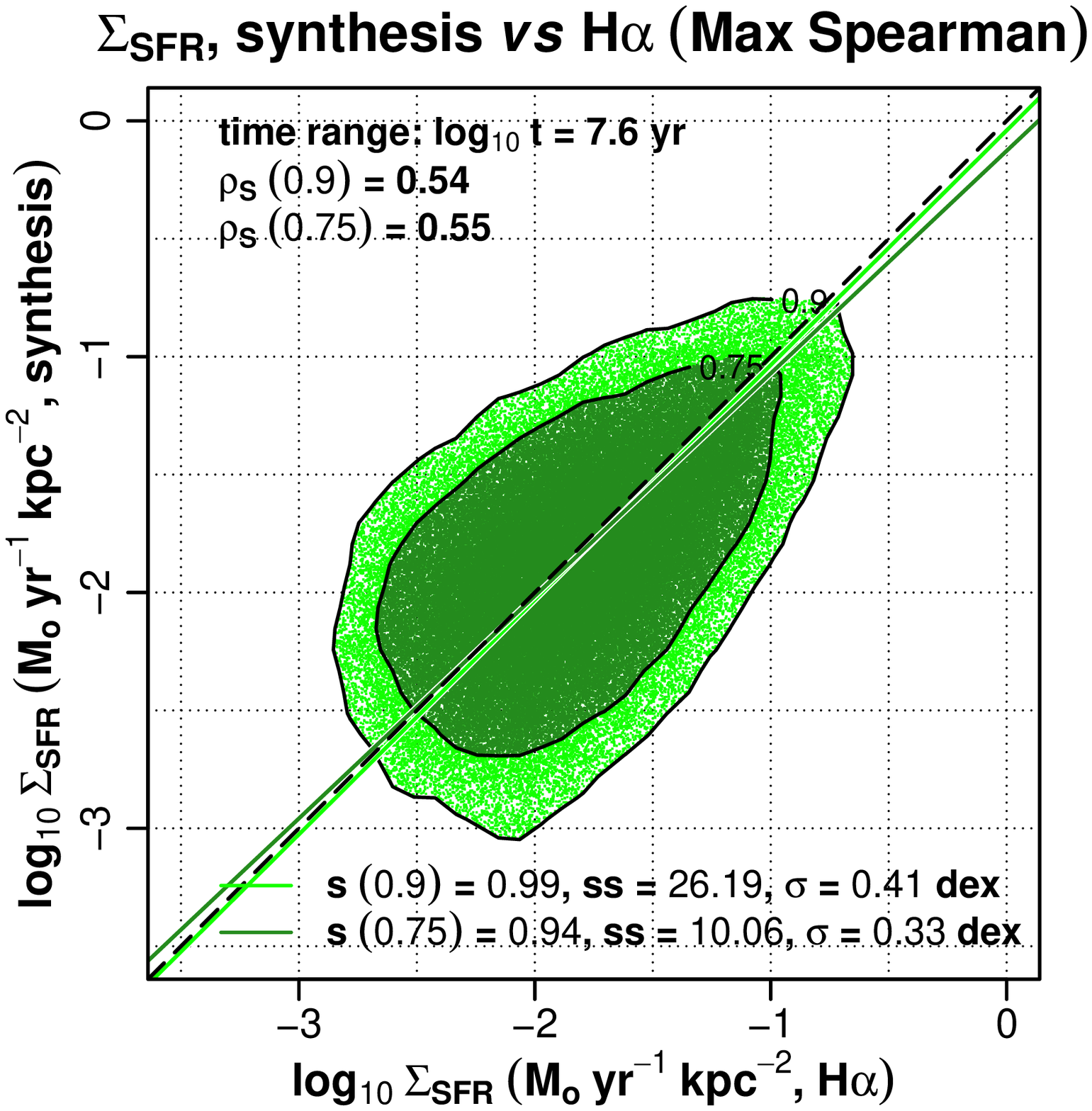}}
   \mbox{\includegraphics[width=.865625\columnwidth]{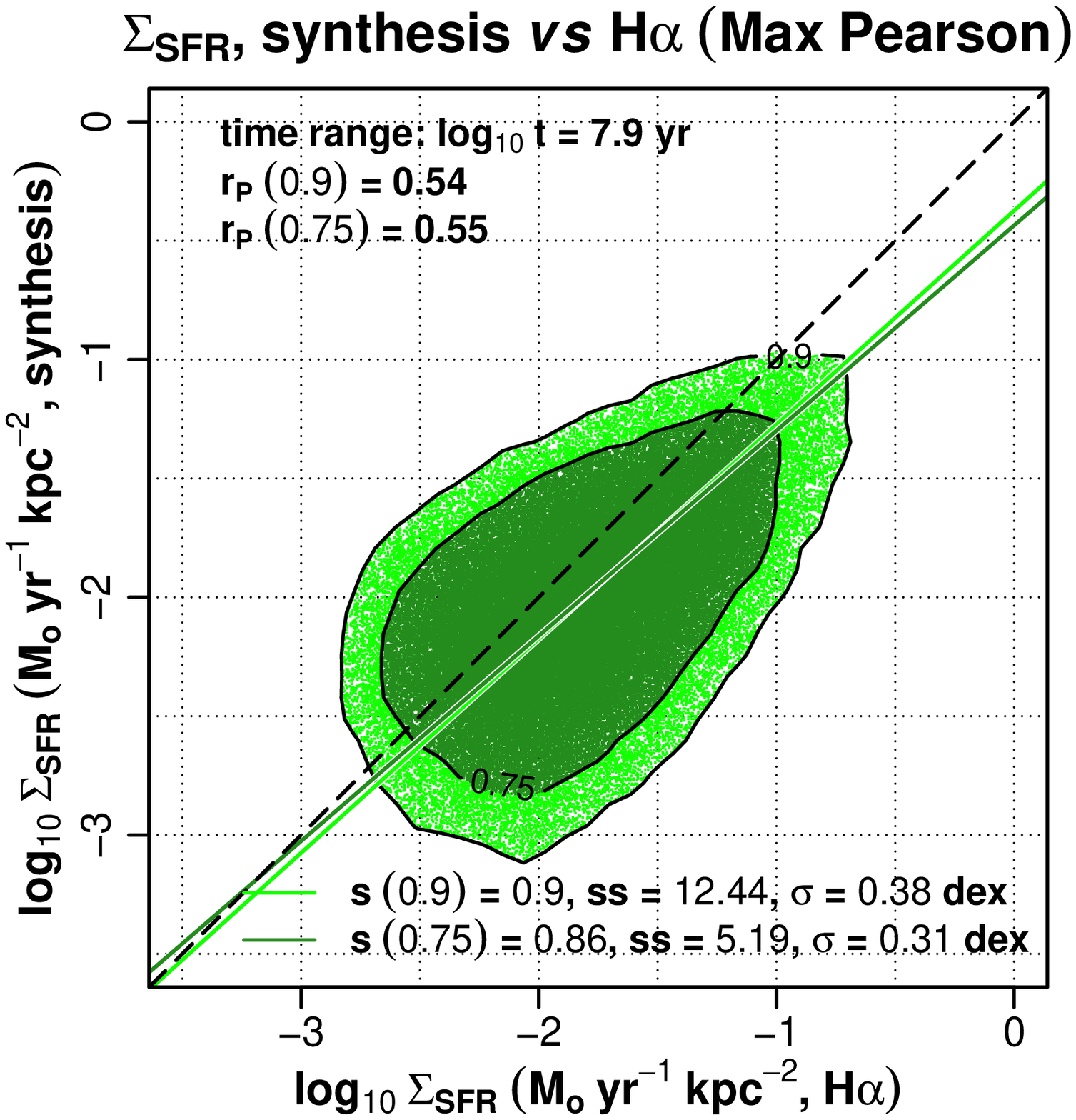}}
   \caption{\scriptsize{The synthesis-based $\Sigma\mathrm{_{SFR}}$ against the H$\alpha$ line-based one for 111\,292 star-forming regions 
   from the 224 galaxies that conform our samples (see Paper1, section 3.3). The synthesis-based values are mean rates in the last log$_{10}$\,t\,$=$\,7.6 
   and 7.9\,yr (left and right, see also Table~\ref{tab:1}). These are the time periods on which the maximum correlations fall according to Spearman 
   ($\rho_{\mathrm{S}}$) and Pearson (r$_{\mathrm{P}}$) (left and right too). The H$\alpha$ line-based $\Sigma\mathrm{_{SFR}}$ follows the conversion of A-07 
   (their equation 9). Contours enclosing 90 and 75\,\% of the data pairs (soft and dark-green, respectively) are shown to improve the tightness along the fits. 
   Orthogonal distance regression is used in this figure since neither predictor nor response variables must be assigned. \textit{s} stands for the slope and 
   \textit{ss} for the sum of squares of orthogonal distances. $\sigma$ indicates the dispersion along each fit. Finally, the dashed line represents the one-to-one 
   relation.}}
   \label{f1} 
\end{figure*}

Similarly to previous works \citep[\textit{e.g.} A-07;][]{Gon16,San18,Riff20}, we compare both SFRs, that which uses the H$\alpha$ line emission as a tracer, and 
that one derived from the synthesis of the stellar component. The presence of massive-young stars characterizes the  H$\alpha$ line-based SFR. That from the stellar 
component is more emblematic because it comes from the light of all stars. Both may be dubbed ``current'', \textit{i.e.}, the former is valid during the net ionization 
time of young stars whereas the latter represents a regular mean over a defined time interval (A-07). 

Our samples and methods differ from those of A-07, \citet{San18} and \citet{Riff20}. First, A-07 use 
both continuum and emission-line S/N ratio thresholds higher than the ones of ours. 
Moreover, their spectra are limited to within the 3-arcsec-SDSS fibre whereas ours span to at least 2 times the 
SDSS Petrosian half-light radius. Second, \citet{San18} implement a spatial binning and tessellation method to increase the S/N of the data to $\geq$ 50 whereas 
our SP synthesis employs a continuum S/N$\,\geq\,$5\footnote{Considering the differences in the SP templates and their assigned weights, \citet{Pet20} show that 
\textsc{starlight} retrieves reliable SFHs from spectra with S/N\,$>$\,5.}. Moreover, they do not consider the ionization source in the spaxels to estimate the  
H$\alpha$ line-based SFR. And third, more than 1/3 of the galaxies analysed by \citet{Riff20} are pure Active Galactic Nuclei (AGNs) while the rest are SFGs and 
Low Ionization Emission Regions (the latter ionized by post-AGB stars). A fifth of our sampled galaxies are AGN-like while the rest are SFGs. Finally, $\sim$40\,\% 
of their galaxies are of early type whereas our fraction with this morphology is negligible. 

Figure~\ref{f1} shows our best correlations between SFR intensities for which we use the corresponding surface correcting 
factor for each galaxy (see~\ref{subsubsec:sf}). We estimate the stellar component $\Sigma_{\mathrm{SFR}}$ by following A-07 (their equations 7 and 10). 
From Fig.~\ref{f1} notice that, according to Spearman and Pearson coefficients, we find two peaks 
of correlation in the last log$_{10}$\,t\,$=$\,7.6 and 7.9 yr. Table~\ref{tab:1} gives basic statistics for the offsets between the two 
SFR intensities at these time periods. The peak in the last log$_{10}$\,t\,$=$\,7.6 yr has the closest-to-one slopes, in contrast to 
those of the peak in the last log$_{10}$\,t\,$=$\,7.9 yr. Moreover, the data pairs of the latter peak are clearly the tightest since the 
orthogonal distances (\textit{ss}) and the dispersions along the fits ($\sigma$) are the smallest. Regarding the distributions of the 
offsets between the SFR intensities (Table \ref{tab:1}), the peak in the last log$_{10}$\,t\,$=$\,7.6 yr has the best medians 
and 3rd quartiles (Qs), and the best rms values whereas its standard deviations ($\sigma$) are very similar to those deviations of the 
peak in the last log$_{10}$\,t\,$=$\,7.9 yr. It is also important to remark that both peaks fall within the 
period range of correlation with the maximum strength of A-07, and within the range of \citet{Gon16} and \citet{San18}. Both peak time ranges, $\sim$40 and 
$\sim$80 Myr, agree then time scales for recent SF. However, they are larger 
than the lifetime of massive ionizing stars. For instance, A-07 and \citet{Gon16} propose 24.5 and 32\,Myr, but also they suggest periods as 
long as 200 and 100 Myr respectively \citep[see also][]{San18}. Unless they refer to continuous SF bursts \citep[\textit{e.g.}][]{Tor12}, these 
timescales and the ones found here exceed the ionization time of massive-young stars. We therefore propose t$\mathrm{_{ionization}}$ $\sim$10 
Myr\footnote{From simulated SFHs of SFGs, \citet{FlV21} report a shorter average timescale for H$\alpha$ line-based SFR.} as the period for a 
valid-current SFR.

\begin{table}%[h]
   \setlength{\tabcolsep}{\tabcolsep}
 \begin{minipage}{\columnwidth}
 \captionsetup{width=\columnwidth}      
 \caption{\scriptsize{Basic statistics for the offsets between the synthesis-based and the H$\alpha$ line-based $\Sigma\mathrm{_{SFR}}$ at each 
 respective timescale peak.}
 \label{tab:1}}
 \centering
  \vspace{0.125cm} 
 \begin{scriptsize}
 \begin{tabular}{@{\hspace{0.1\tabcolsep}}l@{\hspace{2\tabcolsep}}c@{\hspace{2\tabcolsep}}c@{\hspace{2\tabcolsep}}c@{\hspace{2\tabcolsep}}c@{\hspace{2\tabcolsep}}c@{\hspace{2\tabcolsep}}}
 \hline
 $\mathrm{log_{10}}$ &\multicolumn{5}{c}{Offsets\footnote{\label{nota}\scriptsize{Offsets are the synthesis SFR intensities subtracted from the H$\alpha$ line ones.}} (dex)} \\
t\,(yr)             &1st Q      &med    &3rd Q  &rms   &$\mathrm{\sigma}$ \\
\hline
                    &\multicolumn{5}{c}{0.9}                              \\                  
7.6                 &$-$0.252   &0.037  &0.314  &0.407 &0.406             \\                  
7.9                 &$-$0.085   &0.198  &0.475  &0.444 &0.401             \\[1ex]             
                    &\multicolumn{5}{c}{0.75}                             \\     
7.6                 &$-$0.224   &0.032  &0.276  &0.342 &0.341             \\     
7.9                 &$-$0.061   &0.191  &0.437  &0.386 &0.339             \\[1ex]
\hline\\
 \end{tabular}
 \end{scriptsize}
 \end{minipage}
 \end{table}

\begin{figure}\centering
   \mbox{\includegraphics[width=.4944\columnwidth]{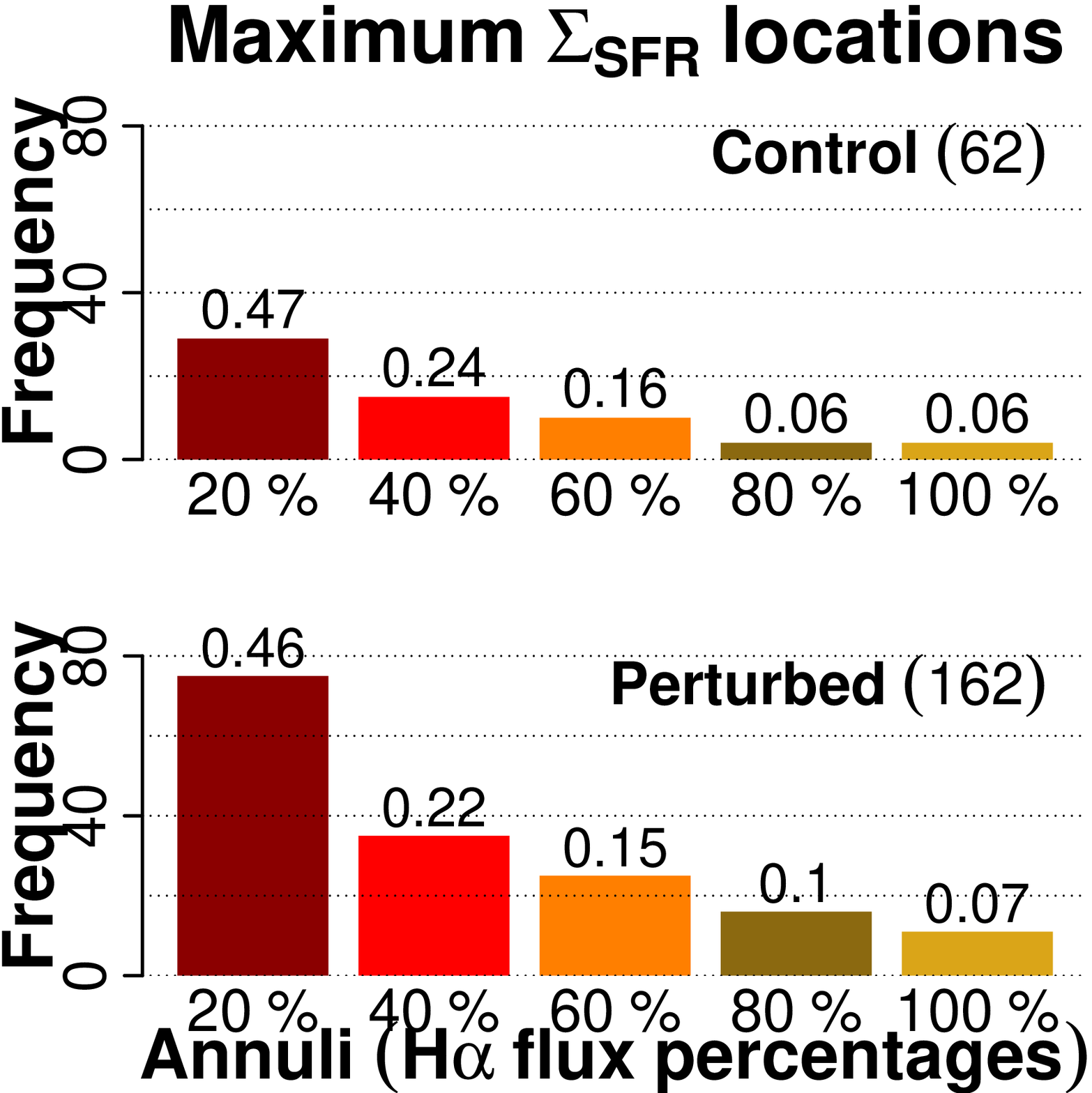}}
   \mbox{\includegraphics[width=.4944\columnwidth]{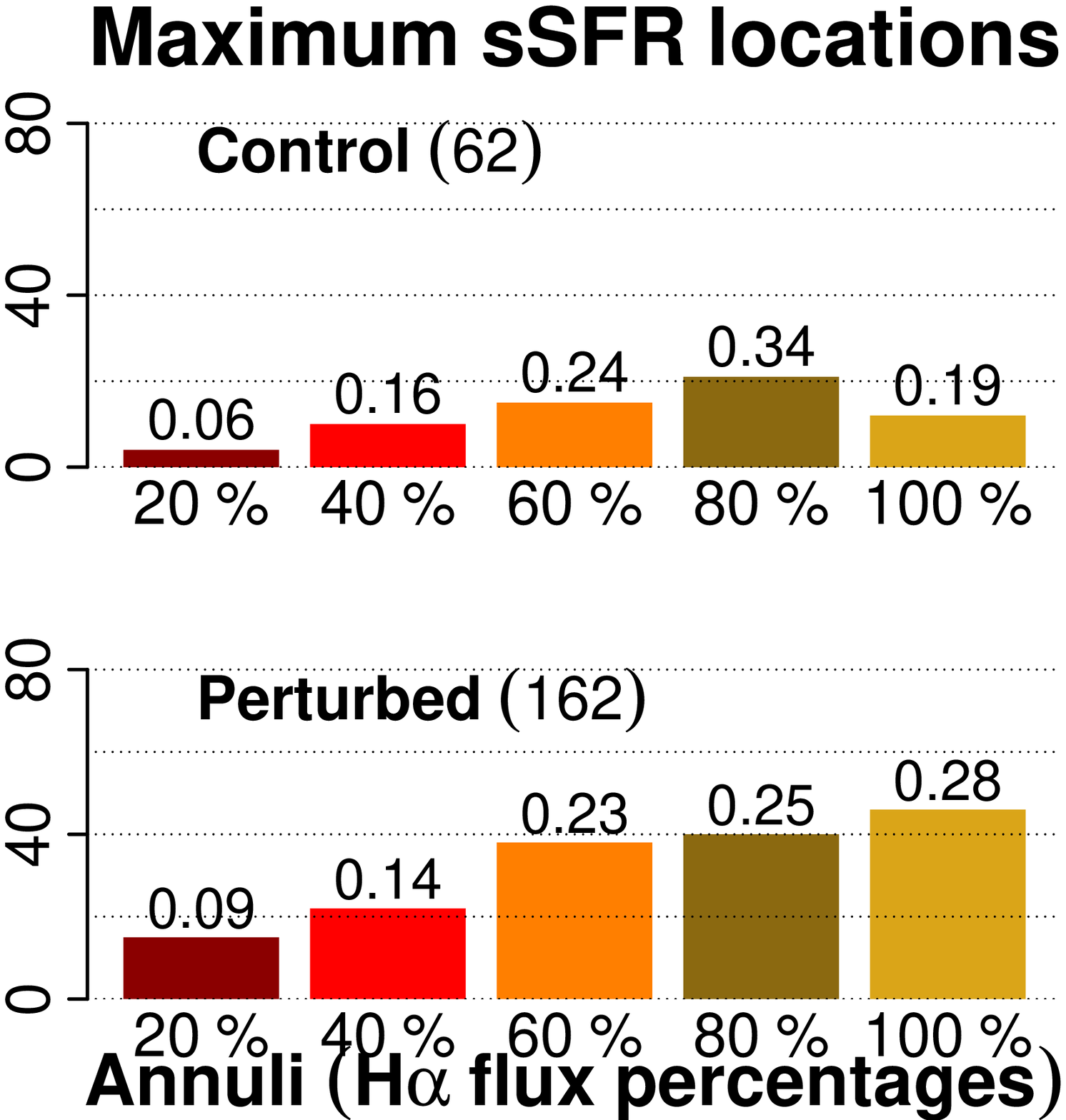}}
   \caption{\scriptsize{Annular locations of the star-forming regions/spaxels hosting the $\Sigma\mathrm{_{SFR}}$ (left) and sSFR (right) current maxima. Frequencies in the 
   y-axis and fractions at the top of the bars. Numbers in parenthesis indicate the net galaxies. The radial direction (annuli) results from H$\alpha$ flux percentages (see 
   Section~\ref{subsubsec:sf}).}}
   \label{f2} 
\end{figure}

\subsubsection{Locations of the SFR intensity and sSFR maxima}
\label{subsubsec:locmax}

Figure~\ref{f2} shows the annuli hosting the star-forming regions with the maxima of the $\Sigma\mathrm{_{SFR}}$ and specific SFR (sSFR, $\Sigma\mathrm{_{SFR}}\,\Sigma_{*}^{-1}$). 
Notice, for both control and perturbed galaxies, decrements and increments from the centre, for the $\Sigma\mathrm{_{SFR}}$ and sSFR respectively. Though frequencies 
favour perturbed galaxies, fractions (tops of bars) are very similar except for the outmost annuli hosting the maxima of the sSFR. In general, these locations of SF 
maxima, for a fraction of the CALIFA survey spiral galaxies, agree with the results of \citet{Gon16} about inside-out quenching, \textit{e.g.} the bulge to disk transition 
(also see \citealt{Bel17,San20}).
 
\begin{figure}\centering
   \mbox{\includegraphics[width=.865625\columnwidth]{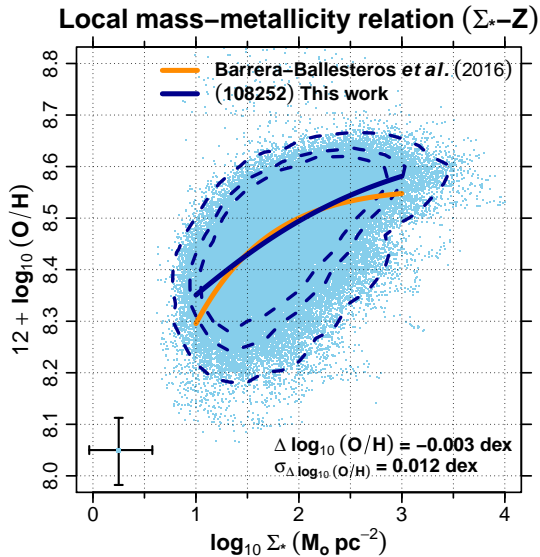}}
   \caption{\scriptsize{Local $\mathrm{\Sigma}_{*}$-Z relation for 108\,252 (blue fit) out of 111\,292 (dots and contours) star-forming regions. It follows the method of 
   BB-16 (their fit in orange). For a comparison free of bias, this figure uses inclination-corrected areas for the $\Sigma_{*}$ values. Such a correction is the reason 
   of regions below the \citet{Can19} threshold (log$_{10}$\,$\Sigma_{*}\,=$\,1.5\,M$_{\odot}$\,pc$^{-2}$) since we opted to avoid observational constraints in the data 
   (see the beginning of \ref{subsec:validation}). The dashed lines are density contours at 0.1, 0.5 and 0.9 from outside-in. Find the interquartile ranges (IQRs, 1st to 
   3rd) of the overall distributions of both $\Sigma_{*}$ and 12$+\mathrm{log}_{10}$\,(O/H) at the bottom-left.}}
   \label{f3} 
\end{figure}

\subsubsection{The local mass-metallicity relation}
\label{subsubsec:relation}

For a valid comparison with the local mass-metallicity relation of \citet[][hereafter BB-16]{BaBa16}, we adopt their method to fit an asymptotic function 
\citep[see also][]{San13,Mou11}. By using the T$_{e}$-based $O3N2$ calibration of \citet{Mar13}, we compute the median abundances for star-forming regions 
with $\Sigma_{*}$ within bins of 0.2 dex width along the 1$\,\leq\,\mathrm{log_{10}\,\Sigma_{*}}\,\leq\,$3 range. The need for an asymptotic function on 
such median values is due to saturation in $12+\mathrm{log_{10}\,(O/H)}$ predicted at the highest $\Sigma_{*}$ values (see \citealt{San13} and references 
therein)\footnote{A similar asymptotic regime, in the global MZR, is attributed to the action of either outflows or inflows of gas \citep[see][]{San20}.}. 
Moreover, the relation of BB-16 uses inclination-corrected areas for the estimation of $\Sigma_{*}$ values. We correct ours only for this case and that of 
\ref{subsubsec:sigmamass} by using the semiminor to semimajor axis ratios obtained for each galaxy by W-14.

Figure~\ref{f3} plots our local $\mathrm{\Sigma}_{*}$-Z measurements and asymptotic fit against that one of BB-16. They find $\sigma\,<$\,0.08 dex for 
their median abundances at the mass density bins. Our abundances show $\sigma\,=$\,0.078 dex along our fit. Also, the comparison of both fits (orange and 
blue solid lines) gives a median 
offset\footnote{\label{off}$\Delta\,\mathrm{log_{10}\,(O/H)}\,=\,\mathrm{log_{10}\,(O/H)}-\mathrm{log_{10}\,(O/H)}_{\mathrm{BB-16}}$.} 
$\Delta\,\mathrm{log_{10}\,(O/H)}\,=\,-$0.003\,dex (1st and 3rd Qs are $-$0.007 and 0.006\,dex respectively) with a dispersion of $\sigma_{\Delta\mathrm{log_{10}\,(O/H)}} 
=$\,0.012\,dex. This is a good agreement considering the differences in sample selection (theirs based on low exponential surface brightness profiles) and $z$ (theirs, 
$\sim$0.03). Finally, our saturation abundance, $\sim$8.66 (parameter of our asymptotic function), is in good agreement with the average value of 8.8\,$\pm$0.15 
\citep[see][]{San13,Pil07} with the differences mainly due to the calibrator used.

\section{Results}
\label{sec:res}

For fair and more detailed comparisons, our samples are split up into subsamples according to the gas excitation source that prevails per galaxy (either AGN-like 
or SFG). Since SFG excitation results dominant (see PaperI), we further split up SFGs into morphological group (either early or late type spiral, ETS, LTS) and 
colour (Red, Green, Blue). This additional subdivision defines our subsamples or galaxy types. Moreover, the star-forming spaxels of the perturbed samples merge 
and we compare them with the control ones. This way shows the average trend of the perturbed samples. We highlight that we do all comparisons at the closest current 
$\Sigma_{*}$ values of the star-forming regions. Minimizing the influence of M$_{*}$ allows us to explore real contrasts between control and perturbed galaxies 
(see also PaperI).

\begin{figure*}\centering
   \mbox{\includegraphics[width=.5066549\columnwidth]{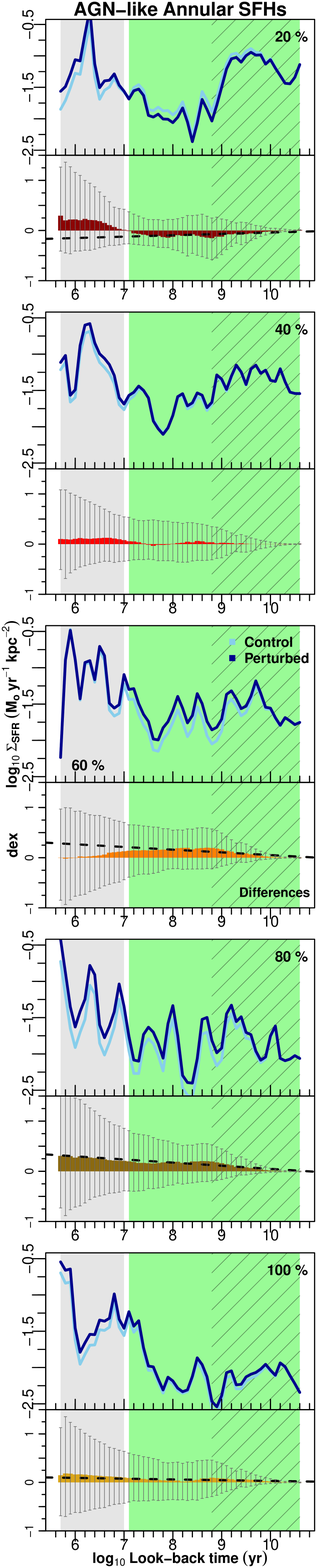}}
   \mbox{\includegraphics[width=.5066549\columnwidth]{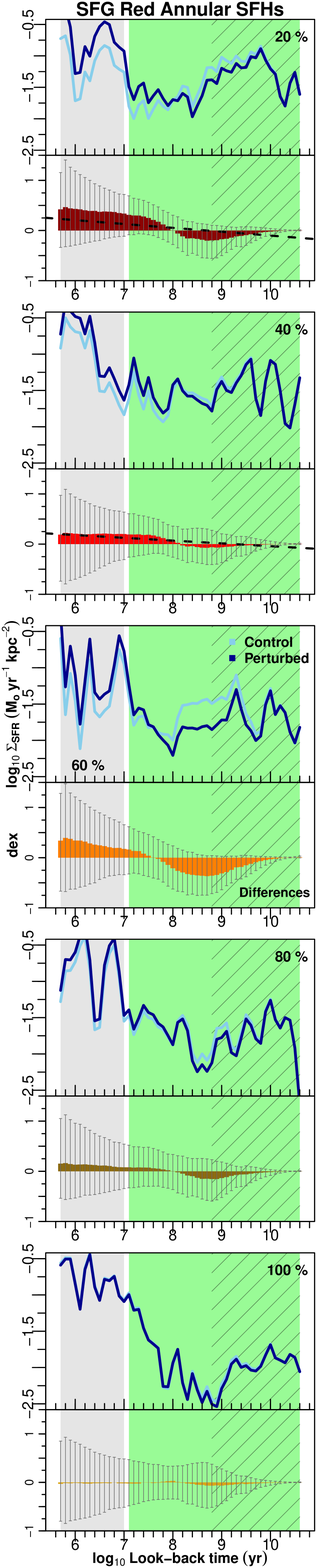}}
   \mbox{\includegraphics[width=.5066549\columnwidth]{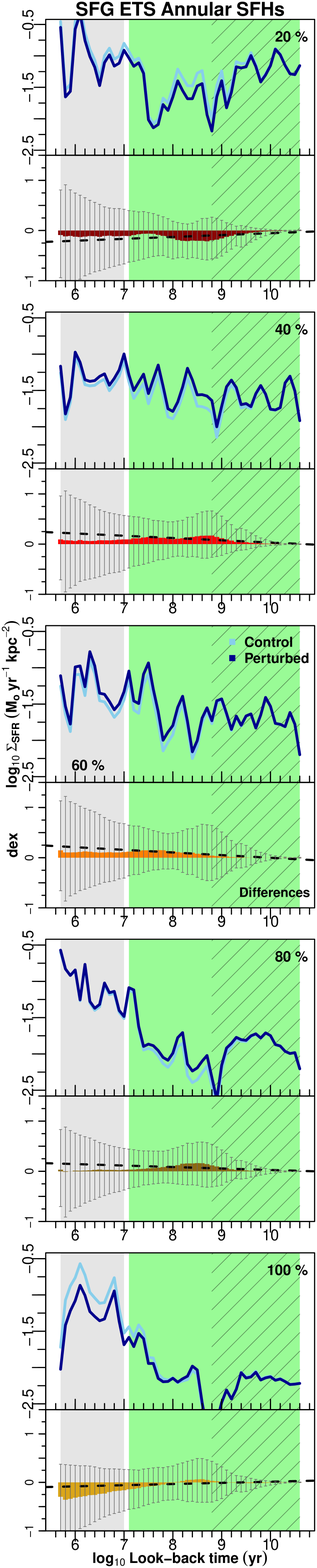}}
   \mbox{\includegraphics[width=.5066549\columnwidth]{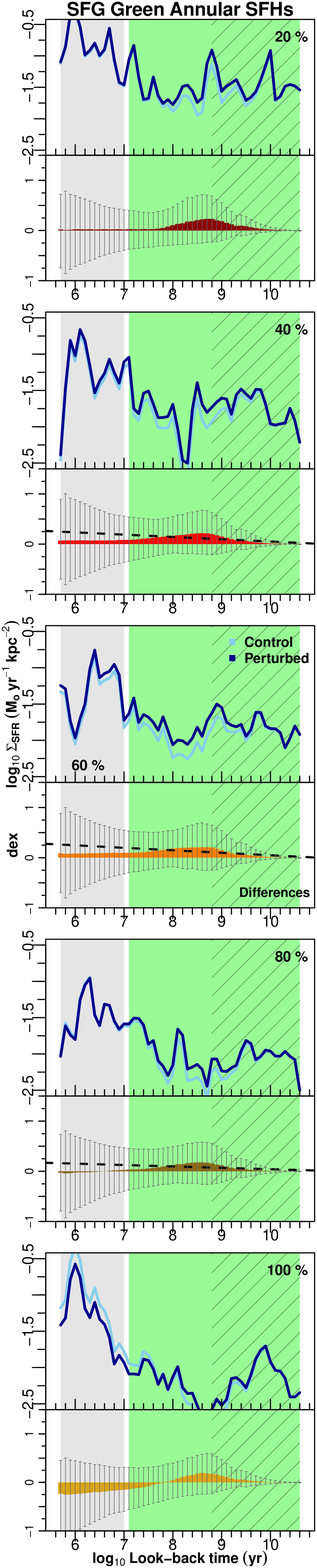}}
\caption{\scriptsize{Annular SFHs for control and perturbed galaxies. For fair comparisons, we keep the subsample division of PaperI \textit{i.e.}, AGN-like, 
SFG Red, SFG ETS and SFG Green from left to right. Five consecutive-outward annuli (``20'', ``40'', ``60'', ``80'' and ``100\,\%'') related to the 
H$\alpha$ flux of each galaxy as the radial extension (see~\ref{subsubsec:sf}). These panels result from pairing star-forming spaxels in all 
perturbed samples to those ones in the control sample by minimizing their differences in current $\Sigma_{*}$. The ``Differences'' in 
the history records (by subtracting the control values from the perturbed ones, secondary plots) depict the medians (bar heights) of the distributions 
of differences and their interquartile ranges (IQRs, 1st to 3rd, bar lines). The dashed lines are linear regression model fits, with statistically 
significant slopes (Pr($>$F)\,$<$\,statistical level), on the Differences within the green background only (the Differences constitute 
the response variable). The histories (main plots) are both control (light blue) and perturbed (dark blue) values corresponding to each $t_{j,s}$ bin median 
Difference. The proposed timescale for a valid-current SFR ($\sim$10\,Myr, light gray, see \ref{subsubsec:synVSHa}) has been disregarded since it matches 
the range characterized by the least accurate values of the histories (\textit{e.g.} compare the extensions of the bar lines between the light gray and 
green backgrounds). In contrast, the green and shaded backgrounds indicate the time ranges, starting at present, at which the more significant mass fractions 
($>$\,0.001 and $>$\,0.01 respectively) were assembled.}}
   \label{f4} 
\end{figure*}

\begin{figure*}\centering
\ContinuedFloat   
   \mbox{\includegraphics[width=.5066549\columnwidth]{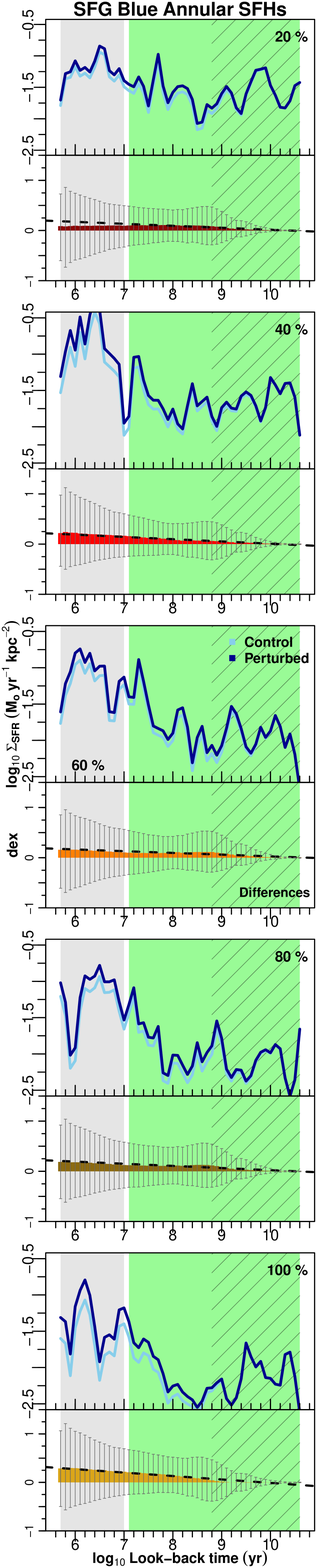}}
   \mbox{\includegraphics[width=.5066549\columnwidth]{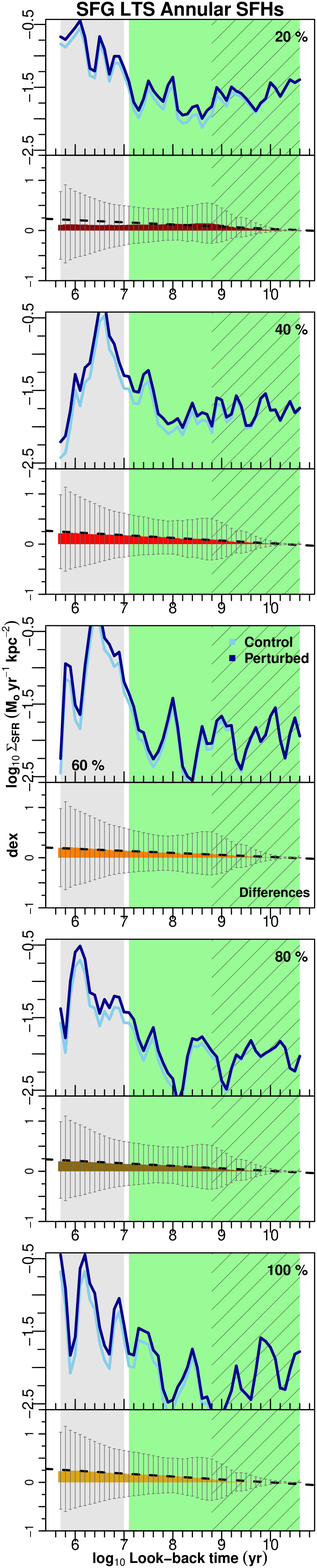}}
   \mbox{\includegraphics[width=.5066549\columnwidth]{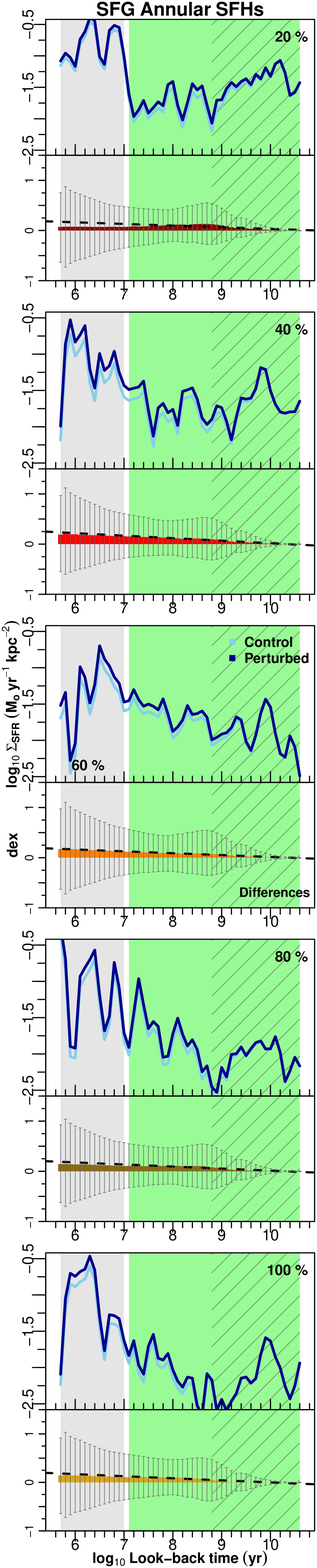}}
   \mbox{\includegraphics[width=.5066549\columnwidth]{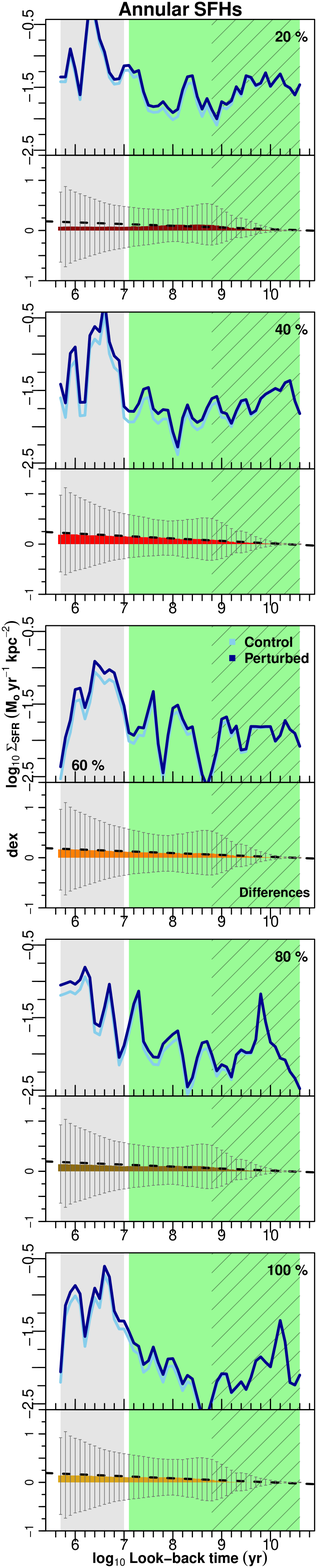}}
\caption{\scriptsize{Annular SFHs for control and perturbed galaxies (cont.). Same caption as above but for SFG Blue, SFG LTS, SFG and all subsamples.}}
   \label{f4} 
\end{figure*}

\begin{table*}
 \setlength{\tabcolsep}{0.15\tabcolsep}
\begin{minipage}{\textwidth}
\centering
 \caption{\scriptsize{First, second and third Qs (1st, 2nd and 3rd) of the medians (``Differences'', bar heights in Fig.~\ref{f4}; ``Q2'' columns here and in 
 Table~\ref{tab:B1}, all them in dex) of the distributions of differences between the annular SFHs of control and perturbed galaxies (same information in bold 
 font from Table~\ref{tab:B1}). We have considered only the time ranges (log$_{10}$\,t\,$>$\,7 yr), starting at present, at which the more significant mass 
 fractions were assembled ($>$\,0.001 and $>$\,0.01, green and shaded backgrounds respectively, see Fig.~\ref{f4}). ``t'' (log$_{10}$\,t, yr), ``Q1'', ``Q3'' 
 (both in dex) and ``ext.'' are the associated time range, 1st and 3rd Qs,  and the fractional extension above zero dex of the IQRs (\textit{i.e.}, from Q1 
 to Q3, the extension of bar lines in Fig.~\ref{f4}). Data in bold font are the annular cases with fractional extensions $>$\,0.5 in all the three Q2 Qs, 
 and the cases where the 2nd Q of Q2 is $>$\,0.06 dex (see text).}
  \label{tab:sum_of_summaryf4}}
 \begin{scriptsize}
 \begin{tabular}{lccccccccccccccccccccccccc}
\hline\\
    &                                                                             \multicolumn{25}{c}{Annuli (H$\alpha$ flux percentages)}                                                                                                                              \\
Q2  &      \multicolumn{5}{c}{20\,\%}                   &      \multicolumn{5}{c}{40\,\%}                   &     \multicolumn{5}{c}{60\,\%}                    &     \multicolumn{5}{c}{80\,\%}                    &     \multicolumn{5}{c}{100\,\%}                   \\ 
Q   &t   &Q1       &Q2            &Q3    &ext.          &t   &Q1       &Q2            &Q3    &ext.          &t   &Q1       &Q2            &Q3    &ext.          &t   &Q1       &Q2            &Q3    &ext.          &t   &Q1       &Q2            &Q3    &ext.          \\
\hline\\                                                                                                                                                                                                                                                          
    &                                                                                                            \multicolumn{25}{c}{AGN-like}                                                                                                                          \\  
1st &8.2 &$-$0.445 &$-$0.105      &0.259 &0.368         &10.1&$-$0.042 &$-$0.001      &0.029 &0.411         &9.7 &$-$0.056 &0.049         &0.173 &\textbf{0.754}&9.7 &$-$0.048 &0.047         &0.149 &\textbf{0.756}&9.9 &$-$0.030 &0.026         &0.097 &\textbf{0.762}\\
2nd &8.4 &$-$0.454 &$-$0.076      &0.298 &0.396         &9.3 &$-$0.150 &0.007         &0.172 &0.533         &7.3 &$-$0.334 &\textbf{0.143}&0.646 &\textbf{0.659}&7.5 &$-$0.342 &\textbf{0.160}&0.711 &\textbf{0.675}&9.3 &$-$0.126 &0.049         &0.234 &\textbf{0.650}\\
3rd &9.8 &$-$0.119 &$-$0.030      &0.089 &0.428         &9.0 &$-$0.272 &0.026         &0.296 &0.521         &8.4 &$-$0.226 &0.166         &0.569 &\textbf{0.716}&8.0 &$-$0.273 &0.183         &0.566 &\textbf{0.674}&7.4 &$-$0.380 &0.072         &0.530 &\textbf{0.583}\\[1ex]
    &                                                                                                            \multicolumn{25}{c}{SFG Red}                                                                                                                           \\  
1st &9.2 &$-$0.376 &$-$0.141      &0.070 &0.157         &8.4 &$-$0.348 &$-$0.040      &0.313 &0.474         &9.1 &$-$0.557 &$-$0.287      &0.027 &0.047         &8.3 &$-$0.514 &$-$0.084      &0.269 &0.344         &9.5 &$-$0.222 &$-$0.035      &0.129 &0.366         \\
2nd &9.8 &$-$0.151 &$-$0.039      &0.062 &0.289         &9.9 &$-$0.083 &$-$0.006      &0.074 &0.472         &9.6 &$-$0.320 &$-$0.123      &0.041 &0.113         &8.1 &$-$0.431 &$-$0.019      &0.310 &0.418         &9.7 &$-$0.138 &$-$0.021      &0.073 &0.347         \\
3rd &10.6&$-$0.007 &0.003         &0.016 &0.682         &8.0 &$-$0.261 &0.067         &0.358 &0.578         &10.3&$-$0.020 &$-$0.002      &0.014 &0.417         &10.6&$-$0.004 &0.003         &0.013 &0.752         &7.5 &$-$0.378 &$-$0.003      &0.383 &0.504         \\[1ex]
    &                                                                                                            \multicolumn{25}{c}{SFG ETS}                                                                                                                           \\  
1st &8.0 &$-$0.565 &$-$0.158      &0.185 &0.246         &9.7 &$-$0.069 &0.020         &0.168 &\textbf{0.709}&9.8 &$-$0.086 &0.001         &0.107 &\textbf{0.555}&9.7 &$-$0.094 &0.005         &0.110 &\textbf{0.540}&9.6 &$-$0.154 &$-$0.015      &0.084 &0.352         \\
2nd &7.7 &$-$0.503 &$-$0.086      &0.305 &0.378         &8.0 &$-$0.253 &\textbf{0.096}&0.445 &\textbf{0.638}&8.8 &$-$0.455 &0.042         &0.643 &\textbf{0.586}&7.3 &$-$0.308 &0.043         &0.415 &\textbf{0.574}&9.2 &$-$0.231 &$-$0.006      &0.191 &0.453         \\
3rd &9.7 &$-$0.184 &$-$0.032      &0.046 &0.201         &7.4 &$-$0.403 &0.131         &0.518 &\textbf{0.563}&7.1 &$-$0.362 &0.120         &0.647 &\textbf{0.641}&7.9 &$-$0.234 &0.099         &0.419 &\textbf{0.642}&9.0 &$-$0.335 &0.004         &0.320 &0.488         \\[1ex]
    &                                                                                                           \multicolumn{25}{c}{SFG Green}                                                                                                                          \\  
1st &7.2 &$-$0.364 &0.025         &0.396 &\textbf{0.521}&9.7 &$-$0.068 &0.036         &0.166 &\textbf{0.710}&9.7 &$-$0.065 &0.034         &0.165 &\textbf{0.718}&9.7 &$-$0.074 &0.023         &0.145 &\textbf{0.660}&10.5&$-$0.005 &$-$0.001      &0.003 &0.355         \\
2nd &7.8 &$-$0.306 &0.051         &0.399 &\textbf{0.566}&7.4 &$-$0.280 &\textbf{0.109}&0.512 &\textbf{0.646}&7.3 &$-$0.277 &\textbf{0.103}&0.516 &\textbf{0.650}&7.5 &$-$0.309 &0.056         &0.440 &\textbf{0.587}&9.7 &$-$0.062 &0.025         &0.134 &0.685         \\
3rd &8.2 &$-$0.273 &0.144         &0.586 &\textbf{0.682}&8.0 &$-$0.179 &0.166         &0.548 &\textbf{0.754}&8.9 &$-$0.239 &0.179         &0.591 &\textbf{0.712}&8.0 &$-$0.238 &0.124         &0.490 &\textbf{0.673}&9.1 &$-$0.146 &0.108         &0.390 &0.727         \\[1ex]
    &                                                                                                           \multicolumn{25}{c}{SFG Blue}                                                                                                                           \\  
1st &9.7 &$-$0.059 &0.014         &0.101 &\textbf{0.631}&9.7 &$-$0.088 &0.019         &0.135 &\textbf{0.606}&9.7 &$-$0.089 &0.023         &0.157 &\textbf{0.638}&9.7 &$-$0.085 &0.019         &0.143 &\textbf{0.629}&10.1&$-$0.028 &0.000         &0.031 &\textbf{0.525}\\
2nd &8.8 &$-$0.285 &\textbf{0.086}&0.469 &\textbf{0.622}&8.7 &$-$0.308 &\textbf{0.062}&0.452 &\textbf{0.595}&7.5 &$-$0.294 &\textbf{0.091}&0.457 &\textbf{0.609}&8.0 &$-$0.252 &\textbf{0.107}&0.478 &\textbf{0.655}&8.8 &$-$0.389 &0.044         &0.465 &\textbf{0.545}\\
3rd &8.4 &$-$0.232 &0.102         &0.448 &\textbf{0.659}&8.0 &$-$0.207 &0.088         &0.407 &\textbf{0.663}&7.1 &$-$0.335 &0.100         &0.543 &\textbf{0.618}&7.4 &$-$0.302 &0.114         &0.519 &\textbf{0.632}&8.0 &$-$0.210 &0.146         &0.519 &\textbf{0.712}\\[1ex]
    &                                                                                                            \multicolumn{25}{c}{SFG LTS}                                                                                                                           \\  
1st &9.7 &$-$0.068 &0.029         &0.154 &\textbf{0.692}&9.7 &$-$0.103 &0.022         &0.155 &\textbf{0.601}&9.7 &$-$0.114 &0.019         &0.156 &\textbf{0.579}&9.7 &$-$0.108 &0.011         &0.136 &\textbf{0.557}&9.7 &$-$0.101 &0.002         &0.116 &\textbf{0.534}\\
2nd &7.5 &$-$0.232 &\textbf{0.116}&0.439 &\textbf{0.654}&8.8 &$-$0.316 &\textbf{0.092}&0.504 &\textbf{0.615}&8.3 &$-$0.304 &\textbf{0.072}&0.471 &\textbf{0.608}&8.8 &$-$0.355 &\textbf{0.075}&0.514 &\textbf{0.592}&8.8 &$-$0.370 &\textbf{0.065}&0.479 &\textbf{0.564}\\
3rd &8.9 &$-$0.240 &0.126         &0.503 &\textbf{0.677}&8.0 &$-$0.199 &0.125         &0.464 &\textbf{0.699}&8.1 &$-$0.271 &0.095         &0.449 &\textbf{0.624}&8.1 &$-$0.259 &0.119         &0.485 &\textbf{0.652}&8.0 &$-$0.214 &0.135         &0.501 &\textbf{0.700}\\[1ex]
    &                                                                                                              \multicolumn{25}{c}{SFG}                                                                                                                             \\  
1st &9.7 &$-$0.070 &0.026         &0.145 &\textbf{0.675}&9.7 &$-$0.096 &0.021         &0.152 &\textbf{0.613}&9.7 &$-$0.108 &0.017         &0.155 &\textbf{0.589}&9.7 &$-$0.101 &0.010         &0.131 &\textbf{0.566}&10.4&$-$0.007 &0.000         &0.005 &0.455         \\
2nd &7.2 &$-$0.315 &\textbf{0.083}&0.455 &\textbf{0.591}&8.8 &$-$0.315 &\textbf{0.097}&0.518 &\textbf{0.622}&8.3 &$-$0.318 &\textbf{0.073}&0.501 &\textbf{0.612}&8.8 &$-$0.350 &\textbf{0.080}&0.515 &\textbf{0.596}&8.8 &$-$0.383 &0.049         &0.469 &0.550         \\
3rd &8.3 &$-$0.244 &0.114         &0.490 &\textbf{0.668}&8.0 &$-$0.217 &0.118         &0.467 &\textbf{0.683}&8.0 &$-$0.272 &0.091         &0.449 &\textbf{0.623}&8.5 &$-$0.325 &0.105         &0.537 &\textbf{0.623}&7.6 &$-$0.285 &0.104         &0.501 &0.637         \\[1ex]
    &                                                                                                              \multicolumn{25}{c}{All}                                                                                                                             \\  
1st &9.7 &$-$0.075 &0.022         &0.141 &\textbf{0.654}&9.7 &$-$0.100 &0.019         &0.149 &\textbf{0.599}&9.7 &$-$0.104 &0.018         &0.155 &\textbf{0.598}&9.7 &$-$0.097 &0.012         &0.134 &\textbf{0.579}&10.1&$-$0.030 &0.000         &0.032 &\textbf{0.512}\\
2nd &7.3 &$-$0.305 &\textbf{0.081}&0.446 &\textbf{0.594}&8.8 &$-$0.322 &\textbf{0.088}&0.509 &\textbf{0.612}&8.3 &$-$0.313 &\textbf{0.077}&0.502 &\textbf{0.616}&8.8 &$-$0.342 &\textbf{0.083}&0.517 &\textbf{0.602}&8.8 &$-$0.381 &0.053         &0.474 &\textbf{0.554}\\
3rd &8.3 &$-$0.250 &0.107         &0.483 &\textbf{0.659}&8.0 &$-$0.224 &0.112         &0.463 &\textbf{0.674}&8.0 &$-$0.273 &0.093         &0.452 &\textbf{0.624}&7.2 &$-$0.323 &0.103         &0.538 &\textbf{0.625}&7.6 &$-$0.292 &0.101         &0.500 &\textbf{0.631}\\[1ex]
\hline\\                                                                                                                                                                                                                                                                         
 \end{tabular} 
 \end{scriptsize}
 \end{minipage}
\end{table*} 

\subsection{SFHs and sSFHs}
\label{subsec:sfh_ssfr}

The SFH plots $\mu_{j,s}$ times the total mass\footnote{Including the mass returned to the interstellar medium due to stellar 
evolution.} turned into stars divided by the whole timescale. With their angular surface, all spaxels have a $\Sigma_{\mathrm{SFR}}$ history record. 
The specific SFH (sSFH or the sSFR history) is the SFR vector as defined above divided by the 
total mass. In this work, both SFHs and sSFHs correspond to regions currently classified as star-forming (Section~\ref{subsubsec:sf}). We then assume that 
these regions have remained star-forming along their whole existence, a quite real fact for spiral galaxies.\footnote{Only \emph{stellar migration} would 
be uncontrolled since nothing ensures that the actual locations of stars are indeed their birthplaces.} We further remark the following. According to all simple SP 
models, for time ranges from the present to say 10 Myr, the fractions of M$_{*}$ formed are really tiny. Tiny indeed if compared with the fractions formed at time 
ranges from the present to $>\,\sim$10 Myr. Consequently, uncertainties associated to the former fractions are significant so that any trend that the SFHs and the 
sSFHs may show in the last log$_{10}$\,t\,$\sim$ 7 yr or less are the least accurate and easily subject to misinterpretations. Hence, with each 
median from the distributions of cumulative functions corresponding to control and perturbed galaxies, we identified the time ranges at which 1$-\eta_{*}(t_{j,s})\,>\,$0.001 
and $>\,$0.01\footnote{1$-\eta_{*}(t_{j,s})$ represents the net fraction of stellar mass per $t_{j,s}$ time range (see A-07 their equation 10).}. These are, for control 
galaxies, log$_{10}$\,t\,$=$\,7.1 and 8.8 yr; and log$_{10}$\,t\,$=$\,6.8 and 8.7 yr for perturbed galaxies. 
As secular processes are the reference of all comparisons throughout this work, we adopted the values for control galaxies as the thresholds for 
more reliable SFHs and sSFHs as well as for their trends. We therefore emphasize our attention to only larger ranges of time, say log$_{10}$\,t\,$>$\,7 yr, where the 
fractions of M$_{*}$ produced are $>$\,0.001.
 
Figures~\ref{f4} and \ref{fA1} respectively show the SFHs and sSFHs. For the sake of briefness, Tables~\ref{tab:sum_of_summaryf4} and \ref{tab:B1} 
show the tendencies of the distributions of differences only between the SFHs. We remark that, in each annular set of spaxel histories, spaxels in 
control and perturbed galaxies are paired by minimizing their differences in current $\Sigma_{*}$. We show the medians of the differences of rates per $t_{j,s}$ bin 
as the bars in the secondary plots (``Differences'' in Figs.~\ref{f4} and \ref{fA1}, and ``Q2'' columns in Tables~\ref{tab:sum_of_summaryf4} and \ref{tab:B1}). The bar 
lines correspond to the respective interquartile ranges (IQRs) and their ends are the quartiles (Q) ``Q1'' and ``Q3''. The dashed lines represent linear model fits 
(``Differences'' as the \emph{response} variable), and we show them only if the slopes of the models are statistically significant. The histories (represented by the 
light/dark blue lines in the main plots) are both control and perturbed values corresponding to each $t_{j,s}$ bin Difference.

From $\sim$2.2 Gyr lookback time (log$_{10}\,\mathrm{t}\,\sim$9.34 yr), \citet{Gon17} find that the $\Sigma_{\mathrm{SFR}}$ increases 
with redshift. From log$_{10}$\,t\,$\sim$9 yr, Fig.~\ref{f4} shows increments with $z$ only in the central annulus. In contrast, 
from log$_{10}$\,t\,$\sim$9 yr to larger cosmic times, rather constant rates appear in the rest annuli. Moreover, \cite{Gon17} find that average 
rates below $\sim$2.2 Gyr lookback time indicate a general rejuvenation for at least low-to-intermediate stellar mass and Sbc-to-Sd galaxies. In 
Fig.~\ref{f4}, within the 7.1\,$<$\,log$_{10}$\,t\,$<\sim$9 yr range, no rejuvenation, or few cases of it, are found from the central to the mid 
annulus. However, for the two outmost annuli, evident SFR increments occur in the same time range.

By looking at the Differences in Fig.~\ref{f4}, we may contrast regions in control and perturbed galaxies. Starting with AGN-like ones, all 
annuli, but that of the 40\,\%, suggest Differences that vary as cosmic time decreases (linear model fits). For the 60, 80 and 100\,\% annuli, the 
Differences indicate higher rates for regions in perturbed galaxies from time ranges within the shaded background towards shorter time ranges 
(green background). For these annuli, most extensions of the IQRs of the Differences are above zero dex. For regions in SFG Red objects, the rates of the 
control sample are clearly higher in the 20, 60 and 80\,\% annuli (the 8\,$<$\,log$_{10}$\,t\,$<$9 yr range). Regions in perturbed SFG ETSs show 
higher rates within the 40, 60 and 80\,\% annuli. Similarly, the extensions of the IQRs of the Differences for these annuli are mostly above zero dex.
Briefly, the regions in the rest subsamples and all galaxies, excluding only the 100\,\% annulus of SFG Green objects, show Differences that are 
always above zero dex and so the major extensions of their respective IQRs. These contrasts are evident ($\geq\,\sim$0.06 dex) within the green 
background in all annuli, and within the shaded one in almost all annuli (except that of the 100\,\%). The linear models suggest increasing Differences, 
\textit{i.e.}, higher rates for regions in perturbed galaxies as the cosmic time decreases (31 out of 40 annuli along all subsamples and all galaxies).

What has been visually perceived from Fig.~\ref{f4} is analytically confirmed by the numbers in Tables~\ref{tab:B1} and \ref{tab:sum_of_summaryf4}. They respectively 
list the quartiles (Qs) of each distribution of differences per time bin, and the Qs of the second Q (\textit{i.e.}, the ``Differences'' or bar heights in the 
plots and the ``Q2'' columns in the Tables) of the same differential distributions. Since any data distribution is nicely characterized by its interquartile 
range (IQR), we calculate what fraction of each IQR is above zero dex. In this way, the trend each distribution of differences has can be identified since 
all rates of control spaxels are subtracted fom those of the perturbed ones. From Table~\ref{tab:B1} we find that the fraction of IQRs with a 
extension above zero dex (``ext.'') of $\geq$\,0.5 is 1123/1440 ($\sim$\,78\,\%). Similarly, the fractions of IQRs with extensions of $\geq$\,0.6 and $\geq$\,0.7 
are 788/1440 and 194/1440 ($\sim$\,55\,\% and $\sim$\,13\,\%). This indicates a clear trend of differences that suggests that the SFRs, through time, have been 
higher for regions in perturbed galaxies. Similarly, from Table~\ref{tab:sum_of_summaryf4}, we identify the annular cases, along all subsamples, where the 
extensions are $>$\,0.5 in all the three Qs (1st, 2nd and 3rd) of Q2. The result is 29/40 cases, $\sim$\,73\,\%, a very similar fraction to 
that obtained from Table~\ref{tab:B1}. These cases are highlighted in bold font in Table~\ref{tab:sum_of_summaryf4} and are distributed as follows: AGN-like and 
SFG ETS with 3, SFG Green and SFG with 4, and SFG Blue, SFG LTS and 
all galaxies with 5 cases each. The only one subsample with no cases is the SFG Red. From these 29 cases, we find those where the 2nd Q of Q2 is $>$\,0.06 
dex, restriction that ensures a Difference of the order of $\sim$15\,\% (that may be considered significant enough). The result is 22/29 (22/40 overall) cases 
($\sim$\,76\,\%, $\sim$\,55\,\% overall). For this fraction, the 1st and 3rd Qs of Q2 fluctuate between $\sim$0.00-0.18 dex. Finally, the 22 cases are highlighted 
in bold font in Table~\ref{tab:sum_of_summaryf4} and are distributed as follows: SFG ETS with 1, AGN-like and SFG Green with 2, SFG Blue, SFG and all galaxies with 4, 
and SFG LTS with 5. 

Figure~\ref{fA1} poorly shows decreasing sSFHs with decreasing cosmic time that end at log$_{10}$\,t\,$\sim$9.3 yr (\textit{e.g.} 
A-07, \citealt{Tho10,Gon17}). Many annuli show flat histories in the very distant past (shaded background). We further notice clear 
rejuvenations, as cosmic time decreases, for SFG Red objects, specifically their 60 and 80\,\% annuli; for the SFG Green, SFG Blue and SFG LTS 
subsamples as well from their 60 to 100\,\% annuli; and for SFG and all galaxies from their 40\,\% annulus outwards. Again, we contrast regions 
from control and perturbed samples through their Differences. Notice that annuli with higher specific rates for regions in perturbed galaxies 
are generally frequent in all subsamples excluding only the SFG Red one. As in the SFHs, all Differences and their respective IQRs in all annuli 
suggest higher values for regions in perturbed galaxies from the SFG Blue subsample to all galaxies. We may say, in summary, that the histories 
of SF for star-forming regions comparable in current stellar mass density show, though sometimes moderate, clear contrasts due to the environment 
(close companions) of the host galaxies (\textit{e.g.} \citealt{Gug15}).

\subsection{Current maxima and all region histories}
\label{subsec:sfr&ssfr}

\begin{figure*}\centering
   \mbox{\includegraphics[width=.6925\columnwidth]{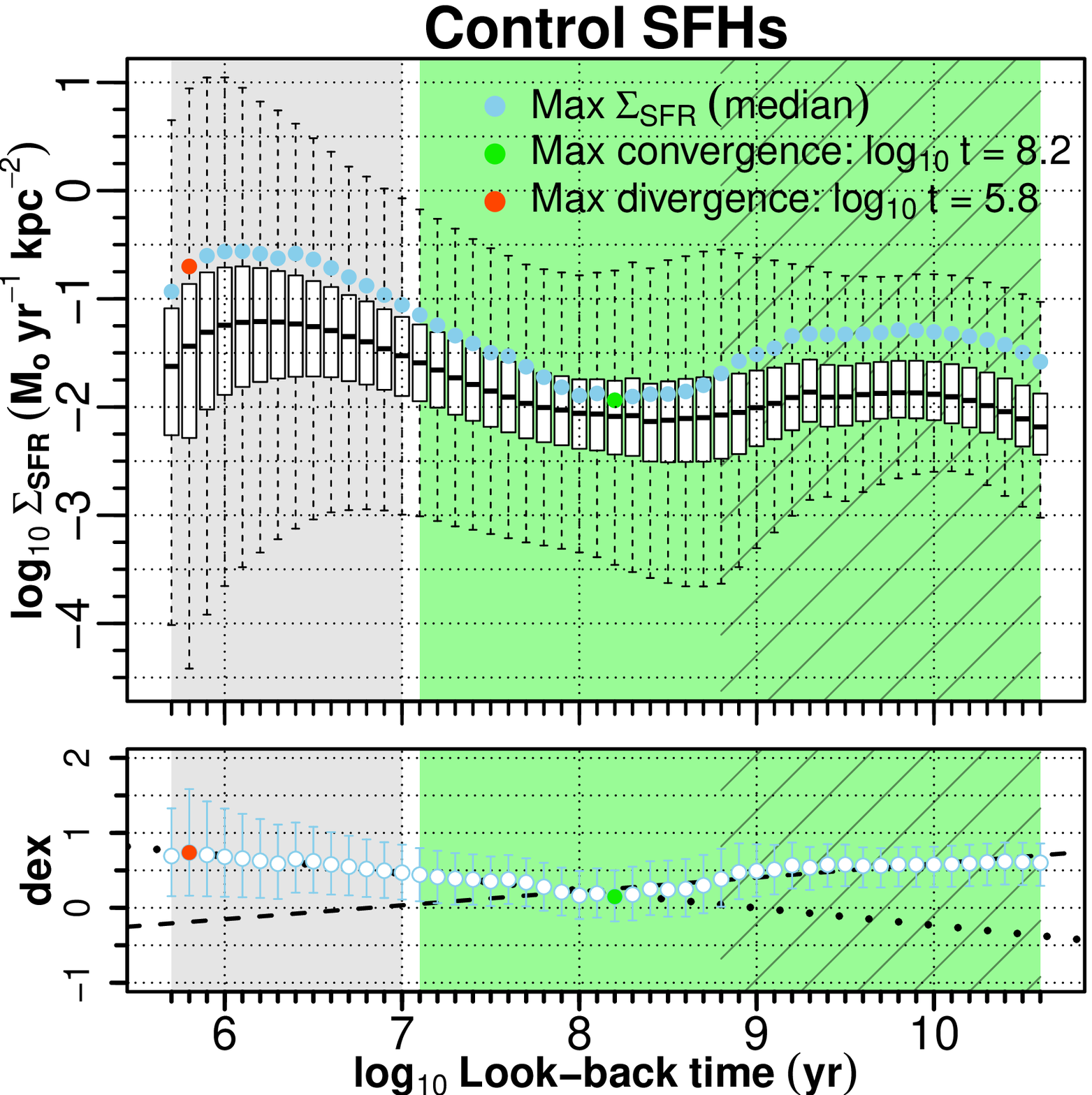}}
   \mbox{\includegraphics[width=.6925\columnwidth]{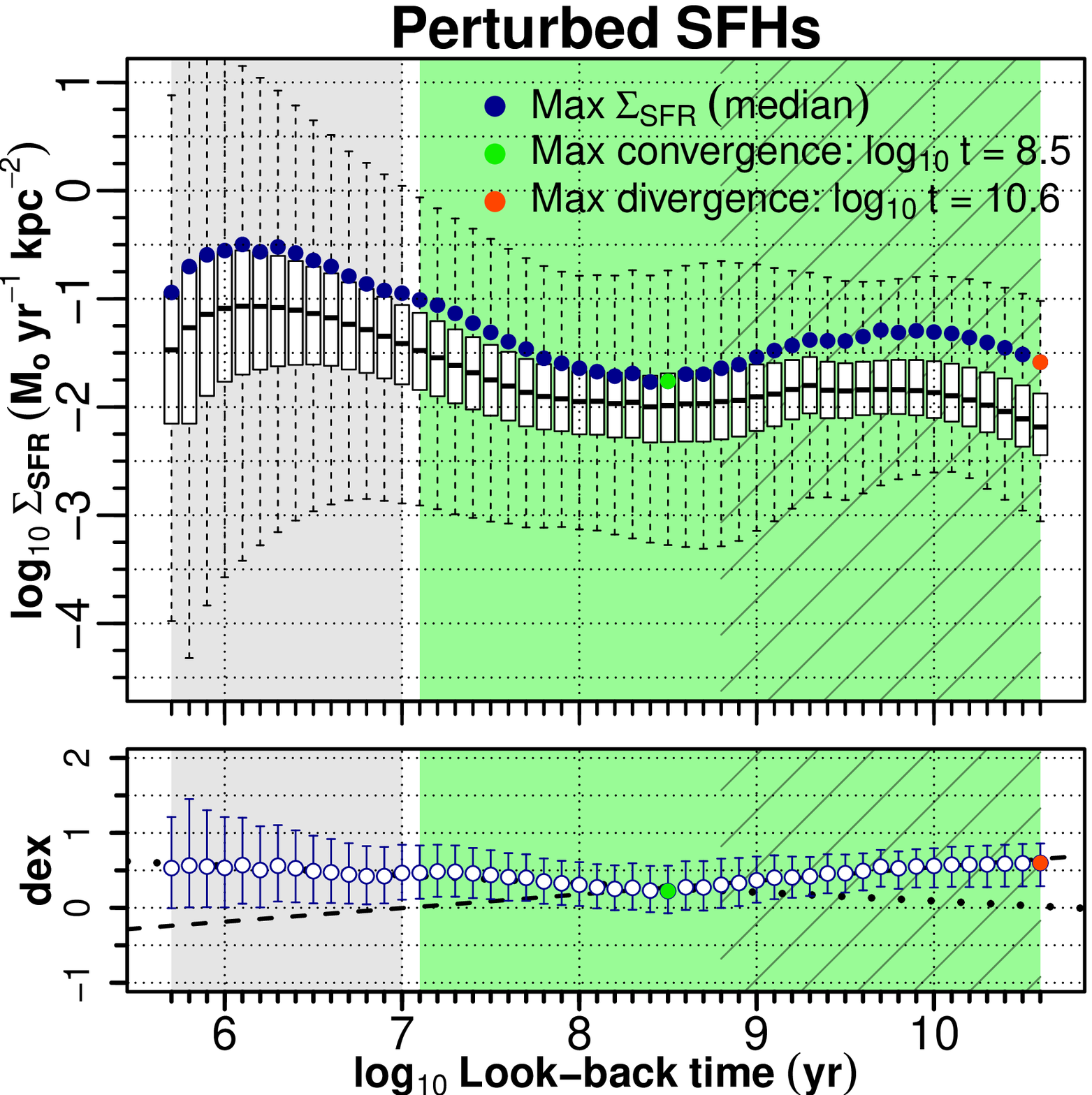}}\\
   \mbox{\includegraphics[width=.6925\columnwidth]{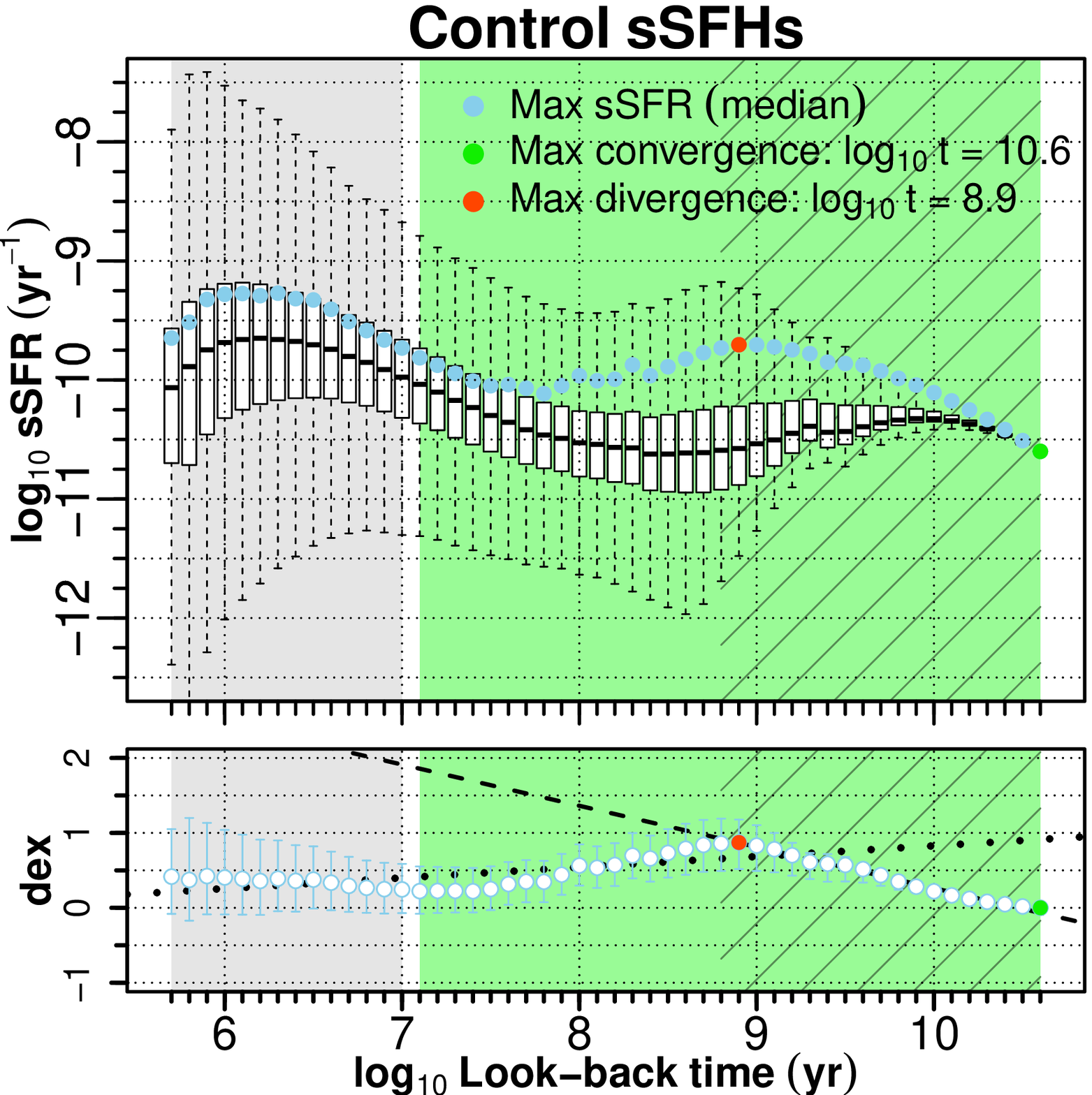}}
   \mbox{\includegraphics[width=.6925\columnwidth]{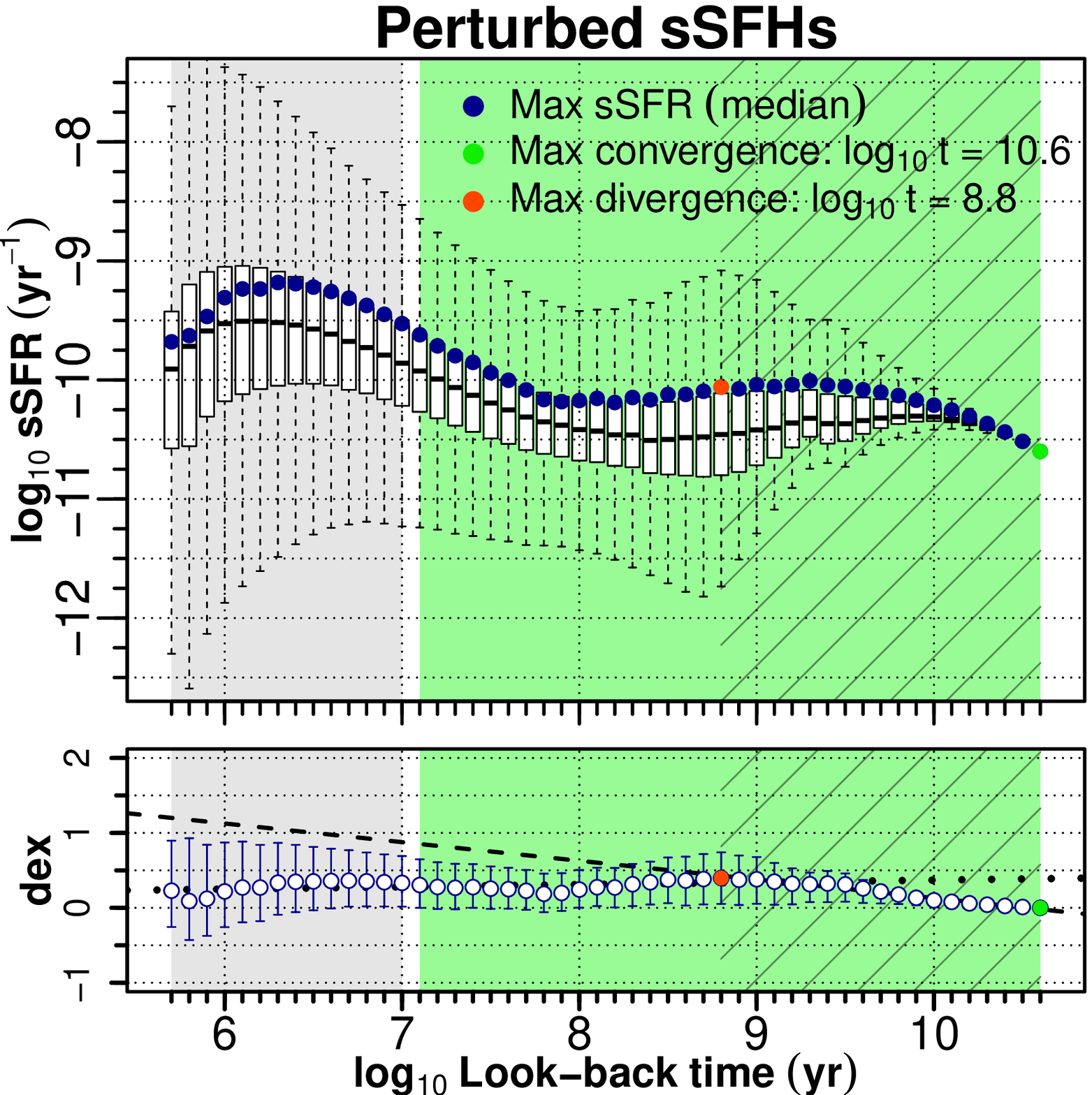}}
   \begin{minipage}[]{\textwidth}
   \caption{\scriptsize{The SFHs and sSFHs. Between control and perturbed galaxies, the distributions of current $\Sigma_{*}$ are similar for both 
   star-forming regions, and star-forming regions hosting the current maxima ($\mathrm{\Sigma_{SFR}}$ and sSFR). Boxplots as the distributions of 
   all star-forming regions. Solid circles as the medians of the distributions of the star-forming regions hosting the current maxima. Green/red 
   solid circles as the times for maximum convergence/divergence between solid circles and midlines of boxes. Empty circles as the differences 
   between both medians, \textit{i.e.}, midline values of boxes are subtracted from solid-circle values. Lines of empty circles are the interquartile 
   ranges (IQRs, 1st to 3rd) of the distributions of differences. Statistically significant linear regression model fits (dashed and dotted lines) 
   help to visualize phases along the time scale. We finally disregard the timescale for a valid-current SFR ($\sim$10\,Myr, light gray) since 
   the uncertainties associated to its fractions of mass formed are the largest (compare the extensions of the boxplots between the light gray and green 
   backgrounds). As in Fig.~\ref{f4}, the green and shaded backgrounds indicate the time ranges, starting at present, at which the more significant 
   mass fractions ($>$\,0.001 and $>$\,0.01 respectively, see Section~\ref{subsec:sfh_ssfr}) were assembled.}}
   \label{f5} 
   \end{minipage}
\end{figure*}

Here we compare the histories of all regions with the histories of the regions that at the current time show the peaks of 
$\mathrm{\Sigma_{SFR}}$ and sSFR (hereafter the current maxima). Describing first the SFHs of all regions, Fig.~\ref{f5} (top row) shows differences, in 
$\Sigma_{\mathrm{SFR}}$, between control and perturbed galaxies 
from log$_{10}\,\mathrm{t}\,\sim$9 yr till more recent times. Medians (midlines of boxes) of regions in perturbed galaxies show SFR intensities of at least 
log$_{10}\,\Sigma\mathrm{_{SFR}}\,\sim$0.1\,M$_{\odot}$\,yr$^{-1}$\,kpc$^{-2}$ higher than the medians of regions in control galaxies. Regarding the medians 
of the $\mathrm{\Sigma_{SFR}}$ current maxima (solid circles), regions in perturbed galaxies have 
higher values only within 7.1\,$\leq$\,log$_{10}$\,t\,$\leq$\,9 yr. The maximum convergence, \textit{i.e.} the smallest difference with the respective box 
midline (green circle), occurred at an earlier time in perturbed galaxies (log$_{10}\,\mathrm{t}\,\sim$8.5 yr). Similarly, the maximum divergence, the maximum 
difference with the respective box midline (red circle), occurred at the very past for both control and perturbed galaxies (log$_{10}$\,t\,$\sim$10.4 and 
$\sim$10.6 yr, respectively). Regarding the differences between both median histories, those of the maxima and those of all regions (empty circles, secondary 
plots), notice declining (dashed-line fit) and increasing phases (dotted-line fit) from the very past time with green circles as the turnover points. Both the 
declining and increasing phases are very similar in slope if each phase is compared between control and perturbed galaxies. That is because 
the differences between the medians of maxima (solid circles) and the medians of all regions (midlines of boxes) are very similar. Along the 
7.1\,$\leq$\,log$_{10}$\,t\,$\leq$\,9 yr range, the influence of close companions gets reflected in the $\mathrm{\Sigma_{SFR}}$ maxima since dark-blue circles 
show higher values, of at least log$_{10}\,\Sigma\mathrm{_{SFR}}\,\sim$0.1\,M$_{\odot}$\,yr$^{-1}$\,kpc$^{-2}$, 
than those of light-blue circles. For log$_{10}$\,t\,$>$\,9 yr, the maxima in control and perturbed galaxies are of the same order.
In general, the same previous notes result regarding the different subsamples (not shown for the sake of briefness). Exceptions 
are the control, SFG Green, SFG Red and SFG ETS galaxies. For the former, there is no increasing phase for the differences between median intensities. Interestingly, 
higher $\mathrm{\Sigma_{SFR}}$ maxima within the 7.1\,$\leq$\,log$_{10}$\,t\,$\leq$\,9 yr and 
within the 7.1\,$\leq$\,log$_{10}$\,t\,$\leq$\,8 yr ranges characterize control SFG Red and SFG ETS galaxies respectively. So, for both types, 
their increasing phases are steeper than the corresponding ones of their perturbed counterparts. Somehow, in these control galaxies, the regions 
hosting the $\mathrm{\Sigma_{SFR}}$ maxima received larger amounts of gas or experienced higher SFEs.

From log$_{10}$\,t\,$\sim$9.5 yr to more present times, the sSFR for all regions in perturbed galaxies (Fig.~\ref{f5} bottom row) is higher, by 
at least log$_{10}$\,sSFR\,$\sim$0.1 yr$^{-1}$, than that for all regions in control objects. Interestingly, the sSFR maxima of regions in control 
galaxies are clearly higher within 8\,$<$\,log$_{10}$\,t\,$<$\,10 yr. This result has consequences in the differences (empty circles) 
which reach $\sim$1 dex compared to those for perturbed galaxies, $\sim$0.5 dex (red circles). The maximum divergence for control galaxies is 
more a turnover point than the one for perturbed galaxies (red circles). Moreover, notice a coincidence in the maximum convergence 
for control and perturbed cases at the very past. That is, however, an omnipresent artefact\footnote{\label{note2}Since there is no accumulated mass 
for the widest bin of $t_{j,s}$, all SFRs are proportional to their related stellar masses so that the sSFR is the inverse of that $\Delta$t timespan. 
The sSFRs are then the same regardless of the stellar masses assigned by the evolutionary tracks. As cosmic time decreases, the stellar mass accumulates 
so the sSFRs start to vary.} in all sSFHs that we ignore from now on. Likewise, regarding the different subsamples, the median sSFRs 
of all regions (midlines of boxes) are slightly higher for perturbed galaxies from log$_{10}$\,t\,$\sim$9 yr to more recent times. Within 
8\,$<$\,log$_{10}$\,t\,$<$\,10 yr, medians of maxima are higher in control galaxies. Control SFG Red and SFG ETS galaxies are still interesting 
cases. Their median maxima are higher than those of their respective perturbed galaxy subsamples, particularly, within the 
7.1\,$\leq$\,log$_{10}$\,t\,$\leq$\,8 yr range. Secular processes seem to have been more effective by clearly increasing the SFE though only in the 
more recent periods and for regions in two similar types of galaxies.

In summary, the effects of close companions affect the histories of SF of all regions in general. However, no clear influence is 
found on the histories of the single regions that host the current maxima of SF. The interactions along the histories of the encounters seem not 
to match, at least, the amount of gas conveyed or the SF efficiency shown for regions in certain galaxy types of the control sample. We might relate 
this lack of influence on the maxima of SF to \textit{high-speed encounters} (also known as ``fly-bys'') that use to last much less than a typical crossing time. High-speed 
encounters use to affect the velocities and internal energies of the stellar orbits and the bulk of gas but they rarely cause disruptions. An 
analysis of the phase-space location (beyond the scope of this work, see \citealt{Gall21} and references therein) might shed light on this issue. 

\subsection{Look-back time annular profiles of SF}
\label{subsec:lbtap}

\begin{figure*}\centering
   \mbox{\includegraphics[width=1.975\columnwidth]{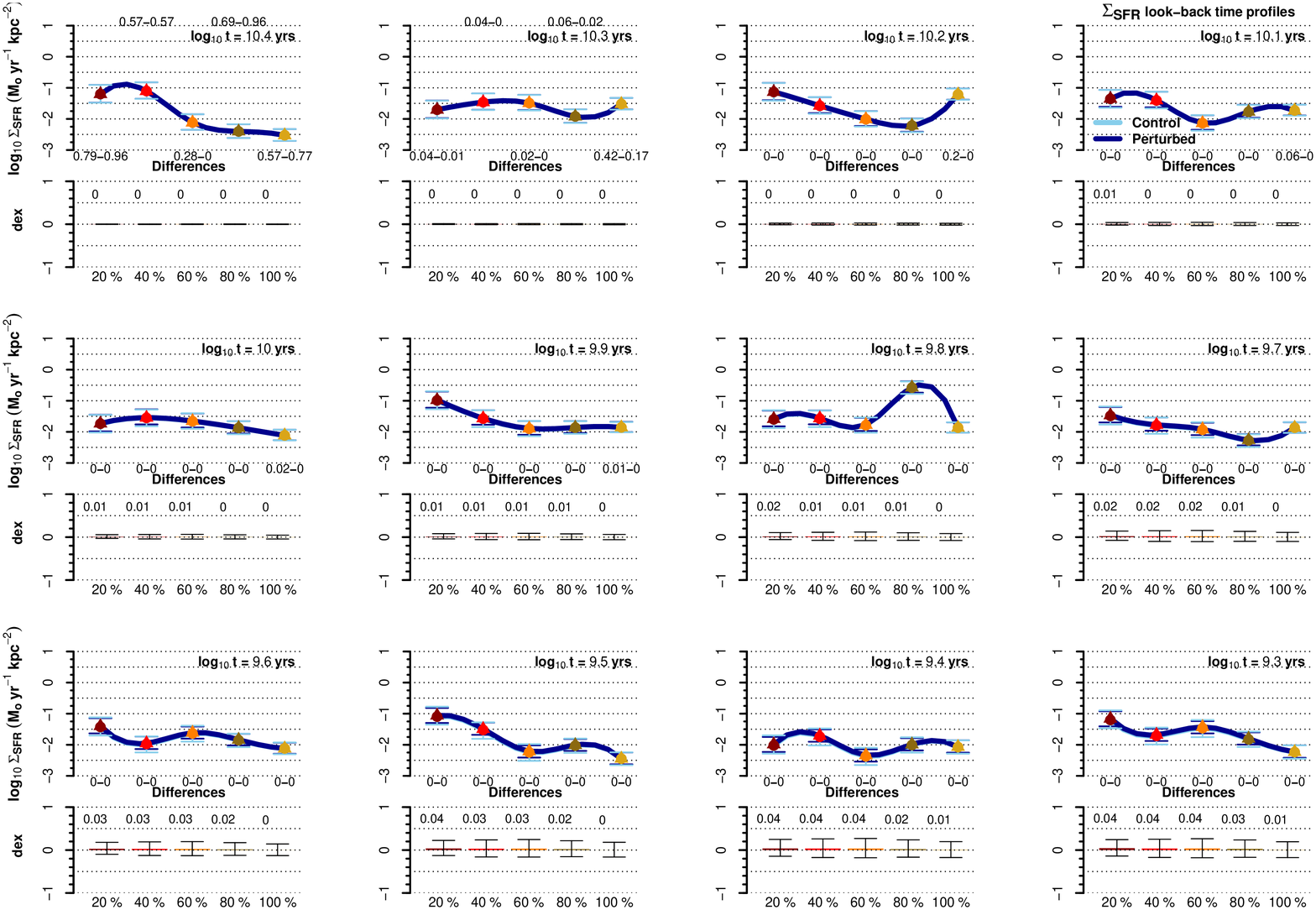}}
   \mbox{\includegraphics[width=1.975\columnwidth]{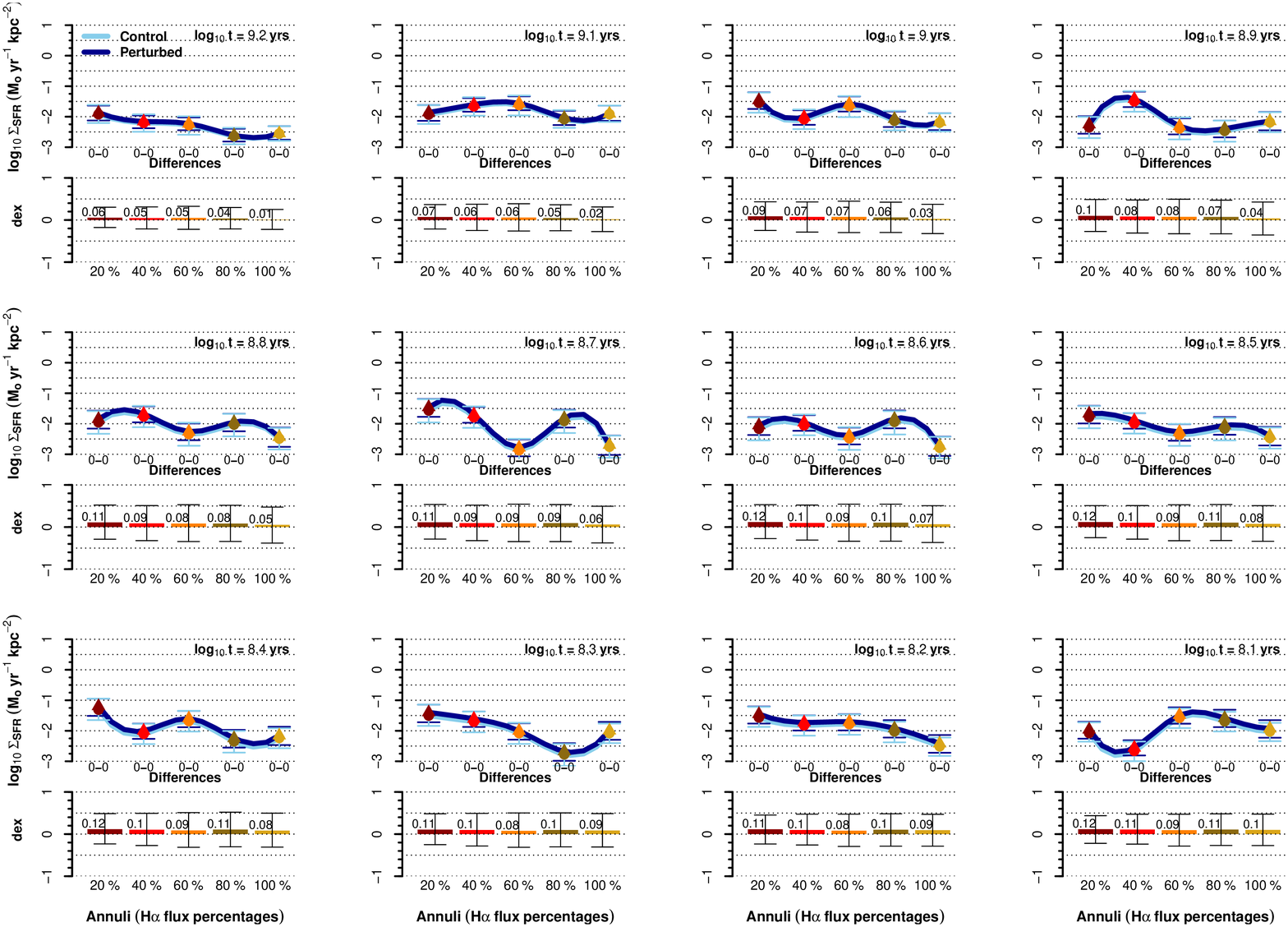}}
 \caption{\scriptsize{$\mathrm{\Sigma_{SFR}}$ look-back time annular profiles from log$_{10}$\,t$\,=\,$10.4 to log$_{10}$\,t\,$=$\,8.1 yr. Five consecutive-outward 
 annuli denote the radial extension (see Section~\ref{subsubsec:sf}). The histories in these profiles correspond to star-forming regions paired at their closest 
 current $\Sigma_{*}$ values. The ``Differences'' (bar plots) are the medians of the annular differences by subtracting the control values from the 
 perturbed ones. Symbols are control and perturbed values corresponding to each Difference. Bar and symbol lines are the interquartile ranges (IQRs, 
 1st to 3rd) of the respective distributions. Find AD and permutation test results for all annular pairs of sample distributions (likelihoods right 
 below the profiles).}}
 \label{f6.1} 
\end{figure*}

\begin{figure*}\centering
\ContinuedFloat   
   \mbox{\includegraphics[width=1.8\columnwidth]{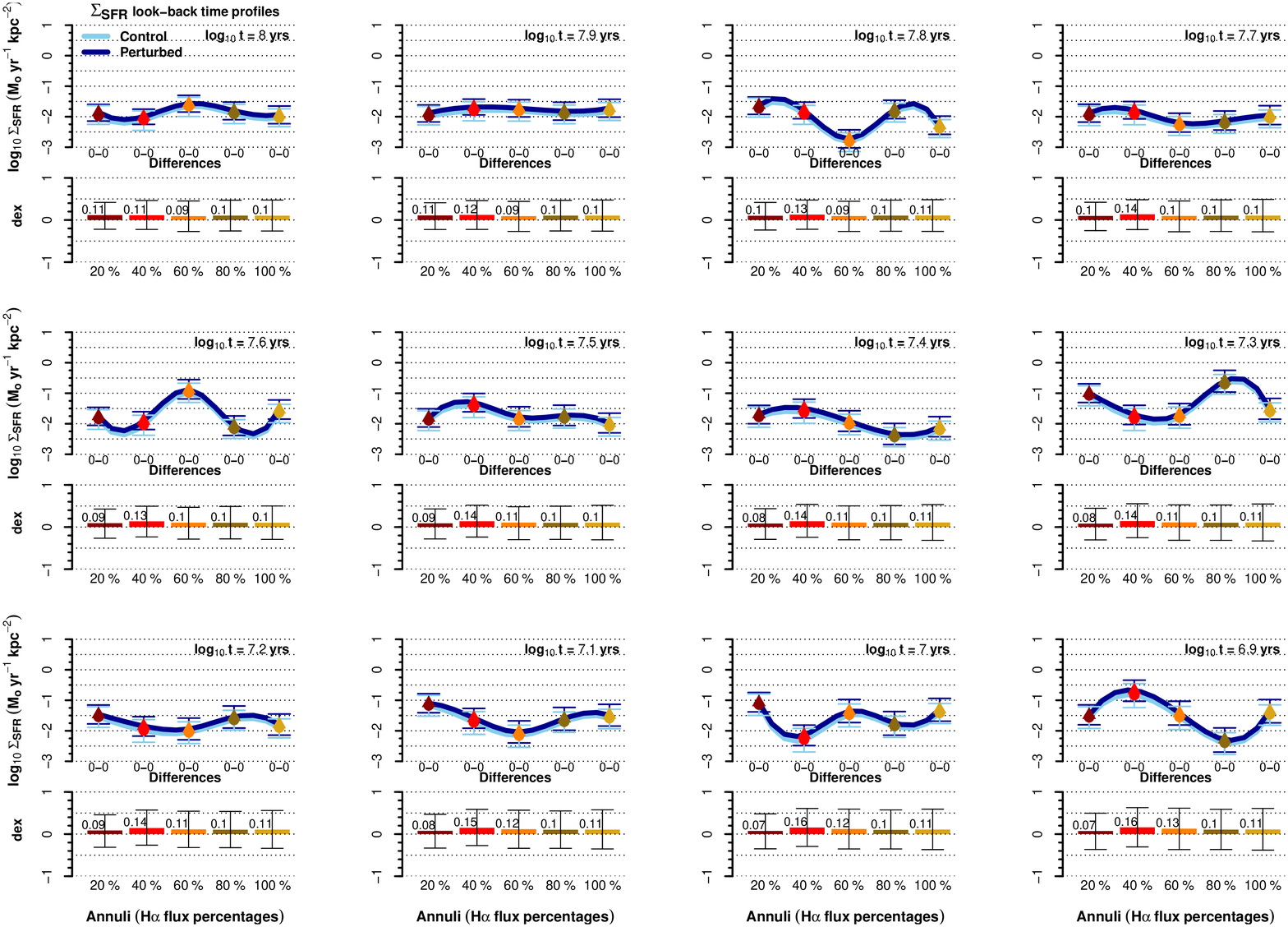}}
 \caption{\scriptsize{$\mathrm{\Sigma_{SFR}}$ look-back time annular profiles (cont.) from log$_{10}$\,t$\,=\,$8.0 to log$_{10}$\,t\,$=$\,6.9 yr. Same 
 caption as above.}}
   \label{f6.1} 
\end{figure*}

\begin{figure*}\centering
   \mbox{\includegraphics[width=1.8\columnwidth]{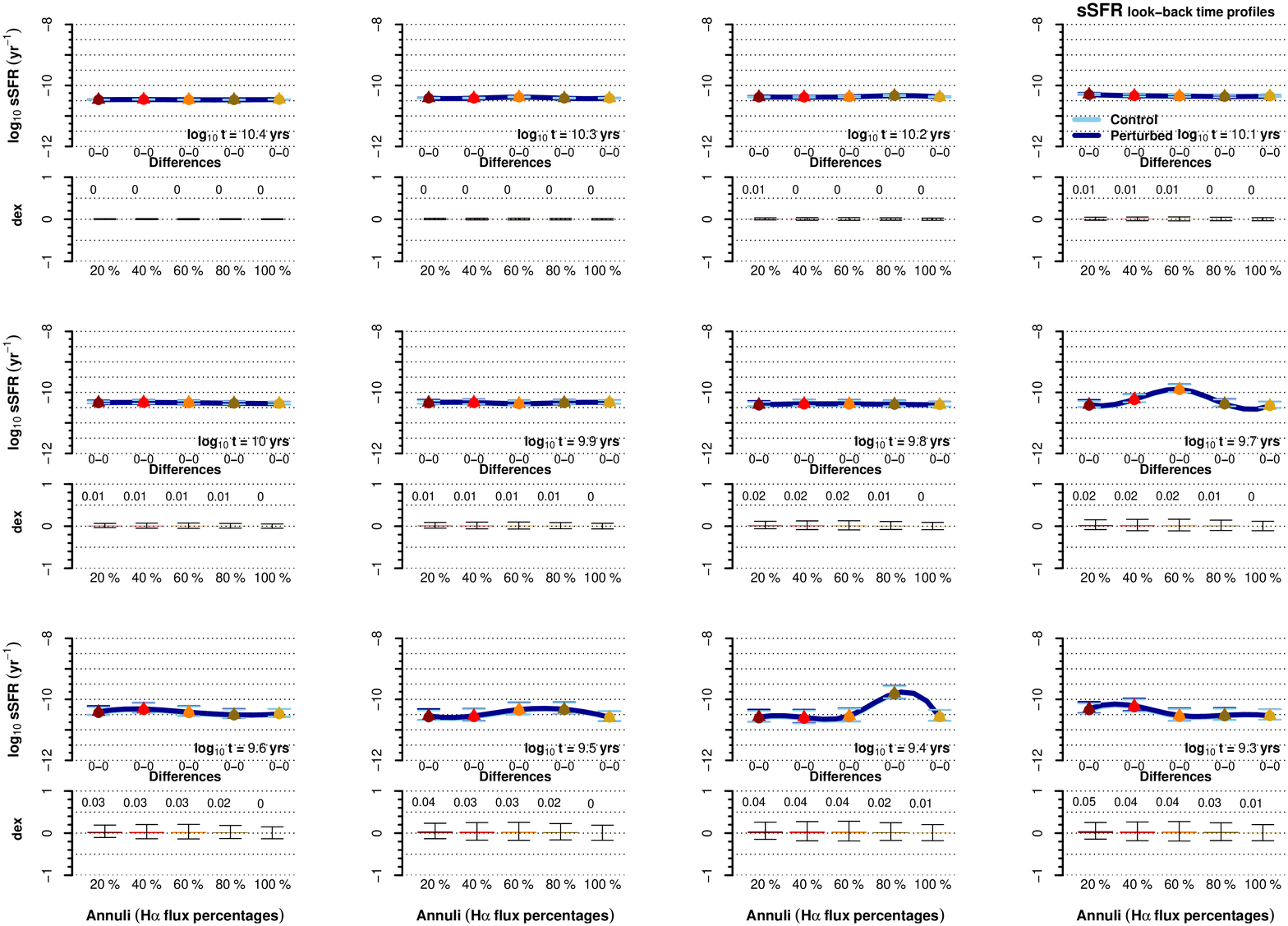}}
 \caption{\scriptsize{sSFR look-back time annular profiles from log$_{10}$\,t$\,=\,$10.4 to log$_{10}$\,t\,$=$\,9.3 yr. Invariant profiles as explained in footnote 
 13. Same caption as in Fig.~\ref{f6.1}.}}
 \label{f6.2} 
\end{figure*}

\begin{figure*}\centering
\ContinuedFloat   
   \mbox{\includegraphics[width=1.975\columnwidth]{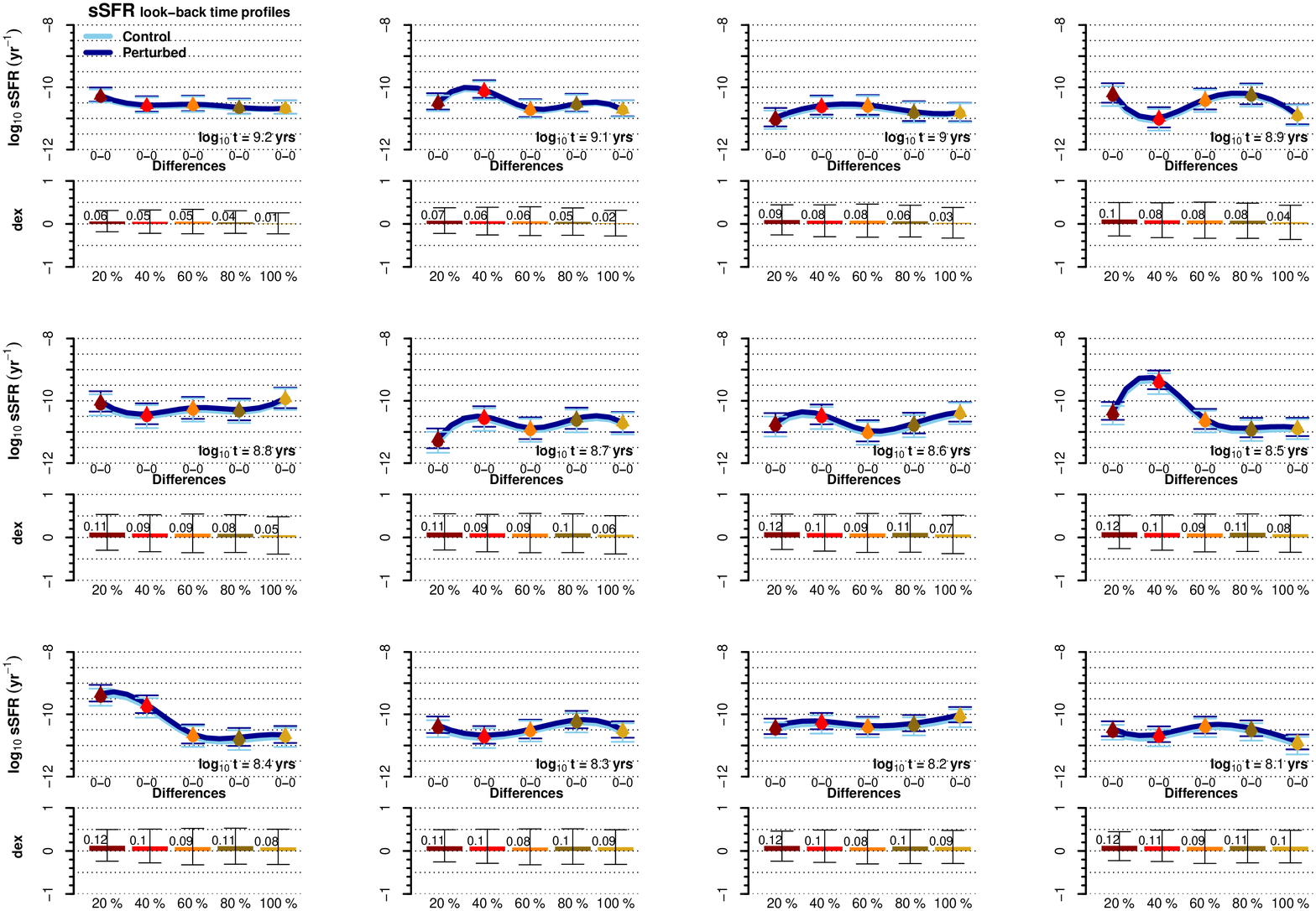}}
   \mbox{\includegraphics[width=1.975\columnwidth]{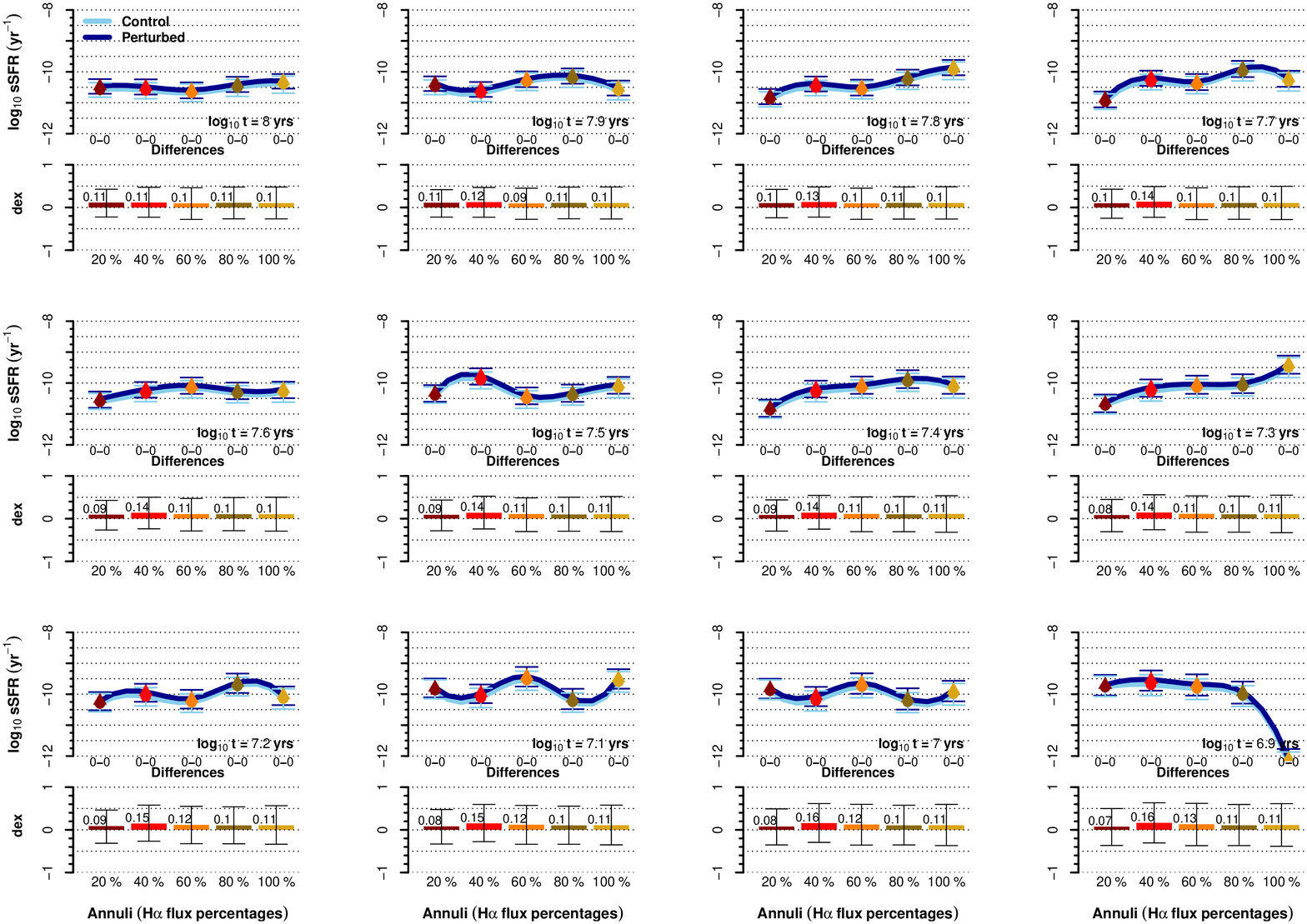}}
 \caption{\scriptsize{sSFR look-back time annular profiles (cont.) from log$_{10}$\,t$\,=\,$9.2 to log$_{10}$\,t\,$=$\,6.9 yr. Same caption as in Fig. 
 \ref{f6.1}.}}
 \label{f6.2} 
\end{figure*}

Figures~\ref{f6.1} and \ref{f6.2} respectively show the annular profiles of the $\mathrm{\Sigma_{SFR}}$ and sSFR at the look-back
times from log$_{10}$\,t\,$=$\,10.4 to log$_{10}$\,t\,$=$\,6.9 yr. From the very present to these times, these profiles contain the information which 
corresponds to the more significant M$_{*}$ fractions assembled, $>$\,0.001 and $>$\,0.01 (see Sections~\ref{subsec:sfh_ssfr} and \ref{subsec:sfr&ssfr}). 
On the other hand, Appendix~\ref{sec:app3} (available in full onlne) shows the look-back time annular profiles within the period that we consider is the 
most appropriate for a valid current SFR (see Section \ref{subsubsec:synVSHa}).

As in Section~\ref{subsec:sfh_ssfr}, we expect these profiles to corroborate that perturbed galaxies experienced, sometime in the past, 
different SFR levels in comparison with their control analogues. Notice that the Differences in Fig.~\ref{f6.1} 
(bar plots) become more obvious as the look-back time decreases. Though improved, the resolution of the SP libraries does not avoid the histories from 
being notable averages at log$_{10}$\,t\,$>$\,9 yr (see A-07, \citealt{Ib16}, \citealt{Gon17} and references therein). Notice that the Differences 
are positive and are $>$\,0.06 dex from log$_{10}$\,t\,$=$\,9.1 yr to more recent times. Moreover, most of the extensions 
of the IQRs are mostly above zero dex for these Differences. This all suggests higher $\mathrm{\Sigma_{SFR}}$ values for regions in perturbed galaxies. 
Regarding the values giving each Difference (symbols), IQRs of the distributions of control galaxies are never biased towards higher values. Also 
notice variant profiles by time bin, specifically, the positions of the central annuli with respect to the rest. In Fig.~\ref{f6.1}, half the 
time bins show the central $\mathrm{\Sigma_{SFR}}$ as the highest one. In this regard, either enhancement or suppression, observed changes in the 
central regions dominate the regulation of SF (\textit{e.g.} \citealt{Ell18}). Figure~\ref{f6.1} continuously shows fluctuations between suppression 
(reduced central $\mathrm{\Sigma_{SFR}}$) and (re)activation (central $\mathrm{\Sigma_{SFR}}$ as the highest one) of SF.

In the same way, the Differences in the sSFR profiles mostly indicate higher values for regions in perturbed galaxies (Fig.~\ref{f6.2}). 
As in the case of the $\mathrm{\Sigma_{SFR}}$, the Differences start to be $>$\,0.06 dex from log$_{10}$\,t\,$=$\,9.1 yr. 
Most of the extensions of the IQRs of the Differences are also above zero dex. Similarly, the IQRs of the distributions of regions in control galaxies 
(light-blue lines) are never biased towards higher sSFRs. As Fig.~\ref{f6.2} shows a poor tendency of pronounced gradients, its profiles confirm the 
sSFR radial distribution of SFGs as found by \citet{San20}, \textit{i.e.}, rather flat gradients.

Finally, regardless of $\Sigma\mathrm{_{SFR}}$ or sSFR, the statistical tests\footnote{The two-sample Anderson-Darling (AD) test shows whether each 
pair of annular distributions arises from a population with a common unspecified distribution function. It advantages the widely used Kolmogorov-Smirnoff 
test by being more sensitive to small differences and to those that might exist at the distribution tails. The permutation test is a non-parametric test 
that compares sets of density estimates by randomly shuffling the data so that the many permutated distribution pairs are compared under the hypothesis 
of equality (\textit{e.g.} an observed tie pattern).} on the pairs of annular distributions result all in the lowest likelihoods. It should be noticed at 
last that, for both $\mathrm{\Sigma_{SFR}}$ and sSFR, at log$_{10}$\,t\,$\geq$\,8 yr, the central Differences are never exceeded by the Differences of the 
rest annuli. Therefore, at the largest look-back times, larger amounts of gas could be conducted to the inner regions or the SF efficiencies could be 
increased consequence of the interactions. In general, within the periods of the most significant stellar masses formed, increased $\mathrm{\Sigma_{SFR}}$ 
and sSFR characterize the regions in perturbed galaxies.

\subsection{Oxygen abundances: stellar mass density \& total gas fraction}
\label{subsec:comp_che_abun}

In regard to stellar evolution, the metal content is regulated by inflows and outflows of gas and they both may be hinted by gas-phase metallicities. We 
indirectly derive the O abundance (Section~\ref{subsubsec:che}) and plot it against the $\Sigma_{*}$ and the total gas fraction ($\mu$, see \ref{subsubsec:gas}). 
Comparisons via the resolved $\mathrm{\Sigma}_{*}$-Z relation make the construction of abundance radial profiles redundant since such a local relation explains 
such profiles (BB-16).

\begin{figure}\centering
   \mbox{\includegraphics[width=1.015625\columnwidth]{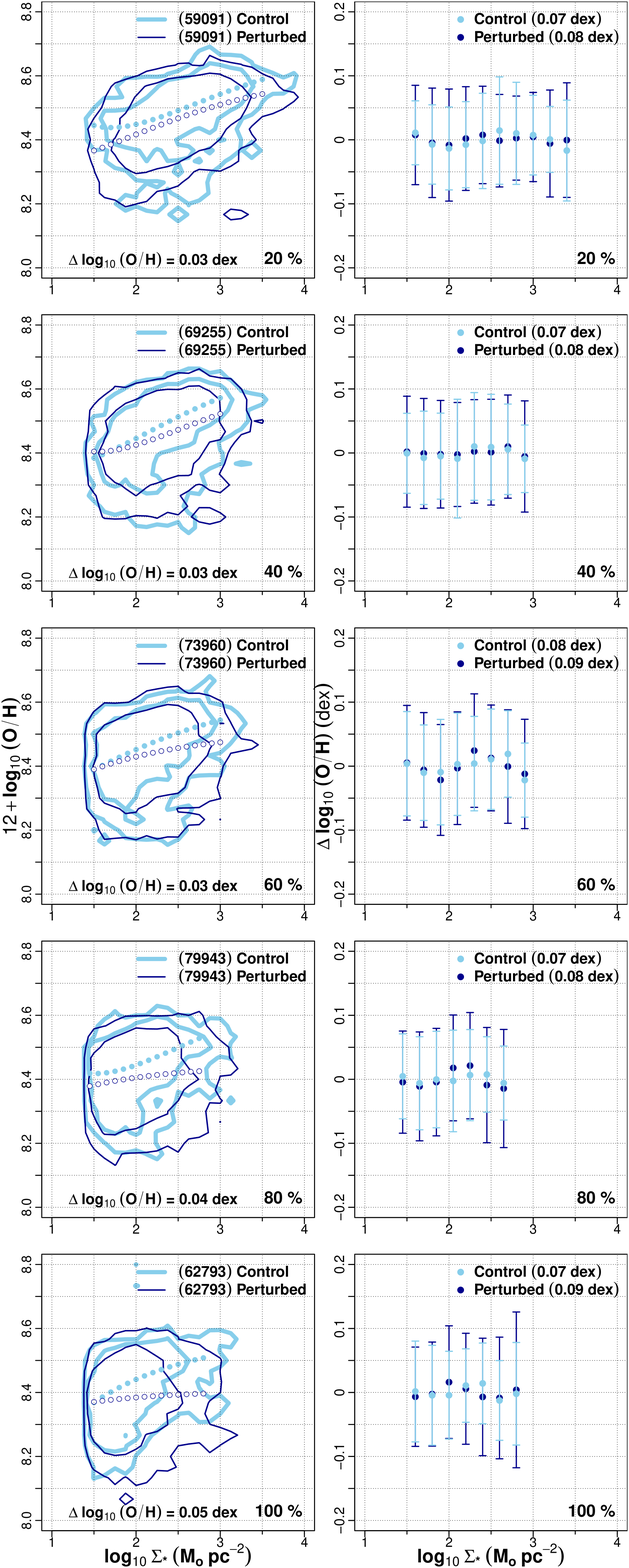}}
       \caption{\scriptsize{Left: annular $\Sigma_{*}$-Z relations. The perturbed samples merge and their star-forming regions form pairs with those of the control 
       sample by minimizing their differences in $\Sigma_{*}$. Kernel density contours show at 0.1 and 0.9 from outside-in. Sequences of solid/empty dots as the best 
       asymptotic fits at paces of 0.1 dex in the given log$_{10}$\,$\mathrm{\Sigma}_{*}$ range. Median offsets between each pair of fits 
       (log$_{10}$\,(O/H)$_{\mathrm{control}}-$log$_{10}$\,(O/H)$_{\mathrm{perturbed}}$) at the bottom. Right: annular median scatters along each fit (dots) in 0.2 
       dex bins of log$_{10}$\,$\mathrm{\Sigma}_{*}$. Vertical positive/negative lines as the standard deviation (sd) in each bin. Medians of these sds within brackets.}}
       \label{f7} 
\end{figure}

\begin{figure}\centering
   \mbox{\includegraphics[width=1.015625\columnwidth]{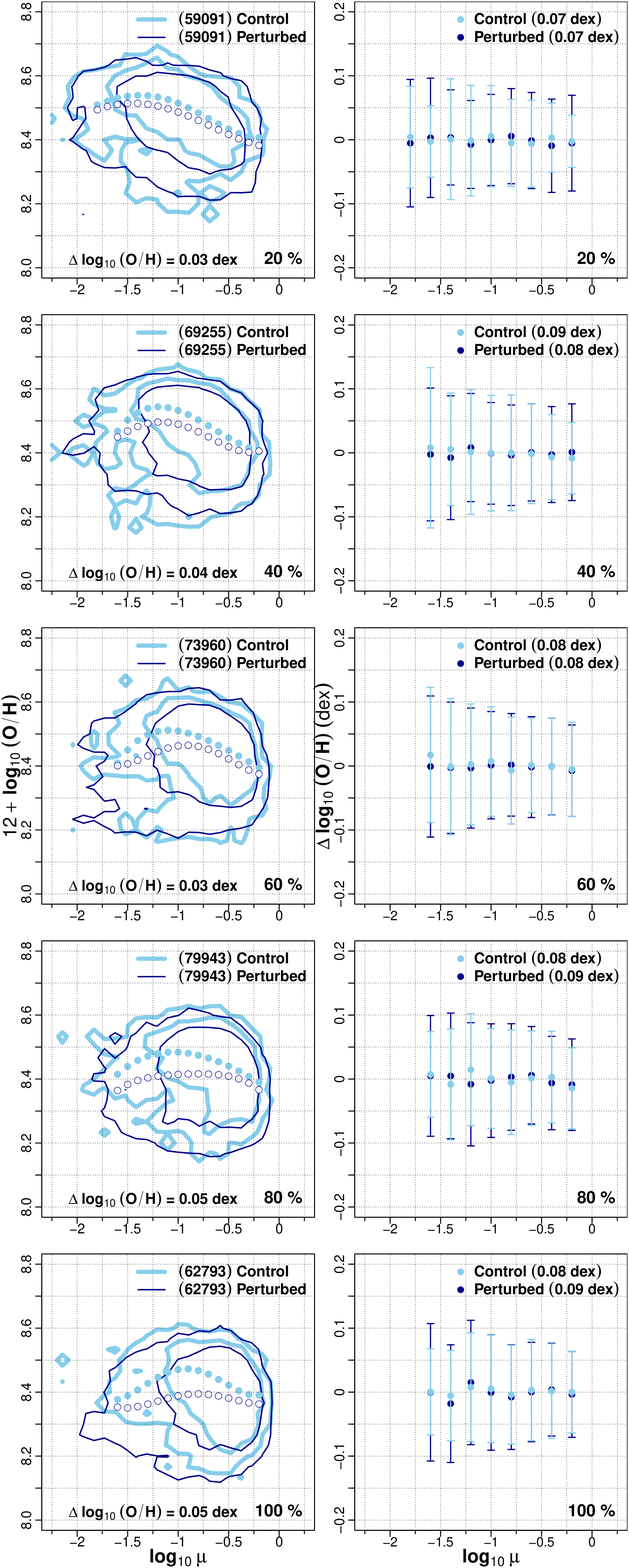}}
       \caption{\scriptsize{Left: annular $\mu$-Z relations. As in Fig.~\ref{f7}, the perturbed samples merge and their regions form pairs at the closest $\Sigma_{*}$ 
       values. Kernel density contours show at 0.1 and 0.9 from outside-in. Sequences of solid/empty dots as fourth-order fits at paces of 0.1 dex in the given 
       log$_{10}$\,$\mu$ range. Median offsets between each pair of fits (log$_{10}$\,(O/H)$_{\mathrm{control}}-$log$_{10}$\,(O/H)$_{\mathrm{perturbed}}$) at the 
       bottom. Right: annular median scatter along each fit (dots) in 0.2 dex bins of log$_{10}$\,$\mu$. Vertical positive/negative lines as the sds in each bin. 
       Medians of these sds within brackets.}}
       \label{f8} 
\end{figure}

Figure~\ref{f7} shows, in density contours, the annular $\mathrm{\Sigma}_{*}$-Z relations for the star-forming regions in all galaxies. 
Notice, from the 20 to 100\,\% annuli, a drecreasing sequence in $\Sigma_{*}$ ($\sim$1 dex offset) and O abundance ($\sim$0.1 dex offset, \textit{i.e.}, from 8.2-8.7 
to 8.1-8.6 12$+$log$_{10}$\,(O/H)). Such decrements agree with BB-16 who show that local terms can explain relations of local and global properties. We also compute 
asymptotic functions from the abundance medians within $\Sigma_{*}$ bins of 0.2 dex width (Section~\ref{subsubsec:relation}) and show the fits at paces of 0.1 dex 
(dot sequences). Notice that control galaxies show slightly larger O contents (offsets of left panels)\footnote{Comparable offsets have been found from galaxy pairs 
and simulations of colliding galaxies (\textit{e.g.} \citealt{Ell08}, \citealt{Torr12}, \citealt{Gar20}).}. These small but detectable differences are at the closest 
$\Sigma_{*}$ values of the star-forming regions. We propose in Section~\ref{subsubsec:inflows} that the influence of close companions can explain such differences.

Moreover, Fig.~\ref{f7} (right) shows median scatters (dots) and their standard deviations (sds, vertical lines) along each annular fit. Though the dots for perturbed 
galaxies (dark-blue) are in some annuli the closest to zero dex (20 and 40\,\%), the sds (dark-blue lines) are in general larger than those of control galaxies 
(light-blue lines). This results even though the medians of these sds are very similar for both samples (compare the values within brackets).

Figure~\ref{f8} shows the annular $\mu$-Z relation. Likewise, we compute median contents within $\mu$ bins of 0.2 
dex width. We also fit fourth-order functions at paces of 0.1 dex. As a function of $\mu$, the control sample still shows 
slightly larger O contents (offsets of left panels). The fits indicate an increment of the abundance with decreasing $\mu$ followed by a decrement 
of the former at the lowest fractions (in partial agreement with BB-18). From the mid annulus outwards, the decrement in perturbed galaxies starts 
at higher gas fractions. We explore this deviation in Section~\ref{subsubsec:f}.

Figure~\ref{f8} (right) shows a tight $\mu$-Z relation: the sds of the scatters rarely exceed $\pm$0.1 dex. It is hard to determine whether control or perturbed 
galaxies are the closest to zero scatter since the medians of the sds of the scatters are very similar (values within brackets). Moreover, the dispersion of the 
offsets along BB-18 fit is just slightly larger than the one of the offsets along BB-16 $\Sigma_{*}$-Z fit ($\sim$0.09 against $\sim$0.08 dex). The  $\Sigma_{*}$-Z 
is then a more robust local empirical relation than the $\mu$-Z or the V$_{\mathrm{esc}}$-Z (escape velocity) relation. Both the latter are shown by BB-18 who 
conclude that the main gas-phase metal tracer is $\Sigma_{*}$ because it indirectly defines $\mu$ and V$_{\mathrm{esc}}$. By comparing the medians of sds of scatters 
(brackets in the right panels of Figs.~\ref{f7} and \ref{f8}), the annuli in which the $\mu$-Z relation has larger values are the 40, 80 and 100\,\% for control, 
and only the 80\,\% for perturbed galaxies.

Following the subsample division, Table~\ref{tab:2} gives a summary of the offsets between the fits (control and perturbed) and their respective sds of their 
scatters. Numbers under ``All'' correspond to Figs.~\ref{f7} and \ref{f8}. Regarding the median offsets (``offset'' rows), notice slightly higher contents for 
perturbed AGN-like and SFG ETS galaxies (negative offsets in 3/5 annuli). Moreover, in SFG Green objects the contents appear to be in balance whereas in the 
rest subsamples they are slightly superior for control galaxies. Still regarding the ``offset'' rows, the subsamples SFG Blue-SFG LTS, and SFG-All, practically 
show the same values. In the case of the median scatters along the fits, relations for the control sample are in general tighter (slightly lower numbers in the 
``sd C'' rows). Similarly, annuli in which the $\mu$-Z fit has higher median scatters are 22/40 for control and only 3/40 for perturbed galaxies.

\begin{table}%[h!]
   \setlength{\tabcolsep}{0.475\tabcolsep}
 \begin{minipage}{\columnwidth}
\caption{\scriptsize{The $\Sigma_{*}$-Z and $\mu$-Z relation summaries for the subsample division of PaperI. As the left panels of Figs.~\ref{f7} and \ref{f8}, 
``offset'' rows give the median offsets between each respective pair of fits (either asymptotic, $\Sigma_{*}$-Z, or fourth-order, $\mu$-Z). As the brackets in 
the right panels of Figs.~\ref{f7} and \ref{f8}, ``sd C'', ``sd P'' rows respectively give the medians of the standard deviations of the scatters along the 
control and perturbed fits.}
 \label{tab:2}}
  \centering
 \begin{scriptsize}
 \begin{tabular}{@{\hspace{0.01\tabcolsep}}lccccccccccc}
 \hline
       &\multicolumn{5}{c}{Annuli (H$\alpha$ flux percentages)}                      &&\multicolumn{5}{c}{Annuli (H$\alpha$ flux percentages)}\\
       &20\% &40\%             &60\%             &80\%             &100\%            &&20\% &40\% &60\% &80\%            &100\%\\
\cline{2-6}\cline{8-12}                                                                                                                                                             \\
       &\multicolumn{11}{c}{$\Sigma_{*}$-Z (dex)}                                                                              \\
       &\multicolumn{5}{c}{AGN-like}                                                 &&\multicolumn{5}{c}{SFG Red}             \\[1ex]
offset &0.01 &\phantom{0}0.00  &$-$0.02          &$-$0.02          &$-$0.03          &&0.05 &0.06 &0.08 &0.06            &0.03 \\
sd C   &0.04 &\phantom{0}0.04  &\phantom{0}0.05  &\phantom{0}0.05  &\phantom{0}0.04  &&0.03 &0.04 &0.05 &0.06            &0.05 \\
sd P   &0.04 &\phantom{0}0.06  &\phantom{0}0.06  &\phantom{0}0.06  &\phantom{0}0.06  &&0.07 &0.07 &0.08 &0.10            &0.11 \\[1ex]
       &\multicolumn{5}{c}{SFG ETS}                                                  &&\multicolumn{5}{c}{SFG Green}           \\[1ex]
offset &0.00 &$-$0.03          &$-$0.01          &$-$0.01          &\phantom{0}0.04  &&0.02 &0.01 &0.01 &$-$0.01         &0.01 \\
sd C   &0.05 &\phantom{0}0.04  &\phantom{0}0.05  &\phantom{0}0.05  &\phantom{0}0.04  &&0.05 &0.06 &0.06 &\phantom{0}0.04 &0.05 \\
sd P   &0.07 &\phantom{0}0.07  &\phantom{0}0.07  &\phantom{0}0.06  &\phantom{0}0.08  &&0.06 &0.07 &0.08 &\phantom{0}0.08 &0.07 \\[1ex]
       &\multicolumn{5}{c}{SFG Blue}                                                 &&\multicolumn{5}{c}{SFG LTS}             \\[1ex]
offset &0.05 &\phantom{0}0.05  &\phantom{0}0.04  &\phantom{0}0.05  &\phantom{0}0.05  &&0.04 &0.05 &0.04 &\phantom{0}0.05 &0.05 \\
sd C   &0.06 &\phantom{0}0.08  &\phantom{0}0.08  &\phantom{0}0.08  &\phantom{0}0.07  &&0.06 &0.07 &0.08 &\phantom{0}0.07 &0.07 \\
sd P   &0.08 &\phantom{0}0.08  &\phantom{0}0.09  &\phantom{0}0.08  &\phantom{0}0.09  &&0.08 &0.08 &0.09 &\phantom{0}0.08 &0.09 \\[1ex]
       &\multicolumn{5}{c}{SFG}                                                      &&\multicolumn{5}{c}{All}                 \\[1ex]
offset &0.03 &\phantom{0}0.03  &\phantom{0}0.03  &\phantom{0}0.04  &\phantom{0}0.05  &&0.03 &0.03 &0.03 &\phantom{0}0.04 &0.05 \\
sd C   &0.06 &\phantom{0}0.07  &\phantom{0}0.08  &\phantom{0}0.07  &\phantom{0}0.07  &&0.07 &0.07 &0.08 &\phantom{0}0.07 &0.07 \\
sd P   &0.08 &\phantom{0}0.08  &\phantom{0}0.09  &\phantom{0}0.08  &\phantom{0}0.09  &&0.08 &0.08 &0.09 &\phantom{0}0.08 &0.09 \\[1ex]
       &\multicolumn{11}{c}{$\mu$-Z (dex)}                                                                                     \\
       &\multicolumn{5}{c}{AGN-like}                                                 &&\multicolumn{5}{c}{SFG Red}             \\[1ex]
offset &0.00 &\phantom{0}0.01  &$-$0.01          &$-$0.02          &$-$0.02          &&0.03 &0.06 &0.08 &0.06            &0.04 \\
sd C   &0.03 &\phantom{0}0.03  &\phantom{0}0.05  &\phantom{0}0.06  &\phantom{0}0.03  &&0.03 &0.04 &0.05 &0.06            &0.04 \\
sd P   &0.04 &\phantom{0}0.04  &\phantom{0}0.06  &\phantom{0}0.05  &\phantom{0}0.05  &&0.06 &0.06 &0.08 &0.09            &0.09 \\[1ex]
       &\multicolumn{5}{c}{SFG ETS}                                                  &&\multicolumn{5}{c}{SFG Green}           \\[1ex]
offset &0.00 &$-$0.03          &$-$0.01          &$-$0.01          &\phantom{0}0.03  &&0.02 &0.01 &0.00 &\phantom{0}0.00 &0.01 \\
sd C   &0.05 &\phantom{0}0.05  &\phantom{0}0.06  &\phantom{0}0.05  &\phantom{0}0.04  &&0.06 &0.06 &0.05 &\phantom{0}0.04 &0.05 \\
sd P   &0.05 &\phantom{0}0.06  &\phantom{0}0.06  &\phantom{0}0.06  &\phantom{0}0.06  &&0.06 &0.07 &0.07 &\phantom{0}0.08 &0.07 \\[1ex]
       &\multicolumn{5}{c}{SFG Blue}                                                 &&\multicolumn{5}{c}{SFG LTS}             \\[1ex]
offset &0.05 &\phantom{0}0.06  &\phantom{0}0.05  &\phantom{0}0.07  &\phantom{0}0.07  &&0.04 &0.05 &0.05 &\phantom{0}0.06 &0.06 \\
sd C   &0.07 &\phantom{0}0.10  &\phantom{0}0.10  &\phantom{0}0.09  &\phantom{0}0.08  &&0.07 &0.10 &0.10 &\phantom{0}0.08 &0.08 \\
sd P   &0.07 &\phantom{0}0.09  &\phantom{0}0.08  &\phantom{0}0.08  &\phantom{0}0.08  &&0.07 &0.08 &0.09 &\phantom{0}0.08 &0.09 \\[1ex]
       &\multicolumn{5}{c}{SFG}                                                      &&\multicolumn{5}{c}{All}                 \\[1ex]
offset &0.03 &\phantom{0}0.04  &\phantom{0}0.04  &\phantom{0}0.05  &\phantom{0}0.06  &&0.03 &0.04 &0.03 &\phantom{0}0.05 &0.05 \\
sd C   &0.07 &\phantom{0}0.09  &\phantom{0}0.09  &\phantom{0}0.08  &\phantom{0}0.08  &&0.07 &0.09 &0.08 &\phantom{0}0.08 &0.08 \\
sd P   &0.07 &\phantom{0}0.08  &\phantom{0}0.08  &\phantom{0}0.09  &\phantom{0}0.08  &&0.07 &0.08 &0.08 &\phantom{0}0.09 &0.09 \\[1ex]
\hline\\
 \end{tabular}
 \end{scriptsize}
 \end{minipage}
 \end{table}

Finally, our metallicity differences are in agreement with analyses on 
close pairs and asymmetric galaxies \citep[\textit{e.g.}][]{Ell08,Row18}. \citet{Row18} 
implicitly treat asymmetric and tidally-perturbed galaxies as alike. Their result of the overall metallicity of spaxels 
being lower in asymmetric galaxies is qualitatively similar to that of \citet{Ell08}. We do not prove our perturbed 
galaxies as asymmetric, however, by the use of concentric annuli, we confirm the result of \citet{Row18}, that of global metallicity differences.
 
\section{Discussion}
\label{sec:dis}

\subsection{SF (re)activation: differences at past times}
\label{subsec:SF(re)}

The SFH allows us to visualize phases and sequences of constant SF and SF bursts. Galaxies with positive SFH slopes, mainly Es 
and early-type Ss, formed most of their stars in the past. Today starbursts are forming theirs in a recent burst or sequence of bursts \citep[\textit{e.g.}][]{Tor12}, 
showing, on the contrary, negative slopes. In other words, more massive galaxies have a stronger SFR at earlier times while less massive ones have peak SFRs at 
lower $z$ \citep{San20}. An activation of SF in recent times in log$_{10}$\,M$_{*}$\,$\sim(10.5$-$10.9)$\,M$_{\odot}$ Sbc-Sc types is detected by \citet{Gon17} and 
related to rejuvenation. The mass-cumulative curve of CALIFA survey spiral galaxies \citep{Gar17} proves this rejuvenation which may be the consequence of either the 
consumption of the cold gas reservoirs driven by self-gravity, or inflow/accretion of low-metallicity gas \citep[\textit{e.g.}][]{Row18} result of interactions. We 
confirm this rejuvenation only in the two outmost annuli of the radial extension within the period where 0.001$<$\,1$-\eta_{*}(t_{j,s})\,\leq\,$0.01 (see Sections 
\ref{subsec:sfh_ssfr} to \ref{subsec:lbtap}). Though this SF (re)activation is detected in both control and perturbed galaxies, statistically significant differences 
indicate higher values for the latter. Moreover, the annular profiles along the 1$-\eta_{*}(t_{j,s})\,>\,$0.001 time scale show superior SF levels for perturbed 
galaxies. Contrary to the overall regular levels of SF, the histories of the regions hosting the current SF maxima in perturbed galaxies seem rather unenhanced, 
perhaps, due to weak interactions along the histories of some encounters (presumably of the type of high-speed).

The case of interest left is that of the contrasts in O content. We look for a relation between these contrasts and inflows of low metallicity gas.

\begin{figure}\centering
   \mbox{\includegraphics[width=.6725\columnwidth]{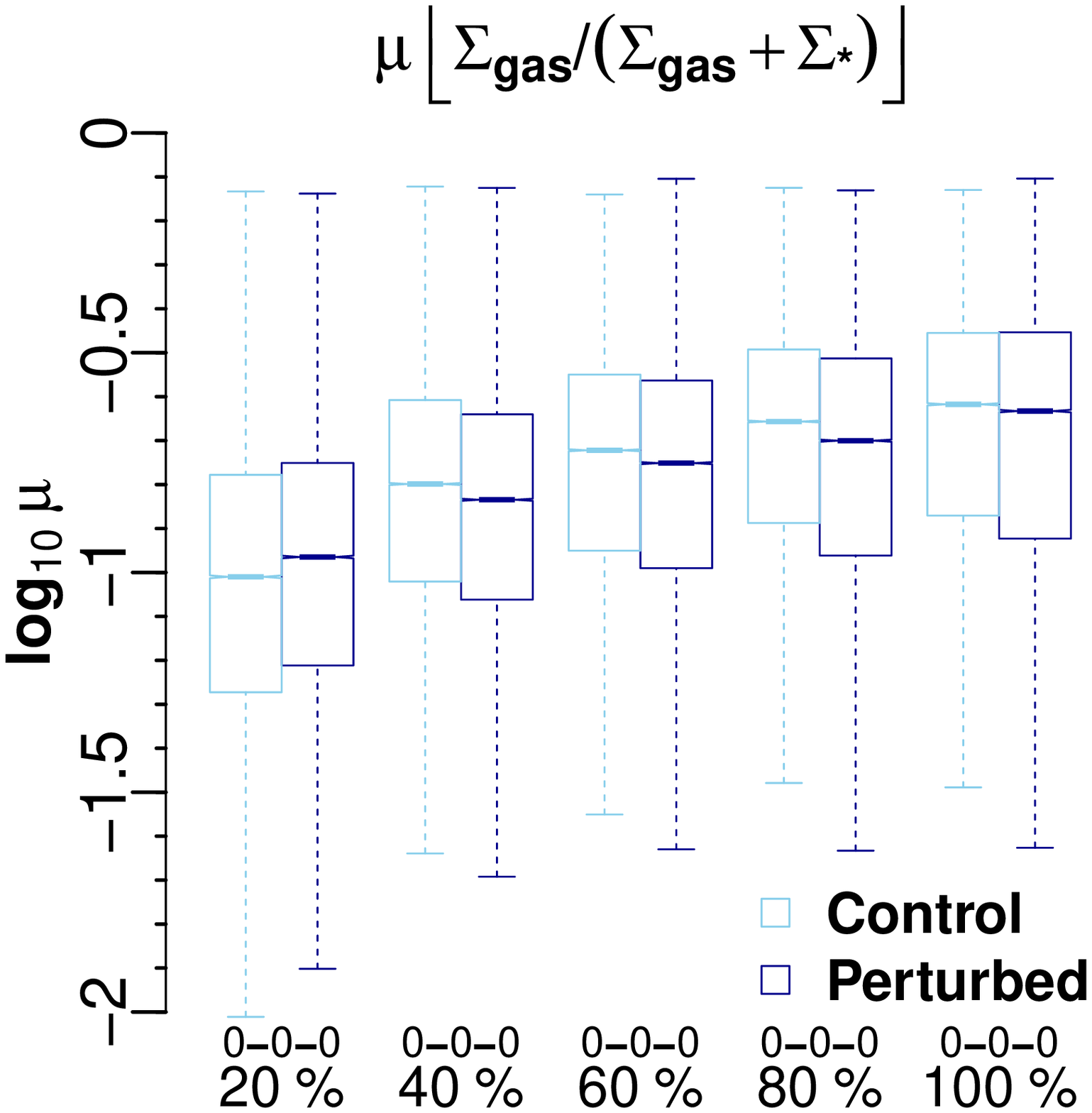}}\\
   \mbox{\includegraphics[width=.6725\columnwidth]{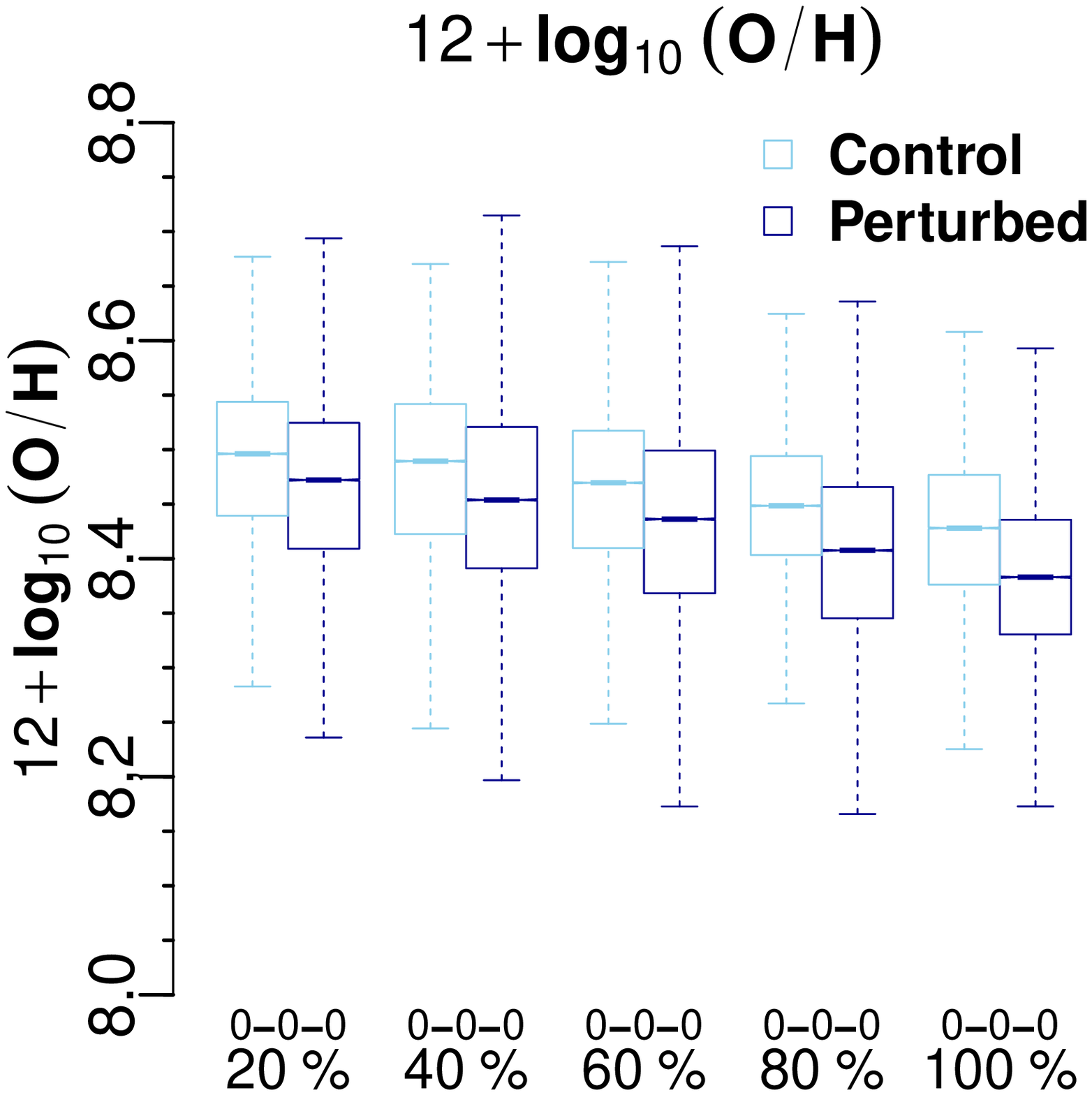}}\\
   \mbox{\includegraphics[width=.6725\columnwidth]{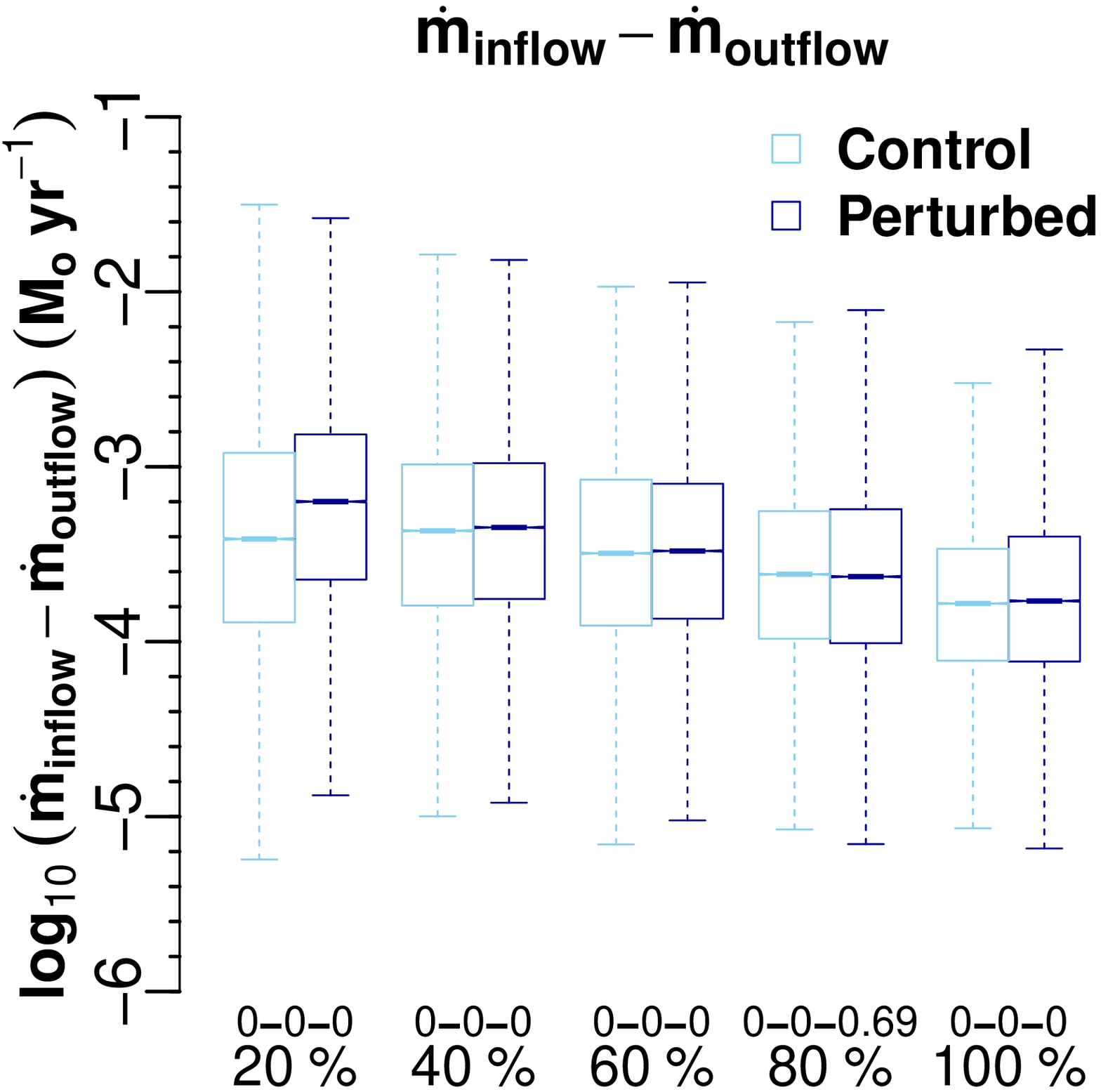}}\\
   \mbox{\includegraphics[width=.6725\columnwidth]{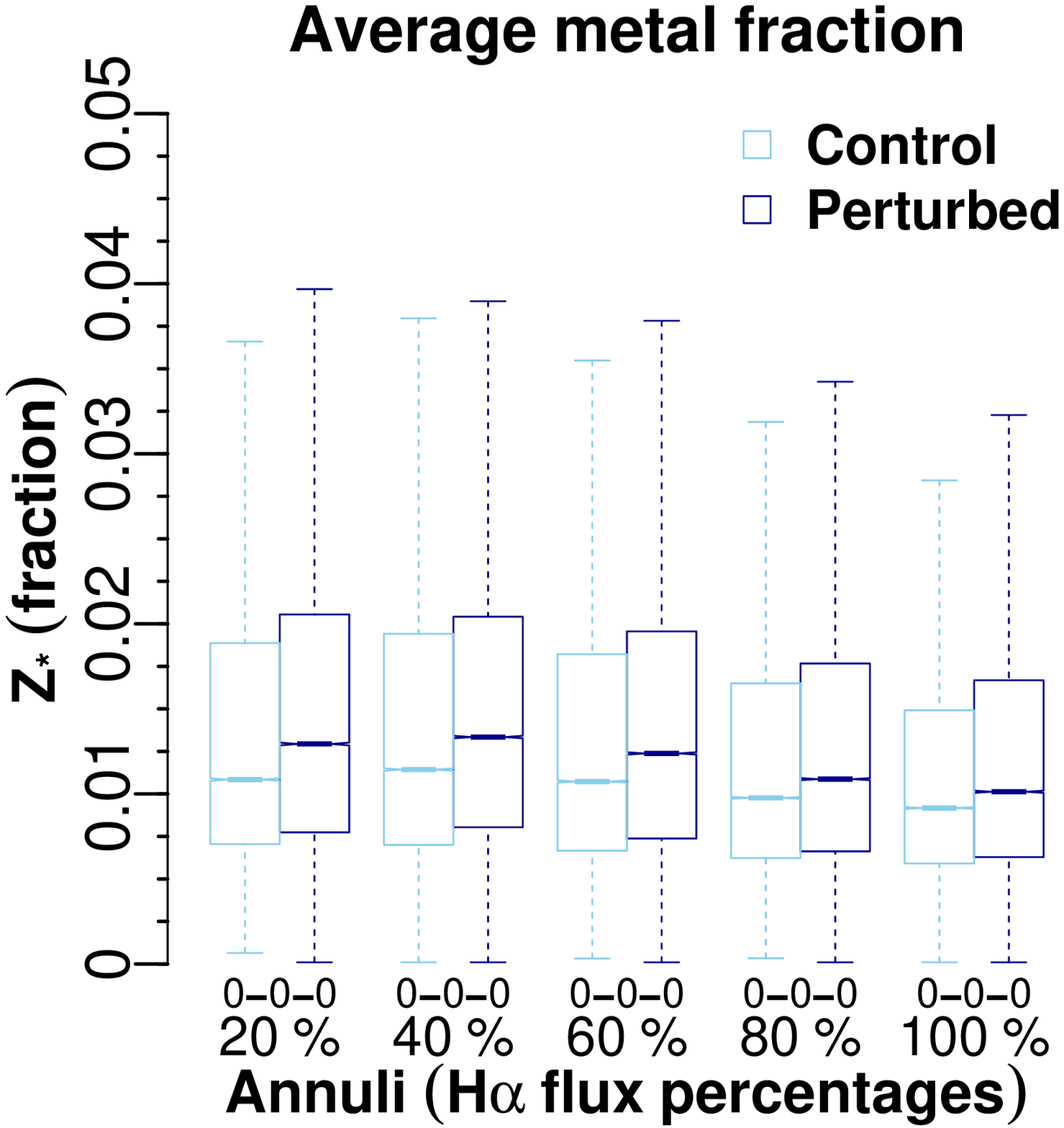}}
   \caption{\scriptsize{Boxplots of the annular distributions of $\mu$, $12+\mathrm{log_{10}\,(O/H)}$, $\mathrm{\dot{m}_{inflow}}-\mathrm{\dot{m}_{outflow}}$, and 
   the stellar metallicity (Z$_{*}$) fraction (the latter two described below). The perturbed samples merge, and their regions form pairs with those of the control 
   sample at the closest $\Sigma_{*}$ values. Find AD, permutation and Mann-Whitney (AD-permutation-MW) test results for all pairs of annular distributions right 
   below the boxplots.}}
   \label{f9} 
\end{figure}

\subsection{Tidal interactions and metal dilutions}
\label{subsec:tid_met}

\subsubsection{The $\Sigma_{*}$-Z relation of perturbed galaxies}
\label{subsubsec:sigmamass}

Considering the shifts in the M$_{*}$-$\Sigma_{*}$ relation, \citet{San13} showed a very consistent shape for both the global and local mass-metallicity relations. 
In a given range of $z$ and M$_{*}$, such a shape does not change with \textit{z}. \citet{Mou11} and \citet{LaLo13} report a shift in the M$_{*}$-Z relation with 
$z$ but towards lower metallicities. Recalling that BB-16 galaxies are biased towards larger redshifts, we separately compare our asymptotic fits with the one of 
BB-16 (as in Section~\ref{subsubsec:relation}). We find $\Delta\,\mathrm{log_{10}\,(O/H)}$\footref{off} distributions (1st, 2nd and 3rd Qs respectively) biased 
towards positive (0.009, 0.012 and 0.018 dex) and negative ($-$0.017, $-$0.012 and $-$0.002 dex) values for control and perturbed galaxies respectively. Notice 
that the O abundances of control galaxies tend to agree with the result of \citet{Mou11} and \citet{LaLo13} whereas those of perturbed galaxies do not. Then, the 
star-forming regions of this work are inclined to be above/below the BB-16 mass-metallicity relation.

\subsubsection{Gas inflows}
\label{subsubsec:inflows}

There seems to be a general agreement in that the flow of gas towards central regions of galaxies provokes a dilution of the metallicity 
\citep[\textit{e.g.}][]{Mon10,Rup10,Per11,Ell13,Ell18,Row18}. Such a dilution is global rather than only central \citep{Row18}. On the other hand, it is well 
established that metallicity in disk galaxies depends most fundamentally on $\Sigma_{*}$ than on other spatially-resolved property (\textit{e.g.} \citealt{Ros12}; 
\citealt{San13}; BB-16; BB-18). It is also well known that this local relation defines the global one. Moreover, models of galaxy evolution have also explained 
the metal content based on secondary properties. For instance, \citet{Lil13} propose a dependency on a metal \emph{yield} (\textit{i.e.} the amount that long-lived 
stars return) and $\mu$. In general, $\mu$ decreases with $\Sigma\mathrm{_{*}}$ and $12+\mathrm{log_{10}\,(O/H)}$ increases with $\Sigma\mathrm{_{*}}$. Then, the 
O abundance increases with decreasing $\mu$. In sum, the lower density regions, rich in gas, are less chemically-evolved (BB-16). This discussion is in reference 
to the lower annular O contents shown by perturbed galaxies.

In this regard, Fig.~\ref{f9} shows, for all galaxies, the annular distributions of $\mu$, $12+\mathrm{log_{10}\,(O/H)}$, $\mathrm{\dot{m}_{inflow}}-\mathrm{\dot{m}_{outflow}}$ and, 
the stellar metallicity (Z$_{*}$) fraction (the latter two described below). We give AD, permutation and Mann-Whitney (MW) test results for each pair of annular distributions. In just 
a few cases the boxplot notches coincide between distributions ($\mathrm{\dot{m}_{inflow}}-\mathrm{\dot{m}_{outflow}}$). That is why we add the MW test.\footnote{The null hypothesis of 
this test is that the distributions to be compared differ by a location shift (mainly the median). Rejection of the hypothesis results if the test \textit{p-values} are higher than the 
used statistical level.} Starting with the $\mu$ distributions, notice that those of perturbed galaxies are lower than those of control ones with the only exclusion of the central annulus. 
AD and permutation tests suggest very low likelihoods of coming from a common distribution and of being equal while MW p-values indicate different medians. The $12+\mathrm{log_{10}\,(O/H)}$ 
distributions confirm the result of Section~\ref{subsec:comp_che_abun}, star-forming regions residing in perturbed galaxies possess lower O abundances: their IQRs are lower than the 
corresponding ones for control galaxies, and the statistical tests suggest different distribution functions and distribution medians. An observed strong inverse correlation between the 
local values of $\mu$ and $\Sigma_{*}$ (BB-16) implies an observed strong inverse correlation between $\mu$ and Z \citep{Car15}. Notice, from the $\mu$ and $12+\mathrm{log_{10}\,(O/H)}$ 
distributions, that the latter note is true only in the central annulus. Higher values in SF properties characterize the central regions of perturbed galaxies (SFG LTS, SFG and all 
galaxies, see PaperI).

BB-18 explore the metallicity dependence on secondary properties. By fitting evolution models of the local metallicity, they highlight the contribution of 
V$\mathrm{_{esc}}$ (outflows driven by momenta supplied by massive stars and supernovae) in setting the O abundance. The model which best describes the data of BB-18 
\citep[the \emph{gas-regulator} model,][]{Lil13,Car15} depends mainly on the total mass of gas. We use this model via their equation 9. It relates the inflow rate of 
metal-poor gas to the rate at which enriched material escapes from any given star-forming region:
\begin{equation} \label{eq:5}
 \mathrm{\dot{m}_{inflow}}-\mathrm{\dot{m}_{outflow}} = \mathrm{SFR} \left[ 0.6 \left( 1 + \frac{\mu}{1-\mu} \right) - 0.25 \right]\,\,\,\mathrm{M}_{\odot}\,\mathrm{yr}^{-1}.
\bigskip
\end{equation}

\noindent This relation is derived from a primary mass conservation equation (see BB-18 for the assumptions taken). To discard the fact that the lower abundances of 
perturbed galaxies might be due to more intense feedback from massive stars and supernovae, we inspect the annular rate-differential distributions 
($\mathrm{\dot{m}_{inflow}}-\mathrm{\dot{m}_{outflow}}$, Fig.~\ref{f9}). For perturbed galaxies, medians of these rate differentials are higher in 4 annuli (20, 40, 
60 and 100\,\%). Their IQRs are also higher in 3 annuli (20, 40 and 100\,\%). The statistical test results validate all this. These rate differentials suggest that 
gas inflows are more prominent than gas outflows in perturbed galaxies. 

On the other hand, \citet{Kan21} model the chemical evolution of SFGs based on the O production by massive, short-lived stars (instantaneous recycling). Inflows and 
outflows are considered proportional to the SFR and, their respective factors, the gas mass accretion ($\omega$) and the load of outflow ($\lambda$) are free parameters. 
\citet{Kan21} prove the importance of Z$_{*}$ to characterize the contributions of $\omega$ and $\lambda$ since the gas-phase metallicity may be affected by degeneracies 
between both factors.\footnote{There are comparisons of the analytical solutions, as function of $\omega$ and/or $\lambda$, that do not suggest a reduction of the 
gas-phase metallicity \citep[see][their figure 1]{Kan21}.} For a given $\lambda$, the average Z$_{*}$ increases with $\omega$, \textit{i.e.}, the larger the $\omega$, 
the larger the fraction of stars eventually formed and hence a higher Z$_{*}$ results. From the SP synthesis, Fig.~\ref{f9} shows the average light-weighted Z$_{*}$ 
fractions for the star-forming regions in our sampled galaxies. Higher fractions correspond to regions in perturbed galaxies. Notice also that the statistical tests 
suggest different parent distributions and different distributions and median values. For the medians, the offsets along the annular sequence are 0.08, 0.07, 0.06, 
0.05 and 0.04 dex from the centre. These Z$_{*}$ results and the SP age profiles of PaperI suggest a faster evolution for perturbed galaxies by having their SF 
processes sped up.

Finally, we propose that the lower O abundances and the higher Z$_{*}$ fractions of regions in perturbed galaxies are the result of metal-poor gas inflows due to 
interactions with close companions.

\subsubsection{Deviations from the $\mu$-Z relation}
\label{subsubsec:f2} 

\begin{figure}\centering
   \mbox{\includegraphics[width=1\columnwidth]{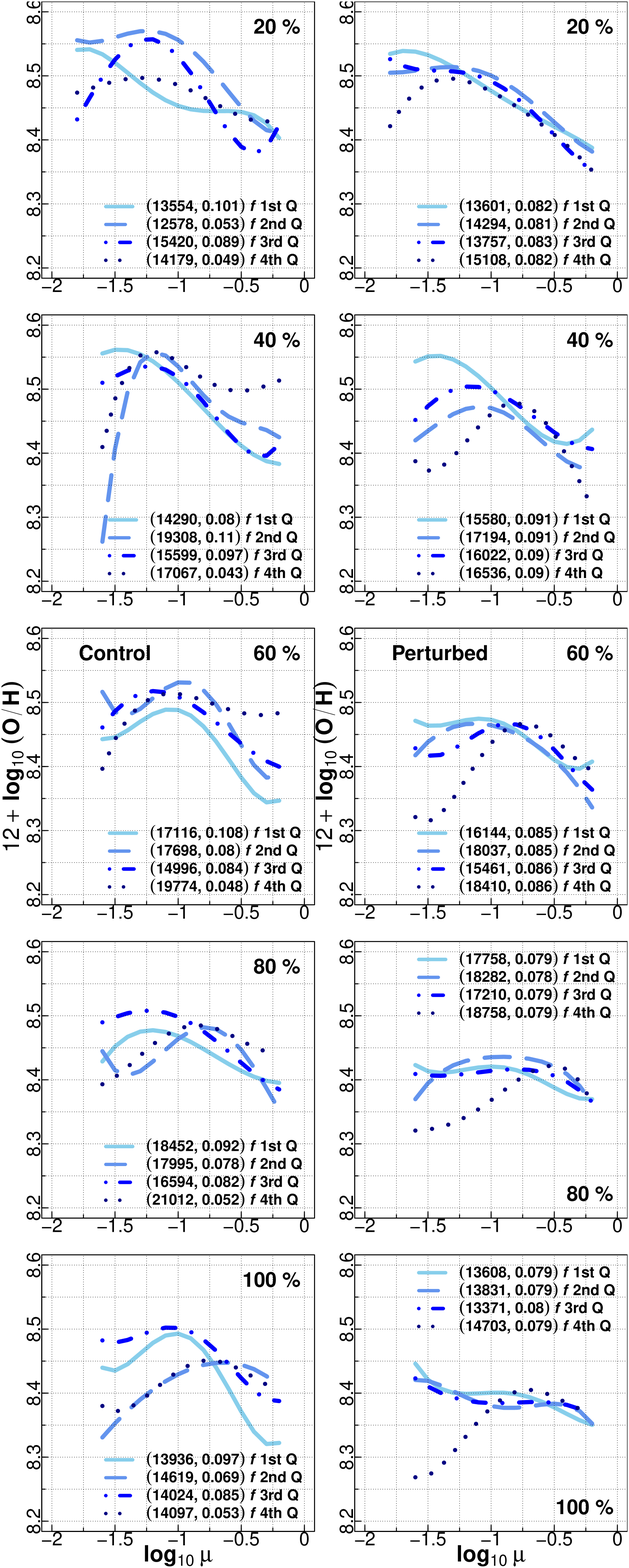}}
   \caption{\scriptsize{Fourth-order fits in the $\mu$-Z plane. The star-forming regions of all perturbed samples form pairs with those of the control sample at 
   the closest $\Sigma_{*}$ values. The fits per annulus correspond to the regions arranged in four groups of their respective tidal perturbation parameters 
   (\textit{f}, see~\ref{subsubsec:f}). The groups are defined from the Q distributions: $f$ 1st Q (min\,$\leq\,f\,<$\,1st Q, solid line), $f$ 2nd Q (1st 
   Q\,$\leq\,f\,<$\,2nd Q, dashed line), $f$ 3rd Q (2nd Q\,$\leq\,f\,<$\,3rd Q, dashed-dotted line) and $f$ 4th Q (3rd Q\,$\leq\,f\,\leq$\,max, dotted line). 
   Totals of star-forming regions used and standard deviations (sds) of the $12+\mathrm{log_{10}\,(O/H)}$ distributions within brackets.}}
   \label{f10} 
\end{figure}

From intermediate to the lowest gas fractions, the O abundances of the star-forming regions in our sampled galaxies decrease, deviating from the form of the 
$\mu$-Z relation (see Fig.~\ref{f8}). Since this deviation is more perceptible for perturbed galaxies, we explore a possible dependence on tidal interactions. 
Figure~\ref{f10} shows fourth-order fits for the annular star-forming regions in the $\mu$-Z plane, split up into four \textit{f} groups (the Qs of the distributions 
of the corresponding \textit{f} values). In general, the fitted functions gradually deviate from the accepted form of the relation from the lowest to the highest 
\textit{f} values (\textit{i.e.} from the \textit{f} 1st Q to the \textit{f} 4th Q group). This deviation is not a consequence of the differences between the totals 
of star-forming regions used (first numbers within brackets). For each annulus, we compare the median of all differences (that can be computed among the 4 totals, 
\textit{i.e.} 6) to the median of the totals. The percentages of the ratios of both medians do not exceed 9\,\%. Another factor that might cause these deviations 
from the accepted form is the dispersion of the data. To revise this, we add the sds of the O-content sets used to estimate each fit (second numbers within brackets). 
Notice that the sds do not increase with increasing Q of the \textit{f} distributions. For control galaxies in 4/5 annuli, the \textit{f} 1st Q group has the highest 
sds whereas the \textit{f} 2nd and 4th Q groups have the lowest ones. For perturbed galaxies, the dispersion is constant. Once the previous is clear, notice, for 
both control and perturbed cases, that the fits of the \textit{f} 2nd and 4th Q groups are the ones that deviate the most in 3/5 annuli. Comparing both group fits, 
that of the \textit{f} 4th Q is the one that deviates the most in 3/5 and 4/5 annuli for control and perturbed galaxies respectively. This fit, with the exclusion 
of the 20\,\% annulus, suggests a decrement rather than an increment of the O content with decreasing fraction of gas. Figure~\ref{f10} hence shows that the form 
of the $\mu$-Z relation depends on the global measure of tidal perturbation.

Finally, the star-forming regions in our sampled galaxies do not seem to deviate from the $\Sigma_{*}$-Z relation (Fig.~\ref{f7}). The reason of this is that the 
main gas-phase metal tracer is $\Sigma_{*}$. This agrees the O content as intrinsically related to massive SF. Since the $\Sigma_{*}$ defines $\mu$ (Section 
\ref{subsec:comp_che_abun}), the $\Sigma_{*}$-Z relation is not a resultant or consequential proportionality as the $\mu$-Z relation is (BB-18).

\section{Summary and conclusions}
\label{sec:conc}

PaperI shows the effects of tidal interactions on SF by comparing SP properties of non-tidally and tidally perturbed CALIFA survey galaxies. To support such 
interactions as modulators of SF, we compare this time the $\Sigma\mathrm{_{SFR}}$ and sSFR histories and look-back time annular profiles between such galaxies. 
We further compare their O contents as function of both stellar mass density and gas fraction to also show the interactions as drivers of gas inflows. In summary: 

\begin{enumerate}
 \item From the annular SFHs and sSFHs, a (re)activation (rejuvenation) of SF characterizes only the two outmost annuli within 
 the period where 0.001$<$\,1$-\eta_{*}(t_{j,s})\,\leq\,$0.01. The Differences in rates, most of them statistically significant, indicate higher 
 values for regions residing in perturbed galaxies almost regardless of the subsample division (excluding only SFG Red types). These Differences tend to increase 
 with decreasing look-back time (Figs.~\ref{f4} and \ref{fA1}).
 \item The median SF histories of regions in perturbed galaxies currently hosting the maxima of the SF are not clearly enhanced 
 by the influence of close companions. Not at least due to the sSFHs shown by control galaxies (Fig.~\ref{f5}). This may be due to either: 1) somehow 
 higher amounts of conveyed gas or higher SF efficiencies along particular time periods for particular galaxy types in the control sample, or 2) 
 the fact that high-speed encounters prevail on perturbed galaxies and their respective close companions.
 \item All along the look-back time, the SF levels of regions within control galaxies are never superior to those of regions within perturbed 
 galaxies (Figs.~\ref{f6.1} and \ref{f6.2}). The shapes of the look-back time profiles vary, mainly for the $\Sigma\mathrm{_{SFR}}$. 
 This reflects fluctuations from suppression to (re)activation of SF and vice versa. 
 \item As function of both $\Sigma_{*}$ and $\mu$, the O abundances are slightly higher for star-forming regions residing in control galaxies 
 (Figs.~\ref{f7} and \ref{f8}). The annuli in which we find the $\Sigma_{*}$-Z relation stronger than the $\mu$-Z one (\textit{i.e.} lower 
 median scatters along the respective fits) are 3/5 for control and 1/5 for perturbed galaxies.
 \item The $\mu$ distributions are biased towards larger values for regions in control galaxies (excluding the central annulus). Their 
 $12+\mathrm{log_{10}\,(O/H)}$ distributions also characterize these regions by higher O abundances. Since a strong inverse correlation 
 between $\mu$ and Z is of general acceptance, it is interesting that the annular lower abundances for perturbed galaxies hold no matter 
 their lower annular gas fractions (Fig.~\ref{f9}). The rate-differential distributions ($\mathrm{\dot{m}_{inflow}}-\mathrm{\dot{m}_{outflow}}$) 
 show higher medians and higher IQRs for perturbed galaxies in 4/5 and 3/5 annuli respectively. The Z$_{*}$ distributions show higher 
 fractions for perturbed galaxies. We then discard that the lower gas-phase and higher stellar abundances for perturbed galaxies are 
 due to more intense stellar feedback. Metal-poor gas inflows due to the interactions with close companions are more likely the cause.
 \item The form of the $\mu$-Z relation is found dependent on the tidal perturbation parameter as the fits derived from the former deviate 
 with increasing \textit{f} (Fig.~\ref{f10}). No deviation from the form of the $\Sigma_{*}$-Z is shown instead. Tidal perturbations can 
 hence affect an empirical relation based on a secondary property such as the gas fraction but not easily affect a dependence on stellar 
 mass, the main driver of SF. 
\end{enumerate}

On the one hand, despite their flattening in the plane of the Star Formation Main Sequence, regions in perturbed galaxies show no reduction of SFR 
properties when compared with regions in control galaxies (PaperI). Also, through time, the former show slightly higher SF levels, signature of the 
influence of their close companions. On the other hand, gas inflows suggest to dilute the gas-phase and to increase the stellar metallicities 
along the annular sequence in perturbed galaxies. The metallicity of their star-forming regions is determined by their local growth of stellar 
mass and also by gas accretion consequence of tidal interactions. 

An analysis of the ionized gas and stellar kinematics is underway to complement this series.

\section*{Acknowledgements}
\footnotesize{
The authors wish to thank the anonymous Referee for her/his comments and suggestions that improved this work. A. Morales-Vargas thanks Assistant Editor 
Bella Lock for her kindness. J. P. Torres-Papaqui and A. Morales-Vargas thank DAIP-UGto for granted support (0173/2019). 

All figures for this paper were possible by the use of \textit{R: A language and environment for statistical computing}\footnote{https://www.R-project.org/}. 

The \textsc{starlight}\footnote{http://www.starlight.ufsc.br/} project is supported by the Brazilian agencies CNPq, CAPES and FAPESP and by the 
France-Brazil CAPES/Cofecub program.

Funding for the SDSS and SDSS-II has been provided by the Alfred P. Sloan Foundation, the Participating Institutions, the National Science Foundation, 
the U.S. Department of Energy, the National Aeronautics and Space Administration, the Japanese Monbukagakusho, the Max Planck Society, and the Higher 
Education Funding Council for England. The SDSS Web Site is http://www.sdss.org/. The SDSS is managed by the Astrophysical Research Consortium for the 
Participating Institutions. The Participating Institutions are the American Museum of Natural History, Astrophysical Institute Potsdam, University of 
Basel, University of Cambridge, Case Western Reserve University, University of Chicago, Drexel University, Fermilab, the Institute for Advanced Study, 
the Japan Participation Group, Johns Hopkins University, the Joint Institute for Nuclear Astrophysics, the Kavli Institute for Particle Astrophysics 
and Cosmology, the Korean Scientist Group, the Chinese Academy of Sciences (LAMOST), Los Alamos National Laboratory, the Max-Planck-Institute for 
Astronomy (MPIA), the Max-Planck-Institute for Astrophysics (MPA), New Mexico State University, Ohio State University, University of Pittsburgh, 
University of Portsmouth, Princeton University, the United States Naval Observatory, and the University of Washington.

This study uses data provided by the Calar Alto Legacy Integral Field Area (CALIFA) survey (http://califa.caha.es/) and it is therefore based on observations 
collected at the Centro Astron\'{o}mico Hispano Alem\'{a}n (CAHA) at Calar Alto, operated jointly by the Max-Planck-Institut f\"{u}r Astronomie and the Instituto 
de Astrof\'{i}sica de Andaluc\'{i}a (CSIC). The CALIFA survey Collaboration\footnote{http://califa.caha.es/?q=content/structure-collaboration} would like to 
thank both as major partners of the observatory and CAHA itself for the unique access to telescope time and support in manpower and infrastructures. The CALIFA 
survey Collaboration thanks the CAHA staff for the dedication to the project.

Para ti Mateo.}

%%%%%%%%%%%%%%%%%%%%%%%%%%%%%%%%%%%%%%%%%%%%%%%%%%

\section*{DATA AVAILABILITY}
The data underlying this article will be shared on reasonable request to the corresponding author.


%\bibliographystyle{mnras}


%%%%%%%%%%%%%%%%%%%%%%%%%%%%%%%%%%%%%%%%%%%%%%%%%%

% Don't change these lines
\bsp	% typesetting comment
% End of mnras_template.tex
%%%%%%%%%%%%%%%%% APPENDICES %%%%%%%%%%%%%%%%%%%%%
\appendix

\section{Annular sSFHs}
\label{sec:app1}

Annular sSFHs (Fig.~\ref{fA1}), see Section~\ref{subsec:sfh_ssfr}.

\begin{figure*}\centering
   \mbox{\includegraphics[width=.5066549\columnwidth]{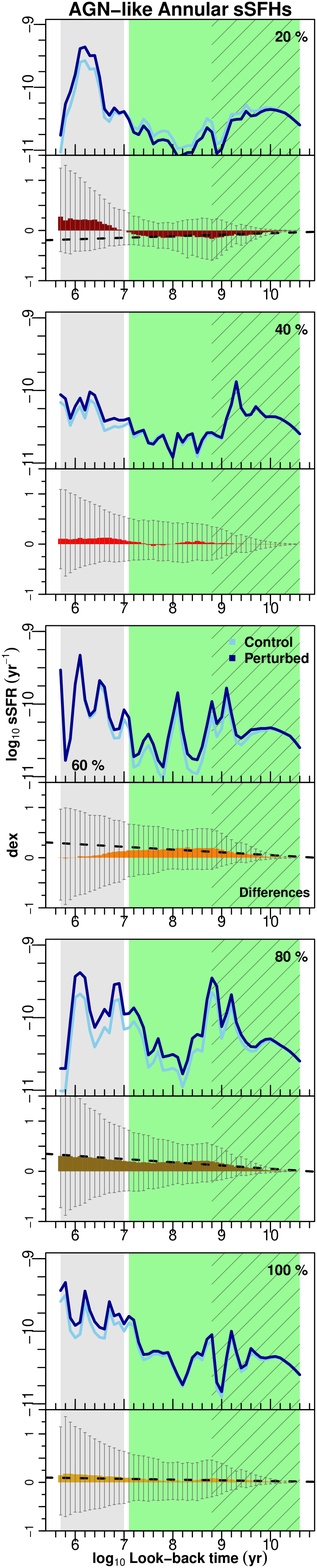}}
   \mbox{\includegraphics[width=.5066549\columnwidth]{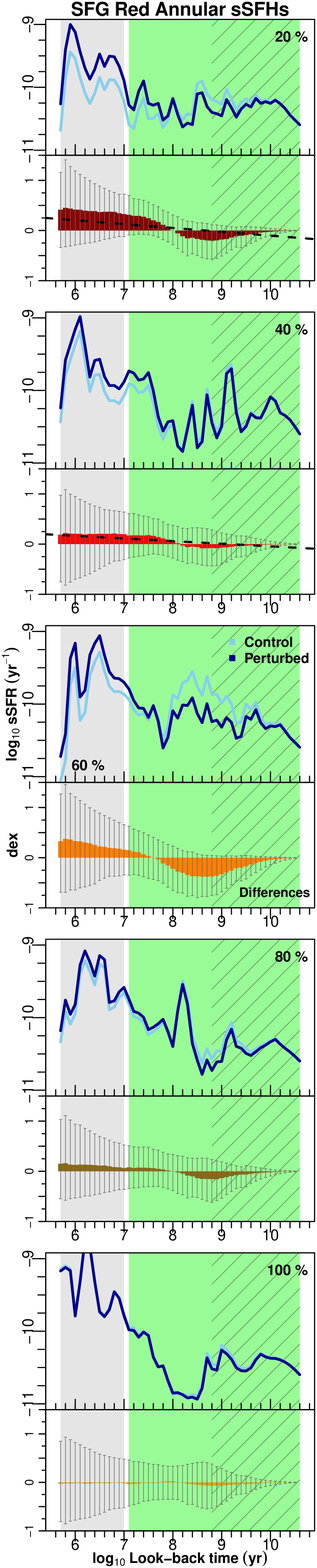}}
   \mbox{\includegraphics[width=.5066549\columnwidth]{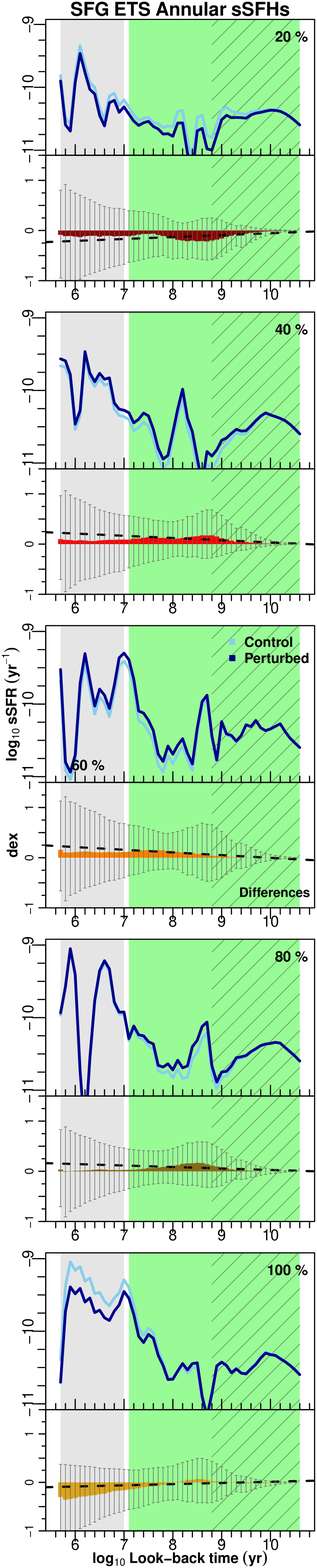}}
   \mbox{\includegraphics[width=.5066549\columnwidth]{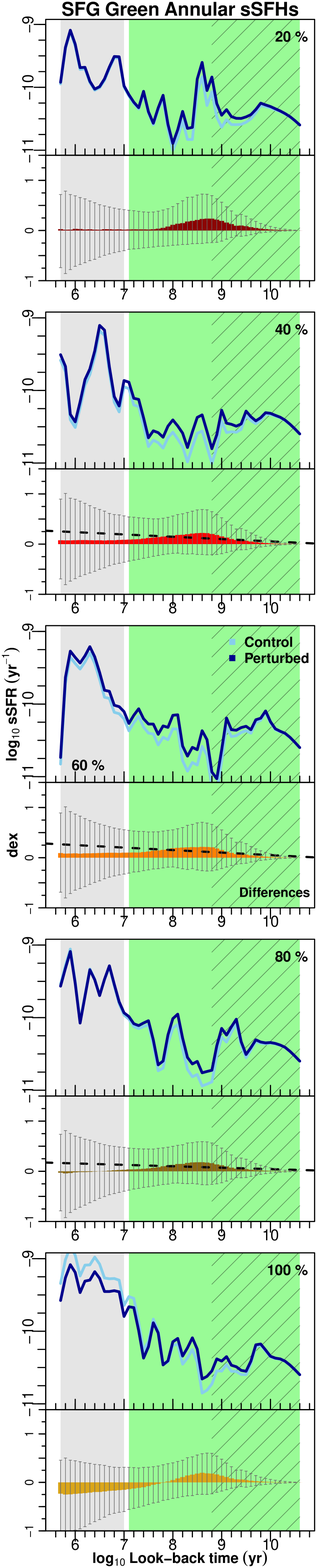}}
\caption{\scriptsize{Annular sSFHs for control and perturbed galaxies. For fair comparisons, we keep the subsample division of PaperI \textit{i.e.}, AGN-like, 
SFG Red, SFG ETS and SFG Green from left to right. Five consecutive-outward annuli (``20'', ``40'', ``60'', ``80'' and ``100\,\%'') related to the 
H$\alpha$ flux of each galaxy as the radial extension (see~\ref{subsubsec:sf}). These panels result from pairing star-forming spaxels in all 
perturbed samples to those ones in the control sample by minimizing their differences in current $\Sigma_{*}$. The ``Differences'' in 
the history records (by subtracting the control values from the perturbed ones, secondary plots) depict the medians (bar heights) of the distributions 
of differences and their interquartile ranges (IQRs, 1st to 3rd, bar lines). The dashed lines are linear regression model fits, with statistically 
significant slopes (Pr($>$F)\,$<$\,statistical level), on the Differences within the green background only (the Differences constitute 
the response variable). The histories (main plots) are both control (light blue) and perturbed (dark blue) values corresponding to each $t_{j,s}$ bin median 
Difference. The proposed timescale for a valid-current SFR ($\sim$10\,Myr, light gray, see \ref{subsubsec:synVSHa}) has been disregarded since it matches 
the range characterized by the least accurate values of the histories (\textit{e.g.} compare the extensions of the bar lines between the light gray and green 
backgrounds). In contrast, the green and shaded backgrounds indicate the time ranges, starting at present, at which the more significant mass fractions 
($>$\,0.001 and $>$\,0.01 respectively) were assembled.}}
   \label{fA1} 
\end{figure*}

\begin{figure*}\centering
\ContinuedFloat   
   \mbox{\includegraphics[width=.5066549\columnwidth]{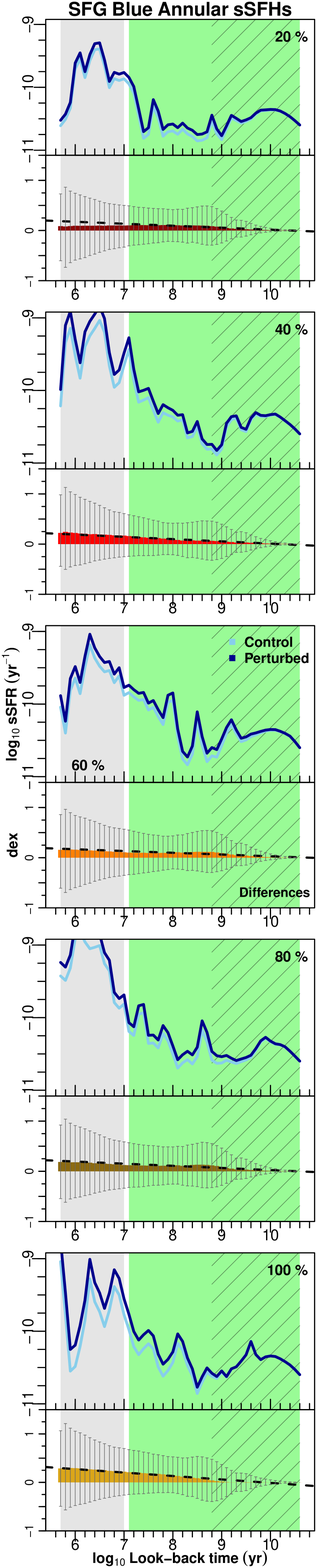}}
   \mbox{\includegraphics[width=.5066549\columnwidth]{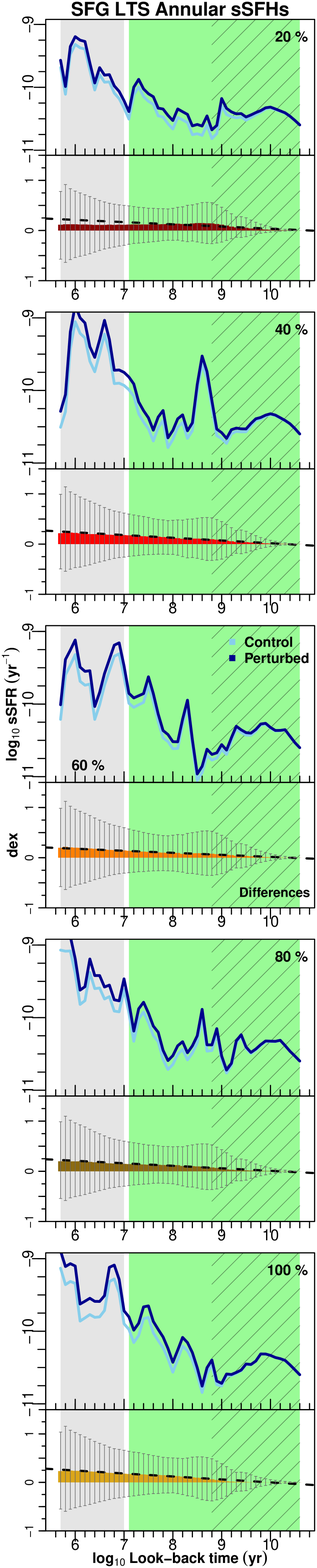}}
   \mbox{\includegraphics[width=.5066549\columnwidth]{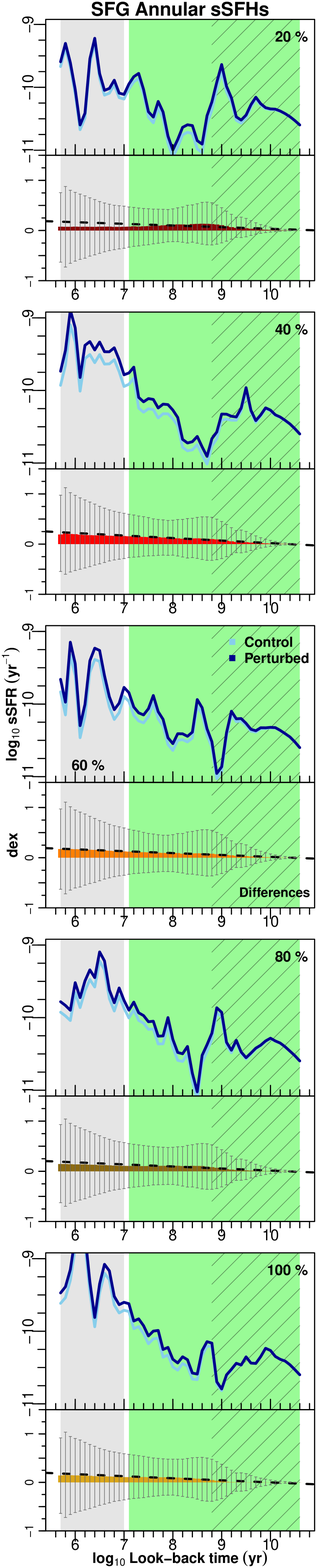}}
   \mbox{\includegraphics[width=.5066549\columnwidth]{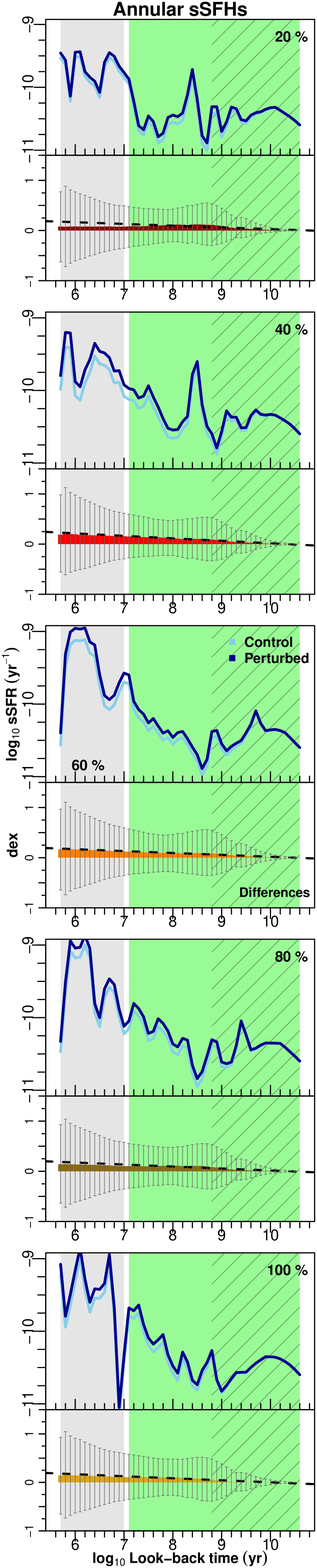}}
\caption{\scriptsize{Annular sSFHs for control and perturbed galaxies (cont.). Same caption as above but for SFG Blue, SFG LTS, SFG and all subsamples.}}
   \label{fA1} 
\end{figure*}

\section{Distributions of differences of the SFHs}
\label{sec:app2}

See Table~\ref{tab:B1} and Section~\ref{subsec:sfh_ssfr}.

\onecolumn
\scriptsize{
\setlength{\tabcolsep}{0.55\tabcolsep}
\begin{longtable}{lcccccccccccccccccccc}%[h!]
 \caption{\scriptsize{Summary of the distributions of differences between the annular SFHs of control and perturbed galaxies. The control values are subtracted from the perturbed ones (see the secondary 
 plots of Fig.~\ref{f4}). ``Q1'' and ``Q3'' (in dex) are the 1st and 3rd quartiles (ends of bar lines) whilst ``Q2'' (also in dex) is the median (``Differences'', bar heights). ``ext.'' represents the fractional 
 extension of the IQRs (extension of bar lines) above zero dex. Per annulus, three rows of data (bold font, see also Table~\ref{tab:sum_of_summaryf4}) indicate each, the time bin that locates the 1st, 2nd (median) and 3rd Qs of 
 ``Q2'', \textit{i.e.}, the IQRs of the ``Differences'' along the look-back time. For this are only considered time ranges (log$_{10}$\,t\,$>$\,7 yr), starting at present, at which the more significant 
 mass fractions were assembled ($>$\,0.001 and $>$\,0.01, green and shaded backgrounds respectively, see Fig.~\ref{f4}).}}
 \label{tab:B1}\\                 
\hline\\
$\mathrm{log_{10}}$&                                                                      \multicolumn{20}{c}{Annuli (H$\alpha$ flux percentages)}                                                              \\
t                  &      \multicolumn{4}{c}{20\,\%}      &      \multicolumn{4}{c}{40\,\%}      &      \multicolumn{4}{c}{60\,\%}      &     \multicolumn{4}{c}{80\,\%}    &     \multicolumn{4}{c}{100\,\%}   \\                                                     
(yr)               &Q1      &Q2        &Q3      &ext.     &Q1      &Q2        &Q3      &ext.     &Q1      &Q2        &Q3      &ext.     &Q1      &Q2        &Q3   &ext.     &Q1      &Q2        &Q3   &ext.     \\[1ex]
\hline\\
\endhead
                   &                                                                                   \multicolumn{20}{c}{AGN-like}                                                                            \\                  
5.7                &$-$0.410&0.295     &1.264   &0.755    &$-$0.507&0.101     &1.081   &0.681    &$-$0.847&$-$0.004  &0.969   &0.534    &$-$0.731&0.307     &1.575&0.683    &$-$0.689&0.147     &1.129&0.621    \\
5.8                &$-$0.446&0.203     &1.363   &0.753    &$-$0.686&0.097     &1.080   &0.611    &$-$0.933&$-$0.018  &0.998   &0.517    &$-$0.884&0.306     &1.726&0.661    &$-$0.704&0.184     &1.354&0.658    \\
5.9                &$-$0.464&0.218     &1.271   &0.733    &$-$0.574&0.090     &1.015   &0.639    &$-$0.841&$-$0.009  &0.992   &0.541    &$-$0.766&0.302     &1.570&0.672    &$-$0.652&0.180     &1.238&0.655    \\  
6                  &$-$0.409&0.221     &1.158   &0.739    &$-$0.511&0.104     &0.975   &0.656    &$-$0.769&$-$0.005  &0.925   &0.546    &$-$0.713&0.272     &1.459&0.672    &$-$0.621&0.164     &1.151&0.650    \\
6.1                &$-$0.408&0.205     &1.098   &0.729    &$-$0.456&0.120     &0.930   &0.671    &$-$0.747&0.025     &0.928   &0.554    &$-$0.690&0.298     &1.430&0.675    &$-$0.623&0.166     &1.094&0.637    \\
6.2                &$-$0.393&0.231     &0.998   &0.718    &$-$0.431&0.110     &0.881   &0.672    &$-$0.731&0.019     &0.888   &0.549    &$-$0.632&0.273     &1.338&0.679    &$-$0.596&0.160     &1.014&0.630    \\  
6.3                &$-$0.394&0.209     &0.943   &0.705    &$-$0.402&0.104     &0.843   &0.677    &$-$0.675&0.026     &0.855   &0.559    &$-$0.564&0.275     &1.263&0.691    &$-$0.572&0.149     &0.954&0.625    \\
\end{longtable}
}              
\twocolumn     
\normalsize{   

\section{Look-back time annular profiles (cont.)}
\label{sec:app3}

Figure~\ref{fC1} shows the $\mathrm{\Sigma_{SFR}}$ and sSFR look-back time annular profiles within the period we consider the current 
SFR as valid (t\,$\leq\,\sim$10 Myr, see Section~\ref{subsubsec:synVSHa}). These profiles should, however, be taken with care since the uncertainties 
associated to the fractions of M$_{*}$ formed in t\,$\leq\,\sim$10 Myr are significant. Consequently, any trend that the SFH and the sSFH may 
show are the least accurate and subject to misinterpretations (see Section~\ref{subsec:sfh_ssfr}).

\newpage
\begin{figure*}\centering
   \mbox{\includegraphics[width=1.975\columnwidth]{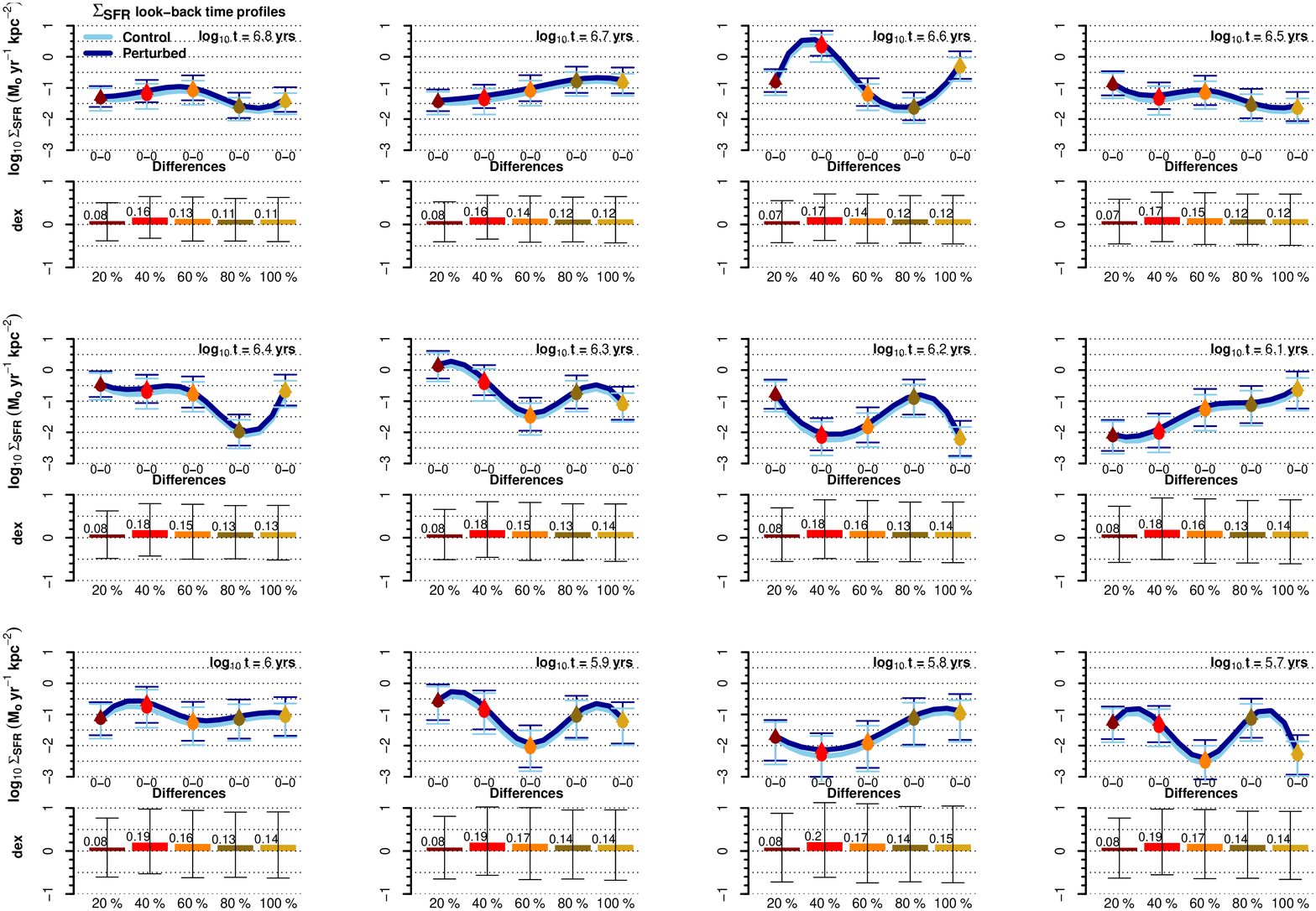}}
   \mbox{\includegraphics[width=1.975\columnwidth]{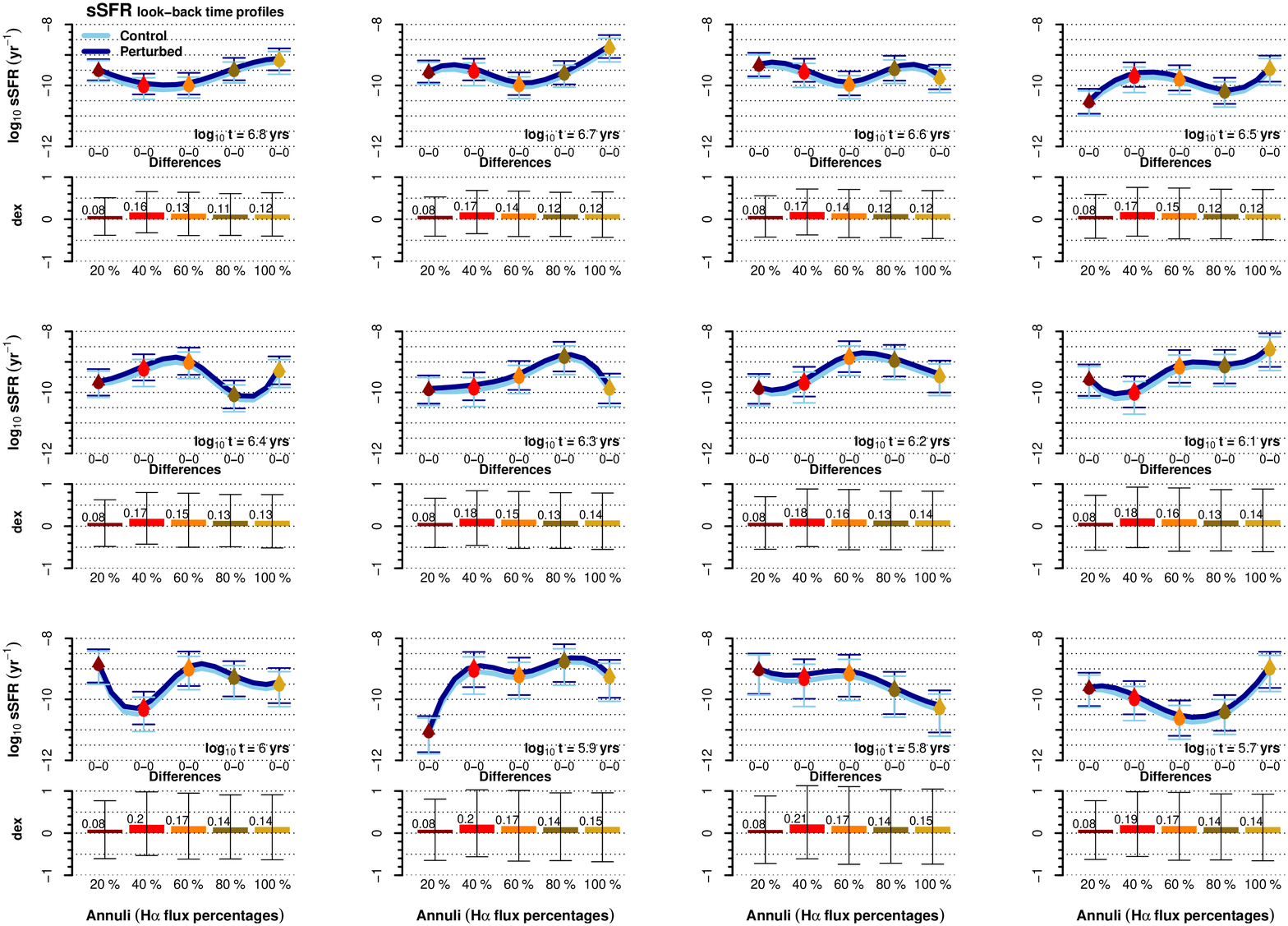}}
 \caption{\scriptsize{The $\mathrm{\Sigma_{SFR}}$ and sSFR look-back time annular profiles within the period for a valid-current SFR 
 (t\,$\leq\,\sim$10 Myr). The annular extension is as same as in Figs. of Section~\ref{subsec:lbtap}. The histories in these profiles correspond to 
 star-forming regions paired at their closest current $\Sigma_{*}$ values. The ``Differences'' (bar plots) are the medians of the annular differences 
 by subtracting the control values from the perturbed ones. Symbols are control and perturbed values corresponding to each Difference. Bar and symbol 
 lines are the interquartile ranges (IQRs, 1st to 3rd) of the respective distributions. Find AD and permutation test results for all annular pairs of 
 sample distributions (likelihoods right below the profiles).}}
 \label{fC1} 
\end{figure*}

%%%%%%%%%%%%%%%%%%%%%%%%%%%%%%%%%%%%%%%%%%%%%%%%%%

\label{lastpage}
\end{document}